\documentclass[10pt,twocolumn,letterpaper,capitalize]{article}

\usepackage[pagenumbers]{cvpr}

\usepackage{booktabs}
\usepackage{tikz}\usetikzlibrary{arrows.meta} %

\usepackage{cuted}

\usepackage{currfile}

\usepackage{caption}

\newif\ifpendingdraft
\pendingdrafttrue
\DeclareRobustCommand{\pending}[1]{\ifpendingdraft{\color{red}#1}\ifinner\else\ifmmode\else\marginpar{\tiny\color{red}PENDING}\fi\fi\else#1\fi}

\newcommand{\code}[1]{\texttt{#1}}

\AtEndPreamble{\AtBeginDocument{%
  \crefname{section}{\S}{\S\S}\Crefname{section}{\S}{\S\S}%
  \crefname{subsection}{\S}{\S\S}\Crefname{subsection}{\S}{\S\S}%
  \crefname{subsubsection}{\S}{\S\S}\Crefname{subsubsection}{\S}{\S\S}%
  \crefformat{section}{\S#2#1#3}\Crefformat{section}{\S#2#1#3}%
  \crefformat{subsection}{\S#2#1#3}\Crefformat{subsection}{\S#2#1#3}%
  \crefformat{subsubsection}{\S#2#1#3}\Crefformat{subsubsection}{\S#2#1#3}%
  \crefrangeformat{section}{\S\S#3#1#4--#5#2#6}\Crefrangeformat{section}{\S\S#3#1#4--#5#2#6}%
  \crefrangeformat{subsection}{\S\S#3#1#4--#5#2#6}\Crefrangeformat{subsection}{\S\S#3#1#4--#5#2#6}%
  \crefmultiformat{section}{\S\S#2#1#3}{ and~#2#1#3}{, #2#1#3}{ and~#2#1#3}%
  \Crefmultiformat{section}{\S\S#2#1#3}{ and~#2#1#3}{, #2#1#3}{ and~#2#1#3}%
  \crefmultiformat{subsection}{\S\S#2#1#3}{ and~#2#1#3}{, #2#1#3}{ and~#2#1#3}%
  \Crefmultiformat{subsection}{\S\S#2#1#3}{ and~#2#1#3}{, #2#1#3}{ and~#2#1#3}%
  \crefrangeformat{subsubsection}{\S\S#3#1#4--#5#2#6}\Crefrangeformat{subsubsection}{\S\S#3#1#4--#5#2#6}%
  \crefmultiformat{subsubsection}{\S\S#2#1#3}{ and~#2#1#3}{, #2#1#3}{ and~#2#1#3}%
  \Crefmultiformat{subsubsection}{\S\S#2#1#3}{ and~#2#1#3}{, #2#1#3}{ and~#2#1#3}%
}}
\usepackage{stfloats} %

\pendingdraftfalse
\usepackage{enumitem}
\setlist{nosep,leftmargin=*}

\definecolor{cvprblue}{rgb}{0.21,0.49,0.74}
\usepackage[pagebackref,breaklinks,colorlinks,allcolors=cvprblue]{hyperref}

\def\confName{CVPR}
\def\confYear{2027}

\title{4D-TEMPO: Differentiable 4D-STEM ptychography via fast Physics Optimizers}

\author{Ramani Duraiswami\\
Computer Science and UMIACS\\
University of Maryland, College Park\\
{\tt\small ramanid@umd.edu}
\and
Yichao Zhang\\
Materials Science and Engineering\\
University of Maryland, College Park\\
{\tt\small zhangych@umd.edu}
}
\AtBeginDocument{%
  \crefname{equation}{Eq.}{Eqs.}\Crefname{equation}{Eq.}{Eqs.}%
  \crefname{figure}{Fig.}{Figs.}\Crefname{figure}{Fig.}{Figs.}%
  \crefname{table}{Tab.}{Tabs.}\Crefname{table}{Tab.}{Tabs.}}

\begin{document}
\twocolumn[{%
\renewcommand\twocolumn[1][]{#1}%
\maketitle
\begin{center}
  \includegraphics[width=\textwidth]{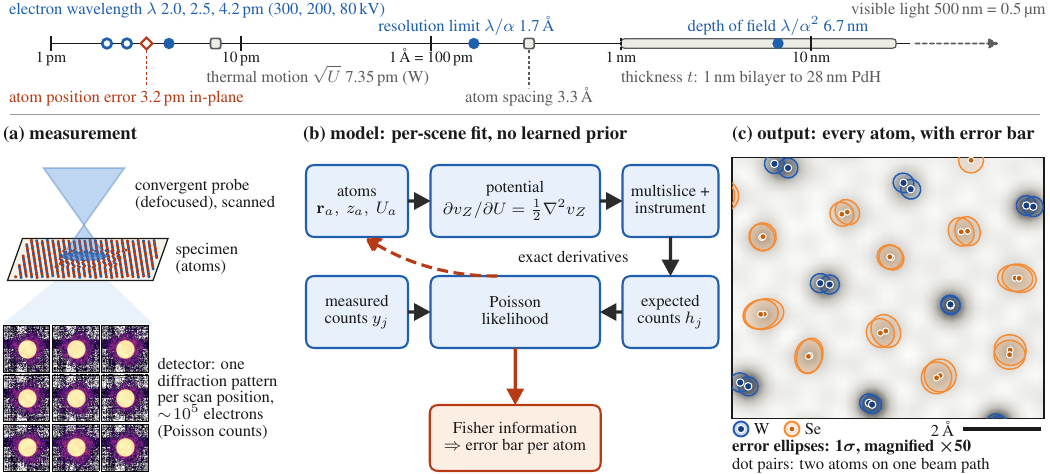}
  \captionof{figure}{4D-TEMPO fits every atom and the microscope directly to electron counts and returns each atom with an error bar.  \textbf{(a)}~A defocused electron probe scans the specimen; the detector records Poisson counts at every position (4D-STEM). \textbf{(b)}~A physical multislice renderer maps the atoms (position $\mathbf r_a, z_a$; thermal width $U_a$) and the instrument to expected counts $h_j$; the Poisson likelihood compares them with the measured counts $y_j$, and the Fisher information gives the error bars. \textbf{(c)}~Real twisted WSe$_2$ bilayer: every fitted atom (dot) on the reconstructed phase, with its $1\sigma$ shot-noise (Fisher) position ellipse (39\% joint coverage) magnified $\boldsymbol{50\times}$; medians 0.5\,pm (W) and 0.8\,pm (Se). Dot pairs are two atoms on one beam path (the two Se of a column, or atoms of the two layers), each with its own ellipse. %
  \textbf{Top:} length scales (log axis; 1\,\AA\ = 100\,pm): electron optics at 25\,mrad (blue; filled: 80\,kV, these data), specimen (gray), and our accuracy against truth on simulated data (red); model error exceeds the shot-noise ellipses of (c) (\protect\cref{sec:bounds}).}
  \label{fig:teaser}
\end{center}
\vspace{0.5em}
}]
\enlargethispage{2\baselineskip}\vspace*{-12pt}\begin{abstract}
Electron ptychography recovers atomic structure from 4D-STEM data: diffraction patterns at multiple positions of a scanned probe. Current codes reconstruct an image and read atoms off it, or fit atoms under an $L_1$ loss; neither gives per-atom uncertainty. We propose 4D-TEMPO, a differentiable physical renderer that fits the atoms, the microscope, and the detector directly to the recorded electron counts under a Poisson likelihood. The output is a list of atoms, each a Gaussian primitive with a 3D position, a thermal width, and a model-conditional error bar from the Fisher information. The Fisher information also conditions the solver, gives the dose (electrons per \AA$^2$) each parameter needs, and separates calibration from imaging. On a simulated WSe$_2$ bilayer, starting from the published reconstruction, fitted atoms cut in-plane error by a third (3.2 vs.\ 4.8\,pm) and depth error by 40--61\% against its 3D peaks. Depth comes from multiple scattering into the dark field, the faint counts outside the direct beam. The two layers, $10\times$ closer than the depth of field, separate; per-atom depth then follows the layer geometry. When the model is well specified, the bars are calibrated (69\% of position coordinates within $1\sigma$). On experimental WSe$_2$, fitted atoms predict held-out patterns better than the three published codes we test. With free probe modes where atoms are incomplete, the same likelihood predicts best on all three held-out benchmarks, matches every WSe$_2$ baseline in 5\,min against up to 20.5\,h, and passes \code{torchslice} at its released setting 180$\times$ sooner on PrScO$_3$. Per-atom error bars, planned doses, and fast reconstruction point to low-dose, real-time 3D imaging of beam-sensitive and dynamic materials.
\end{abstract}
\vspace*{-6pt}
\vspace*{-6pt}\section{Introduction}
\label{sec:intro}
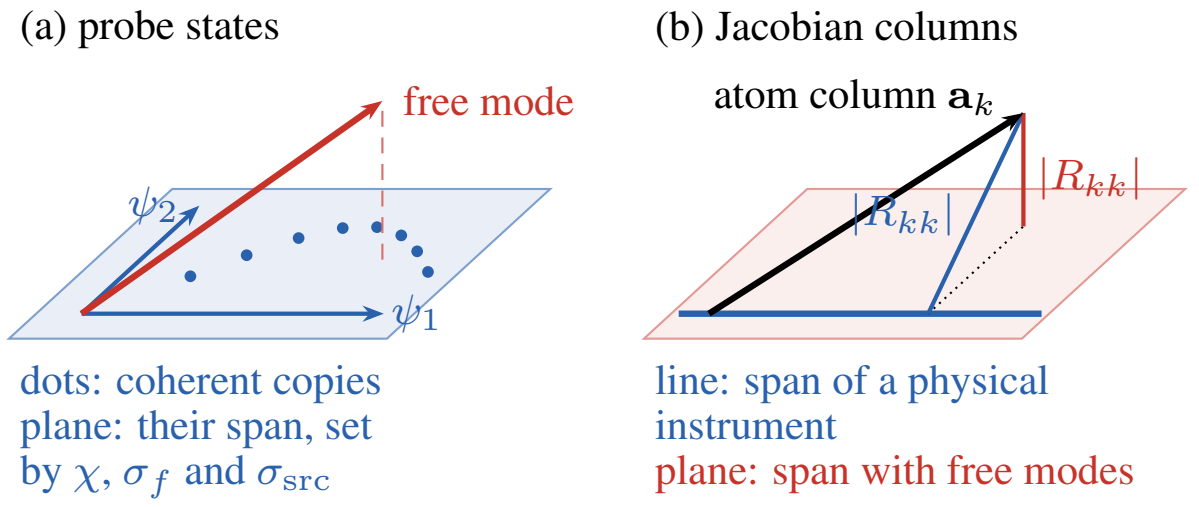
\begin{figure}[hbt]
  \centering
  \begin{tikzpicture}[x={(1cm,0cm)},y={(0.6cm,0.55cm)},z={(0cm,1cm)},
    font=\footnotesize, >={Stealth[length=4pt]}]
  \definecolor{phys}{rgb}{0.16,0.38,0.68}
  \definecolor{free}{rgb}{0.78,0.20,0.16}
  \begin{scope}
    \fill[phys!10] (0,0,0) -- (2.5,0,0) -- (2.5,1.8,0) -- (0,1.8,0) -- cycle;
    \draw[phys!60] (0,0,0) -- (2.5,0,0) -- (2.5,1.8,0) -- (0,1.8,0) -- cycle;
    \coordinate (O) at (0.3,0.3,0);
    \foreach \t in {0,...,7}{
      \pgfmathsetmacro{\xx}{0.75+0.22*\t}
      \pgfmathsetmacro{\yy}{0.75+0.6*sin(25*\t)}
      \fill[phys] (\xx,\yy,0) circle (1.1pt);
    }
    \draw[->,thick,phys] (O) -- (2.3,0.3,0) node[right,inner sep=1pt] {$\psi_1$};
    \draw[->,thick,phys] (O) -- (0.3,1.6,0) node[left] {$\psi_2$};
    \draw[densely dashed,free!70] (1.9,0.95,0) -- (1.9,0.95,1.05);
    \draw[->,very thick,free] (O) -- (1.9,0.95,1.05) node[right] {free mode};
    \node[anchor=south west] at (-0.05,0,1.8) {(a) probe states};
    \node[phys,anchor=north west,align=left] at (-0.05,0,-0.05)
      {dots: coherent copies\\[-1pt]plane: their span, set\\[-1pt]by $\chi$, $\sigma_f$ and $\sigma_{\mathrm{src}}$};
  \end{scope}
  \begin{scope}[shift={(4.2,0,0)}]
    \fill[free!8] (0,0,0) -- (2.5,0,0) -- (2.5,1.8,0) -- (0,1.8,0) -- cycle;
    \draw[free!50] (0,0,0) -- (2.5,0,0) -- (2.5,1.8,0) -- (0,1.8,0) -- cycle;
    \draw[very thick,phys] (0.05,0.3,0) -- (2.45,0.3,0);
    \coordinate (O) at (0.25,0.3,0);
    \coordinate (P) at (1.7,1.35,0.75);
    \draw[densely dotted] (1.7,0.3,0) -- (1.7,1.35,0);
    \draw[phys,thick] (P) -- (1.7,0.3,0);
    \draw[free,thick] (P) -- (1.7,1.35,0);
    \draw[->,very thick] (O) -- (P) node[left,xshift=-1pt,yshift=3pt] {atom column $\mathbf a_k$};
    \node[phys,anchor=east,inner sep=1pt] at (1.62,0.8,0.38) {$|R_{kk}|$};
    \node[free,anchor=west,inner sep=1pt] at (1.75,1.35,0.32) {$|R_{kk}|$};
    \node[anchor=south west] at (-0.05,0,1.8) {(b) Jacobian columns};
    \node[anchor=north west,align=left] at (-0.05,0,-0.05)
      {\textcolor{phys}{line: span of a physical}\\[-1pt]\textcolor{phys}{instrument}\\[-1pt]\textcolor{free}{plane: span with free modes}};
  \end{scope}
\end{tikzpicture}
  \caption{{\bfseries\boldmath Free probe modes make each atom less determined.} (a)~A physical microscope makes probes only from a small family set by its aberrations $\chi$ and coherence ($\sigma_f$, $\sigma_{\mathrm{src}}$): the dots, which lie in a plane (\protect\cref{eq:coherence}). A free probe mode can be any wave, so it can leave the plane. (b)~Each arrow is how the predicted counts change when one parameter moves, in units of the noise. Only the part of an atom's arrow $\mathbf a_k$ that no instrument change can imitate, its height $|R_{kk}|$ above the instrument's span, informs that atom and sets its error bar (\protect\cref{sec:analysis}). Free modes widen the span to the red plane, imitate more of the atom's signal, and shrink $|R_{kk}|$.}
  \label{fig:subspace}
\end{figure}

Can a microscope report every atom's position with an error bar (\cref{fig:teaser})? Electron ptychography comes closest. A focused electron probe scans a thin specimen, and a detector records a diffraction pattern of Poisson counts at every position: a four-dimensional scanning transmission electron microscopy (4D-STEM) dataset (\cref{fig:teaser}a). Current codes reconstruct a pixel image from these patterns and add freedom until the fit is good: free probe modes -- mutually incoherent wavefields of unconstrained pixels -- absorb what the physics misses, including part of the atoms' own signal (\cref{fig:subspace}). The fit improves, but it can no longer say how well it knows any one atom. \mbox{4D-TEMPO} puts {\em atoms before pixels:} it fits atoms, the microscope, and the detector directly to the counts, as crystallographic refinement fits a structure to diffraction data~\cite{murshudov1997refinement}. In vision terms, this is inverse rendering via ``analysis by synthesis'': a physical renderer -- the multislice model of electron scattering -- maps Gaussian atoms to expected patterns, much as NeRF fits a radiance field to photos~\cite{mildenhall2020nerf}. The differentiable renderer gives its Jacobian and the Fisher information (FI): an error bar for every atom, which parameters the data determine, and the dose an experiment needs, in advance (\cref{fig:teaser}c).

A 300\,kV electron has wavelength $\lambda=2$\,pm, but lens aberrations limit the beam to a cone of semi-angle $\alpha$ of tens of milliradians, setting the diffraction-limited resolution $\lambda/\alpha$ to 0.66\,\AA\ (1\,\AA\ = 100\,pm) at 30\,mrad. Electron ptychography shifts this limit from optics to computation, reaching 0.39\,\AA\ in a 2D material~\cite{jiang2018ptycho} and the thermal-vibration limit (20\,pm resolution)~\cite{chen2021latticevib}, recovering depth, and enabling low-dose imaging of frozen biological specimens~\cite{zhou2020low}. The dose, the number of electrons sent through each \AA$^2$ of specimen, is capped by the specimen itself: each electron can deposit energy that breaks bonds or knocks atoms out. Frozen biological specimens tolerate tens of electrons per \AA$^2$; in the WSe$_2$ bilayer studied here, selenium vacancies appear by about $5\times10^5$~\cite{huang2026moire3d}. Every electron must therefore count. 3D atom positions and vibrations reveal how twisted 2D crystal layers relax~\cite{huang2026moire3d}, how moir\'e phasons dominate atomic vibrations~\cite{zhang2025phasons}, how atoms vibrate under heating~\cite{bae2026thermal}, and where hydrogen sits in metals~\cite{shi2025pdh}. Each question demands quantitative numbers with error bars; beam-sensitive and changing specimens also demand low dose and reconstruction fast enough to follow them.

Three computational obstacles stand in the way. {\bf 1)}~A dataset contains up to $10^9$ measurements, and multiple scattering makes the renderer strongly nonlinear. {\bf 2)}~Microscope aberrations, coherence, and detector response are known only approximately and must be fitted with the specimen. {\bf 3)}~Existing codes reconstruct millions of pixels per slice with free probe modes~\cite{thibault2013reconstructing}, introduced for partial coherence but tied to neither source nor lens, so they absorb detector blur, drift, and model error with specimen signal.

Maximum-likelihood ptychography under Poisson or Gaussian noise~\cite{thibault2012maximum,odstrcil2018iterative} reconstructs these pixel grids and extracts atom positions and uncertainties post hoc, as in the published pipeline for twisted tungsten diselenide (WSe$_2$) bilayers~\cite{huang2026moire3d}. Yet atoms are far fewer than pixels, and an atom's potential is a known function of position, species, and thermal width. Diederichs \etal\ put atoms directly into the inversion, fitting atoms and a physical instrument to experimental 4D-STEM data via differentiable multislice under an $L_1$ loss~\cite{diederichs2024exact}. Meanwhile, the FI has designed STEM experiments for atom-column positions~\cite{vanaert2002optimal} and bounded 4D-STEM precision~\cite{dwyer2024fisher}. We take the next step: atoms, instrument, and pixels share one Poisson likelihood with a physical detector model. We add free modes only to test the likelihood and where atoms are incomplete.

\vspace*{-6pt}
\paragraph{Contributions and gains over the state of the art.}
\begin{itemize}[leftmargin=*,itemsep=1pt,topsep=2pt,parsep=0pt]
  \item \textbf{Atoms and instrument fitted to the counts.} Atoms, a physical instrument, and the detector share one Poisson likelihood, so atoms come out as numbers. On a simulated bilayer, starting from the published reconstruction and using no structural priors, fitted atoms cut in-plane error by a third (3.2 vs.\ 4.8\,pm) and depth error by 40--61\% against its 3D peaks (\cref{tab:accsim}). Multiple scattering into the dark field separates two layers $10\times$ closer than the depth of field. The solver is quasi-Newton in scaled units on a deterministic objective, and atom depth is continuous across slices (\cref{sec:model,sec:inference}).
  \item \textbf{Every atom with an error bar.} The FI gives each atom a shot-noise precision, calibrated in simulation (69\% of position coordinates within $1\sigma$, nominal 68\%), identifiability via pivoted QR, and the dose required per parameter (\cref{sec:bounds,sec:analysis}; cf.\ Laplace approximations of neural fields~\cite{goli2024bayesrays}). Calibrate once, image at low dose: a Schur complement separates calibration from imaging, cutting the parameters a low-dose scan must fit by $3\times$ (\cref{fig:calib}). On an independent simulator, planted 5--20\,pm displacements are recovered (median 1.6\,pm) with no false ones (\S S4).
  \item \textbf{The same model predicts best.} We score every code by held-out Poisson deviance per pixel, $D_H$ (lower is better). Atoms alone beat every baseline on unseen bilayer patterns (0.450 vs.\ 0.455 for \code{fold\_slice}); a calibrated 1\% detector distortion widens the lead to 0.024. With free probe modes, our fit leads on all three benchmarks: by 0.053 on the WSe$_2$ bilayer, by 0.07--0.33 on a 21\,nm praseodymium scandium oxide (PrScO$_3$) crystal, where one baseline runs $5\times$ longer, and by 0.46--0.53 on a lead zirconate titanate (PZT) film (\cref{sec:heldout}; \S\S S2, S9).
  \item \textbf{Fast mode.} The free-mode fit matches every WSe$_2$ baseline after 5\,min, vs.\ 22\,min for \code{PtyRAD} and 20.5\,h for \code{fold\_slice}, and passes \code{torchslice}'s best PrScO$_3$ score 180$\times$ sooner; a 1\,h atom refinement of the published reconstruction beats its in-plane error (\cref{sec:heldout}).
\end{itemize}
Sections S0--S18 are in the supplement.\vspace*{-6pt}
\section{Related work}\vspace*{-6pt}
\label{sec:related}
\textbf{Phase retrieval and ptychography.} Detectors record intensity without phase. Gerchberg and Saxton posed this for electron micrographs~\cite{gerchberg1972practical}, and Fienup needed only a support constraint~\cite{fienup1982phase}. Ptychography supplies measurement diversity physically, as every point lies in tens to hundreds of overlapping patterns. The ptychographical iterative engine (PIE)~\cite{rodenburg2004phase}, difference map~\cite{thibault2008high}, and its extension ePIE~\cite{maiden2009improved} recover object and probe; probe modes model partial coherence~\cite{thibault2013reconstructing}; Fourier ptychography swaps probe and pupil~\cite{zheng2013wide}; and multislice electron ptychography reaches the thermal-vibration limit~\cite{jiang2018ptycho,chen2021latticevib}. These reconstruct pixels, as do our baselines: \code{fold\_slice}, built on \code{PtychoShelves}~\cite{wakonig2020ptychoshelves,chen2021latticevib} and Odstr\v{c}il \etal's maximum-likelihood solver~\cite{odstrcil2018iterative}, and the autodiff package \code{PtyRAD}~\cite{lee2025ptyrad}; \code{py4DSTEM} is another~\cite{savitzky2021py4dstem,varnavides2023iterative}.
\begin{figure*}[b]
  \centering
  \includegraphics[width=\textwidth]{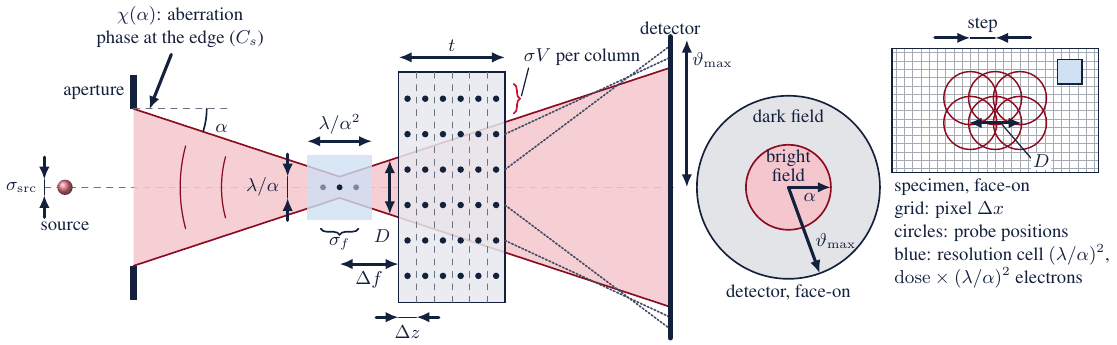}
  \caption{Every quantity the model fits, along one electron's path. Source size $\sigma_{\mathrm{src}}$ and focal spread $\sigma_f$ blur the probe. The aperture sets convergence semi-angle $\alpha$; the focus is $\lambda/\alpha$ wide (lateral resolution) and $\lambda/\alpha^2$ long (depth of field). A probe of diameter $D$ enters at defocus $\Delta f$ and traverses slices ($\Delta z$; thickness $t$): each shifts phase (by the projected phase $\sigma V$ for a column) and propagates the wave, causing multiple scattering. Electrons cross vibrating atoms in $\sim\!10^{-17}$\,s, seeing frozen snapshots averaged in the pattern. The detector records the bright-field disc (the direct beam, radius $\alpha$) and the dark field (scattered electrons) out to the detector edge $\vartheta_{\max}$; charge spreads to neighboring pixels before readout. Independent arrivals make counts Poisson. Inset: pixels $\Delta x$, scan step, resolution cell $(\lambda/\alpha)^2$; values in \protect\cref{tab:quantities}.}
  \label{fig:quantities}
\end{figure*}

\textbf{Atoms fitted to diffraction data.} Diederichs \etal\ first fitted atoms and a physical instrument, with frozen phonons, to experimental 4D-STEM data~\cite{diederichs2024exact}; Herdegen \etal\ recover thermal displacements in simulation~\cite{herdegen2024thermal}. We share their autodiff through multislice, pathwise width gradients, and atomic parameterization, but fix phonon configurations so our objective is deterministic rather than redrawing them at each Adam/SGD step. Singh \etal\ and Mani \etal\ map pixels to Gaussian atoms for tomography~\cite{singh2025gaussian,mani2025anyatom}, while crystallography refines positions and Gaussian displacements via maximum likelihood~\cite{murshudov1997refinement,afonine2018real}.

\textbf{Physics-based inverse imaging in vision.} Cloud tomography inverts multiple scattering~\cite{Levis_2015_ICCV,Ronen_2021_ICCV}, and neural fields image black holes~\cite{Levis_2022_CVPR,Feng_2026_CVPR}. Cryo-electron microscopy pairs single scattering with a neural volume~\cite{Zhong_2021_ICCV,Levy_2022_ECCV}, NFH-SEM learns a scanning electron microscope forward model~\cite{Chen_2026_CVPR}, and generative priors~\cite{mccray2025dgp} and neural fields~\cite{lim2025inr,chien2023spacetime} regularize electron ptychography and tomography. Single-photon imaging similarly fits an inhomogeneous Poisson model (with dead time and jitter) to raw timestamps of single-photon avalanche diodes (SPADs) rather than histogramming detections~\cite{wei2023passive}. %
None reports per-parameter uncertainty; of these, only cloud tomography inverts multiple scattering. Chien \etal\ leave validating the implicit prior open; held-out deviance and bar coverage (\S\S S4, S7) test it.

\textbf{Representations and priors.} Learned reconstructors~\cite{ongie2020deep,loehr2026ptyrannosaur}, deep image priors~\cite{ulyanov2018dip}, and diffusion priors~\cite{chung2023dps,belardi2025generative} regularize inversions with networks; in ptychography, they provide fast initialization but import training biases and hold only near training conditions, so we experiment with one solely as a warm start (\S S9). Neural fields and Gaussian splatting fit one scene via differentiable rendering (\cref{sec:intro}). Our atoms are physical Gaussians, each equipped with an FI error bar.
\vspace*{-6pt}

\section{Formulation: creating the data by scattering}\vspace*{-6pt}
\label{sec:model}
The forward model follows the electron in \cref{fig:quantities} with parameters $\boldsymbol\theta=[\boldsymbol\theta_A,\boldsymbol\theta_\nu]$: specimen $\boldsymbol\theta_A$ (atoms) and instrument $\boldsymbol\theta_\nu$ (beam, optics, scan, detector), each fixed from prior knowledge or fitted per scan; on the 25\,mrad bilayer, 16 numbers describe the optics and detector against 475 for the specimen and 1{,}804 for the scan (Tab.~S1).
\\
\textbf{Image formation.}
Let $\mathbf r$ be position transverse to the beam, $z$ depth along the beam through the specimen, $\mathbf k$ conjugate spatial frequency (scattering angle $\vartheta=\lambda\vert{}\mathbf k\vert{}$). The probe is set by aperture $A(\mathbf k)$, equal to 1 for $\lambda\vert{}\mathbf k\vert{}\le\alpha$ and 0 otherwise, and by the aberration phase $\chi$ with defocus $\Delta f$ and spherical aberration $C_s$~\cite{kirkland2020advanced}:
\begin{gather}
  P(\mathbf r)=\mathcal F^{-1}\big[A\,e^{-i\chi}\big](\mathbf r),
  \label{eq:probe}\\
  \chi(\mathbf k)=\pi\lambda|\mathbf k|^2\big(\tfrac12 C_s\lambda^2|\mathbf k|^2-\Delta f\big).
  \label{eq:chi}
\end{gather}
A specimen of thickness $t$ and potential $V(\mathbf r,z)$ is split into $S$ slices of thickness $\Delta z$. Slice $n$ has projected potential $V_n(\mathbf r)=\int_{z_n}^{z_n+\Delta z}V\,dz$ and transmission function
\begin{equation}
  O_n(\mathbf r)=\exp\big(i\sigma V_n(\mathbf r)\big),\qquad \sigma=\frac{m e\lambda}{2\pi\hbar^2},
  \label{eq:transmission}
\end{equation}
where $\sigma$ is the interaction constant ($m$ the relativistic mass) and $\sigma V$ the column phase (\cref{fig:quantities}). Cowley and Moodie's multislice~\cite{cowley1957multislice} (the split-step beam propagation of optics) alternates transmission with the Fresnel propagator~\cite{kirkland2020advanced} $p_{\Delta z}(\mathbf k)=e^{-i\pi\lambda\,\Delta z\vert{}\mathbf k\vert{}^2}$ from $\psi_{j,0}=P(\mathbf r-\mathbf r_j)$:
\begin{equation}
  \psi_{j,n+1}=\mathcal F^{-1}\big[p_{\Delta z}\,\mathcal F[O_n\psi_{j,n}]\big].
  \label{eq:multislice}
\end{equation}
A far-field detector records intensity at every scan position:
\begin{equation}
  I_j(\mathbf k)=\big|\mathcal F[\psi_{j,S}](\mathbf k)\big|^2 .
  \label{eq:intensity}
\end{equation}
Source size $\sigma_{\mathrm{src}}$ and focal spread $\sigma_f$ (\cref{fig:quantities}) produce an incoherent average $\langle I_j\rangle$ of \cref{eq:intensity} over probe shifts and defoci; the detector applies point-spread function $d$ and background $b$. With $N_e$ the mean measured count per pattern and a unit-intensity probe, counts $y_j(\mathbf k)$ are Poisson with mean $h_j$, giving negative log-likelihood
\begin{gather}
  h_j=N_e\,d*\langle I_j\rangle+b,
  \label{eq:counts}\\
  \mathcal L=\textstyle\sum_{j,\mathbf k}\big[h_j(\mathbf k)-y_j(\mathbf k)\log h_j(\mathbf k)\big].
  \label{eq:likelihood}
\end{gather}
The fit uses dimensionless units (\cref{fig:quantities}): lateral lengths in $\lambda/\alpha$, depths in $\lambda/\alpha^2$, defocus in Scherzer units $\sqrt{C_s\lambda}$, and aberrations in phase radians at the aperture edge.
\begin{table*}[t]
  \centering\small
  \caption{Datasets of tungsten diselenide (WSe$_2$), praseodymium scandium oxide (PrScO$_3$), and palladium hydride (PdH), with step in $\lambda/\alpha$, thickness $t$ in depths of field $\lambda/\alpha^2$, e$^-$ per resolution cell $(\lambda/\alpha)^2$, and projected phase $\sigma V$ band-limited to $2\alpha$ (more in \S\S S9, S10).}
  \label{tab:quantities}
  \footnotesize\setlength{\tabcolsep}{3.5pt}
  \resizebox{\textwidth}{!}{\begin{tabular}{@{}lllcccccc@{}}
    \toprule
    dataset (kV, mrad) & source & use & $\lambda/\alpha$ (\AA) & $\lambda/\alpha^2$ (\AA) & step & $t$ & e$^-$ per $(\lambda/\alpha)^2$ & $\sigma V$, W\,/\,Se$_2$ (rad)\\
    \midrule
    WSe$_2$ twisted bilayer (80, 25) & \cite{zhang2025phasons} & main: held-out, bars & 1.68 & 67 & 0.26 & 0.15 & $1.5\times10^6$ & 0.89\,/\,1.02\\
    WSe$_2$ twisted bilayer, published (80, 25) & \cite{huang2026moire3d} & vs.\ published recon.\ (\S S15) & 1.68 & 67 & 0.26 & 0.15 & $1.8\times10^6$ & 0.89\,/\,1.02\\
    \quad simulated, same conditions & \cite{huang2026moire3d} & accuracy, coverage & 1.68 & 67 & 0.26 & 0.15 & $6.3\times10^5$ & 0.89\,/\,1.02\\
    WSe$_2$ monolayer (80, 18.4) & unpubl. & warm start outside training range (\S S9) & 2.26 & 123 & 0.13 & 0.05 & $4.9\times10^6$ & 0.64\,/\,0.74\\
    WSe$_2$ bilayer (200, 7.5--22.5) & \cite{nguyen2024uncorrected} & uncorrected lens (\S S10) & 1.11--3.34 & 50--446 & 0.16--0.36 & 0.01--0.13 & 0.9--$4.6\times10^6$ & 0.28--0.95\,/\,0.34--1.06\\
    PrScO$_3$ crystal (300, 21.4) & \cite{chen2021latticevib} & thick: held-out & 0.92 & 43 & 0.45 & 4.88 & $1.0\times10^6$ & --\\
    PdH particle (300, 30) & \cite{ptychohub} & thick: pixels (\S S10) & 0.66 & 22 & 0.64 & 12.80 & $2.3\times10^5$ & --\\
    \bottomrule
  \end{tabular}}
\end{table*}
\vspace*{-6pt}
\paragraph{The specimen as atoms.}
Each atom $a$ has in-plane position $\mathbf r_a$, depth $z_a$, species $Z_a$, and thermal width $U_a=\langle u_a^2\rangle$ (mean-square displacement per axis). In the atom model, the projected potential of slice $n$ is
\begin{equation}
  V_n(\mathbf r)=\textstyle\sum_{a\in n} v_{Z_a}(\mathbf r-\mathbf r_a;\,U_a),
  \label{eq:atomsum}
\end{equation}
modeled with the five-Gaussian electron scattering factors $f_Z(\mathbf k)=\sum_i a_i^Z e^{-b_i^Z\vert{}\mathbf k\vert{}^2/4}$ of Peng \etal~\cite{peng1996robust}. A Gaussian displacement of variance $U$ multiplies the transform by the Debye--Waller factor $e^{-2\pi^2U\vert{}\mathbf k\vert{}^2}$:
\begin{equation}
  \hat v_Z(\mathbf k;U)\!=\!\tfrac{2\pi\hbar^2}{m_0e}\textstyle\sum_{i=1}^5\!a_i^Z\exp\!\big[{-}\big(\tfrac14 b_i^Z{+}2\pi^2U\big)|\mathbf k|^2\big],
  \label{eq:atompot}
\end{equation}
with rest mass $m_0$; in real space, each term is a Gaussian of per-axis variance $b_i^Z/(8\pi^2)+U$. Each atom is thus five physical Gaussian splats~\cite{kerbl2023gaussian}. Widening an atom obeys the heat equation (the scale space of vision~\cite{koenderink1984structure}),
\begin{equation}
  \partial v_Z/\partial U=\tfrac12\nabla^2 v_Z ,
  \label{eq:heat}
\end{equation}
making the mean-field width derivative exact and cheap; an anisotropic covariance $\mathbf U_a$ gives $\partial v/\partial\mathbf U_a=\tfrac12\nabla\nabla^{\!\top}v$.
We synthesize $V_n$ in Fourier space on an oversampled grid via one matrix product per slice.
In layered specimens, atoms sit in their atomic-plane slices and layer planes shift together via a fitted interlayer distance. For general depths, each atom carries a Gaussian depth profile of fixed width $\sigma_z\ll\lambda/\vartheta_{\max}^2$ ($\vartheta_{\max}$ is the maximum retained angle), moving rigidly so its depth derivative converges under slice refinement (\S S1).
An atom has four parameters $(\mathbf r_a,z_a,U_a)$ -- six with in-plane covariance, and three, $(\mathbf r_a,U_a)$, in layered specimens, where a fixed plane index and one shared interlayer distance set $z_a$.
The physics is built in: atoms cannot amplify the beam, a width $U$ is strictly positive (\S S6), and no unphysical spatial frequency appears.
\vspace*{-6pt}
\paragraph{Thermal motion.}
\Cref{eq:atompot} is the mean-field potential: each atom is smeared by its own vibration.
Frozen phonons~\cite{loane1991thermal,kirkland2020advanced} average intensities over $K$ static configurations,
\begin{gather}
  \langle I_j\rangle=\frac1K\textstyle\sum_{\kappa=1}^K I_j\big[V^{(\kappa)}\big],
  \label{eq:phonons}\\
  V_n^{(\kappa)}=\textstyle\sum_{a\in n} v_{Z_a}\big(\mathbf r-\mathbf r_a-\sqrt{U_a}\,\boldsymbol\epsilon_{a\kappa};\,0\big),
  \label{eq:configs}
\end{gather}
with fixed standard normal displacements $\boldsymbol\epsilon_{a\kappa}$ drawn in antithetic pairs.
Since $U_a$ enters through the displacement, its gradient is the pathwise derivative of Diederichs \etal~\cite{diederichs2024exact} rather than \cref{eq:heat}; the widths we report use this model. \Cref{eq:likelihood} is evaluated once, on the average of the $K$ patterns.
\begin{figure}[t]
  \centering
  \includegraphics[width=\linewidth]{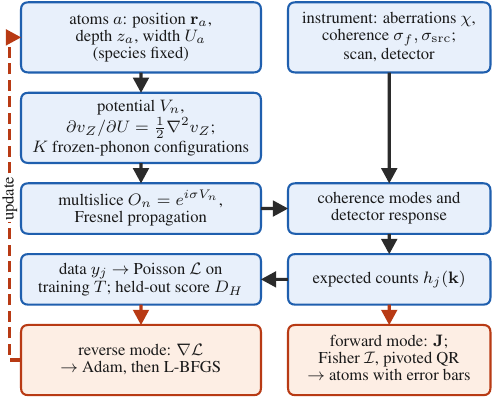}
  \caption{One differentiable model yields both the fit and its error bars. Atom and instrument parameters render expected counts through the potential (\protect\cref{eq:atompot,eq:heat}, or \protect\cref{eq:configs} for frozen phonons), multislice (\protect\cref{eq:multislice}), coherence (\protect\cref{eq:coherence}), and detector (\protect\cref{eq:detector}). Reverse-mode gradients minimize the Poisson likelihood (\protect\cref{eq:likelihood}); forward-mode Jacobian columns give the FI and an error bar on every parameter (\protect\cref{sec:inference}). Held-out positions score the model (algorithms: \S S17).}
  \label{fig:pipeline}
\end{figure}\vspace*{-6pt}
\paragraph{The instrument.}
The aberration function extends \cref{eq:chi} (Eq.~(S1)) with astigmatism $\mathbf A_1$, coma $\mathbf B_2$, and three-fold astigmatism $\mathbf A_2$; higher orders are optional, and the aperture radius is fitted.
Partial coherence is an incoherent sum over defocus offsets $\sigma_f x_q$ and source shifts $\sigma_{\mathrm{src}}\mathbf s_q$, at the nodes $q$ of a product grid spaced evenly over $\pm3.5$ standard deviations, with Gaussian weights $w_q$:
\begin{equation}
  \langle I_j\rangle\!=\!\textstyle\sum_{q}\!w_q I_j\big[A e^{-i\chi}e^{i\pi\lambda\sigma_fx_q|\mathbf k|^2}e^{-2\pi i\sigma_{\mathrm{src}}\mathbf k\cdot\mathbf s_q}\big].
  \label{eq:coherence}
\end{equation}
The copies span a low-rank mixed state, so we replace them by the leading eigenmodes of their Gram matrix—at least 32 and enough to capture all but $10^{-6}$ of the trace~\cite{thibault2013reconstructing} (81 on the 25\,mrad bilayer);
the modes stay differentiable in $\chi$ (exactly), $\sigma_f$, and $\sigma_{\mathrm{src}}$ (\S S17), and free probe modes can be added for what the physics misses; they are off in every atom fit.
The scan positions are $\mathbf r_j=\mathbf r_j^0+\sum_l c_l\,\boldsymbol\beta_l(\mathbf r_j^0)+\boldsymbol\delta_j$, with a degree-2 polynomial distortion basis $\boldsymbol\beta_l$ and per-position offsets $\boldsymbol\delta_j$.
The descan—a scan-wide shift $\mathbf g$ of the diffraction pattern—is a phase ramp $e^{2\pi i\,\mathbf g\cdot\mathbf r}$ on the exit wave, shifting $I_j$ by any fraction of a pixel while keeping it nonnegative.
The detector integrates the pattern over the area of each pixel (a box), applies a Gaussian point-spread function of variance $\sigma_d^2\ge0$, and adds a flat background $b$ subtracted from the scattered total. \Cref{eq:counts} becomes
\begin{gather}
  h_j=(N_e-N_{\mathrm{px}}b)\,\eta\,(d*\langle I_j\rangle)+b,
  \label{eq:detector}\\
  \hat d(\boldsymbol\xi)=\operatorname{sinc}\xi_x\,\operatorname{sinc}\xi_y\,e^{-2\pi^2\sigma_d^2|\boldsymbol\xi|^2},
  \nonumber
\end{gather}
with $\operatorname{sinc}x=\sin(\pi x)/(\pi x)$, $\sigma_d$ in pixels, $\boldsymbol\xi$ in cycles per pixel, $N_{\mathrm{px}}=128^2$ pixels, and a dose scale $\eta$.

\section{Inference and error bounds}
\vspace*{-6pt}\label{sec:inference}

\paragraph{Fitting.}
Over training positions $T$ and pixels in the band-limited disc, we minimize \cref{eq:likelihood} plus weak Gaussian priors on atom displacements (0.3\,\AA), scan polynomial, $\log\sigma_f$, $\log\sigma_{\mathrm{src}}$, and non-defocus aberrations. Per-position offsets are bounded by 1\,\AA, and the 12 gauge directions mimicking other parameters (mean offset and scan-polynomial projection) are removed exactly. Gradients use reverse-mode autodiff through the potential and FFT multislice \cref{eq:multislice}, a network whose layers are slices~\cite{vandenbroek2012method,diederichs2024exact}.
Fits from a coarse start run Adam~\cite{kingma2015adam} on mini-batches of positions; every fit then polishes with full-batch L-BFGS~\cite{liu1989limited} and strong-Wolfe line search. Each polish freezes the probe-mode basis of \cref{eq:coherence} to minimize a smooth objective, refreshing between polishes until a refresh moves the objective by less than 1 nat (optimizer ablation: \S S17(g)).
On real data the FI also sets the optimization metric, scaling every parameter by its Fisher standard deviation (FI computed once near the optimum); simulated-bilayer fits use fixed per-group scales.
Convergence is tested via the squared Newton decrement $\mathbf g^{\!\top}\mathcal I^{-1}\mathbf g$ ($\mathcal I$: FI plus prior precision) on training patterns, in squared error-bar units; we stop below 0.1 with every step below 0.1\,sd (\S S17).
\vspace*{-6pt}
\paragraph{Held-out protocol.}
We score models on withheld scan-grid indices $(j_1,j_2)$ with $(j_1+3j_2)\bmod 8=0$ (one in eight). Each held-out pattern sets per-position parameters (\eg, $\boldsymbol\delta_j$) to its four scan neighbors' mean so it never informs its own prediction. The score is half the Poisson deviance per pixel inside the band-limited disc,
\begin{equation}
  D_H=\frac{1}{|H|\,N_{\mathrm{disc}}}\sum_{j\in H}\sum_{\mathbf k}\Big[h_j-y_j+y_j\log\frac{y_j}{h_j}\Big],
  \label{eq:score}
\end{equation}
which equals \cref{eq:likelihood} minus its value at $h_j=y_j$. We report $D_H$ and the generalization gap $D_H-D_T$ (\S S7).

\begin{table}[t]
  \centering\footnotesize
  \caption{Atoms beat every baseline on unseen bilayer patterns, and the calibrated detector ellipse widens the lead; with free probe modes, the same likelihood predicts best on all three datasets. Held-out deviance $D_H$ (identical counts and scoring per block) on 128 held-out patterns of the 25\,mrad WSe$_2$ bilayer, 512 of a 21\,nm PrScO$_3$ crystal, and 613 of a 20\,nm PbZr$_{0.2}$Ti$_{0.8}$O$_3$ (PZT) film (estimated gain, \S S10). WSe$_2$: exact detector-pixel integration ($^\ast$point-sampled). Free modes: three-seed mean. Baselines at published settings; \code{torchslice}~\cite{diederichs2024exact}: released setup, 50 epochs (longer runs: \S S9). Amp.: amplitude least squares; swap: $D_H$ with only the loss swapped ($L_1$ for ours, Poisson for \code{torchslice}; \S S2). Gap: $D_H-D_T$, held-out minus training deviance. Time: one A6000 (A5000: \code{PtyRAD}, \code{fold\_slice}; $^\dagger$2--5 cards), A100 (PrScO$_3$).}
  \label{tab:heldout}
  \label{tab:heldout_prsco3}
  \setlength{\tabcolsep}{3pt}
  \renewcommand{\arraystretch}{0.9}
  \resizebox{\columnwidth}{!}{\begin{tabular}{@{}llcccc@{}}
    \toprule
    model & loss & $D_H\!\downarrow$ & gap\,$\downarrow$ & time\,$\downarrow$ & swap \\
    \midrule
    \multicolumn{6}{@{}l}{\emph{WSe$_2$ twisted bilayer, 25\,mrad}}\\
    ours, atoms, physical instrument & Poisson & 0.4496 & \textbf{0.005} & 42\,h$^\dagger$ & -- \\
    \quad + calibrated detector ellipse & Poisson & 0.4312 & \textbf{0.005} & +24\,h$^\dagger$ & -- \\
    ours, 8 probe modes & Poisson & \textbf{0.4025} & 0.027 & 53\,min & -- \\
    \code{fold\_slice} (400 + 125 it.) & amp. & 0.4552 & 0.007 & 20.5\,h & -- \\
    \code{PtyRAD}, sub-scan & amp. & 0.4735 & 0.012 & 22\,min & -- \\
    \code{PtyRAD}, full field & amp. & 0.4817 & 0.014 & 50\,min & -- \\
    \code{torchslice}$^\ast$ & $L_1$ & 1.105 & 0.007 & 4\,min & 0.594 \\
    \midrule
    \multicolumn{6}{@{}l}{\emph{PrScO$_3$ crystal, 21\,nm}}\\
    ours, 8 probe modes & Poisson & \textbf{0.4157} & 0.129 & 13\,min & 0.4184 \\
    ours, 4 probe modes & Poisson & 0.4224 & 0.131 & 7\,min & 0.4265 \\
    \code{PtyRAD} (200 it.) & amp. & 0.5586 & 0.070 & 34\,min & -- \\
    \code{PtyRAD} (1{,}000 it.) & amp. & 0.4877 & 0.094 & 2.8\,h & -- \\
    \code{torchslice} & $L_1$ & 0.745 & 0.094 & 4.7\,h & 0.754 \\
    \code{fold\_slice} (200 + 200 it.) & amp. & 0.7359 & 0.051 & 95\,min & -- \\
    \midrule
    \multicolumn{6}{@{}l}{\emph{PZT film, 20\,nm (data of~\cite{diederichs2024exact})}}\\
    ours, 8 probe modes & Poisson & \textbf{0.5754} & 0.146 & 30\,min & -- \\
    ours, 4 probe modes & Poisson & 0.5981 & 0.151 & 15\,min & 0.5826 \\
    \code{torchslice} & $L_1$ & 1.106 & 0.078 & -- & 1.035 \\
    \bottomrule
  \end{tabular}}
\end{table}\vspace*{-6pt}
\paragraph{Error bounds.}
For Poisson counts, the FI is
\begin{equation}
  \mathcal I(\boldsymbol\theta)=\textstyle\sum_{j\in T}\mathbf J_j^{\!\top}\operatorname{diag}(1/h_j)\,\mathbf J_j,\qquad \mathbf J_j=\partial h_j/\partial\boldsymbol\theta ,
  \label{eq:fisher}
\end{equation}
the expected Hessian of \cref{eq:likelihood} (\ie, its Gauss--Newton approximation); at the optimum, $\mathcal I^{-1}$ bounds any unbiased estimator's covariance (Cram\'er--Rao~\cite{kay1993fundamentals}). We report the Laplace covariance, the inverse of $\mathcal I$ plus the prior precision, with clamped parameters held fixed.
A pixel with expected count $h$ contributes $(\partial h/\partial\theta_k)^2/h=h\,(\partial\log h/\partial\theta_k)^2$ to $\theta_k$ -- its counts times squared relative sensitivity. Faint pixels with strong relative response can carry more information than bright ones; on the bilayer, atoms draw most of their information from the dark field (\S S13).

Since the instrument is fitted alongside the atoms, we take the Schur complement of the instrument block,
\begin{equation}
  \mathcal I_{A\mid\nu}=\mathcal I_{AA}-\mathcal I_{A\nu}\,\mathcal I_{\nu\nu}^{-1}\,\mathcal I_{\nu A},
  \label{eq:schur}
\end{equation}
where $\mathcal I_{A\nu}$ has rows $\boldsymbol\theta_A$ and columns $\boldsymbol\theta_\nu$. Per-atom bounds are the diagonal blocks of $\mathcal I_{A\mid\nu}^{-1}$, with $\mathrm{sd}(\sqrt{U_a})=\sqrt{U_a}\,\mathrm{sd}(\log U_a)/2$; widening from $\mathcal I_{AA}^{-1}$ to $\mathcal I_{A\mid\nu}^{-1}$ measures what exact instrument calibration would recover. Each error bar is one parameter's standard deviation with all others profiled out, conditioned on fixed or clamped parameters: detector blur $\sigma_d^2$ (at its lower bound of 0 on the 80\,kV data), fixed exterior atom positions, and clamped $\log U$ values.

Each column of $\mathbf J$ takes one forward-mode pass, computed in parallel in single precision and accumulated in double precision; on the 25\,mrad sub-scan, $\mathbf J$ is formed densely with 1{,}636 columns over 896 training patterns; larger fields use local patches (\S S16).
\vspace*{-6pt}
\section{Experiments}\vspace*{-6pt}
\label{sec:experiments}

\subsection{Data}
\label{sec:data}\vspace*{-6pt}
We test accuracy against truth (\cref{sec:depth}), precision under the model (\cref{sec:bounds}), and prediction of unseen experimental patterns and speed (\cref{sec:heldout}) on the datasets of \cref{tab:quantities}. Our main dataset is a twisted WSe$_2$ bilayer recorded at 80\,kV and 25\,mrad on an aberration-corrected microscope with a pixelated detector~\cite{zhang2025phasons}.
For accuracy, we use the simulated twisted bilayer of Huang \etal~\cite{huang2026moire3d}: 1{,}928 atoms rendered with \code{Prismatic}~\cite{ophus2017prism} under matching conditions, tilted 12.7$^\circ$, with Poisson counts at $2.2\times10^5$\,e/\AA$^2$. Our thick benchmark is the public PrScO$_3$ [001] crystal of Chen \etal~\cite{chen2021latticevib,paradim_ssmm2j11} (21\,nm, or 4.9 depths of field at 300\,kV).
\begin{table}[hbt]
  \centering\small
  \caption{On simulated twisted WSe$_2$ bilayer, our atom fit achieves the lowest in-plane error without structural (bond-length or interlayer) priors. Within 1\,h our fit already beats the published in-plane error; truth enters only through the alignment, depth gauge and atom identity (\S S6). Rms error (pm) after similarity alignment (depth: W\,/\,Se); time: refinement on one RTX A6000 after the published recon. $^\ddagger$475 \code{fold\_slice} iterations at 14.7\,s each per released files (GPU unknown), excluding peak finding/refinement.}
  \label{tab:accsim}
  \setlength{\tabcolsep}{4pt}
  \begin{tabular}{lcccc}
    \toprule
    method & in-plane\,$\downarrow$ & depth\,$\downarrow$ & priors & time\,$\downarrow$ \\
    \midrule
    start: peaks on 2 planes & 4.9 & 49\,/\,46 & no & -- \\
    ours, atoms & \textbf{3.2} & 50\,/\,49 & no & 18\,h \\
    ours, atoms, 1\,h budget & 3.5 & 45\,/\,44 & no & 0.9\,h \\
    published, peaks~\cite{huang2026moire3d} & 4.8 & 84\,/\,127 & no & 1.9\,h$^\ddagger$ \\
    published, refined~\cite{huang2026moire3d} & 4.4 & \textbf{5.3}\,/\,\textbf{5.5} & yes & ${>}$1.9\,h$^\ddagger$ \\
    \bottomrule
  \end{tabular}
\end{table}\vspace*{-6pt}
\subsection{Accuracy against truth}
\begin{figure*}[t]
  \centering
  \includegraphics[width=\textwidth]{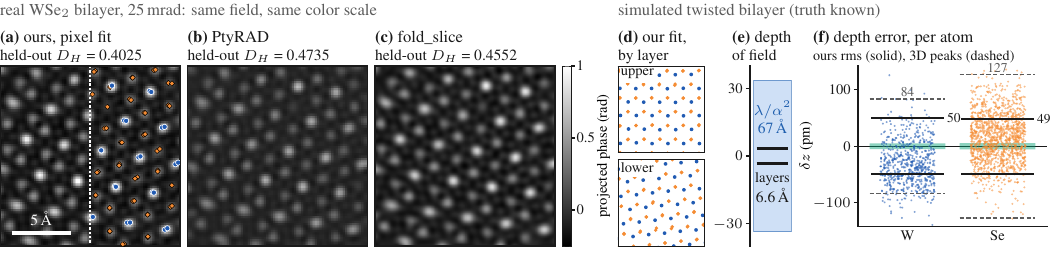}
  \caption{Fitted atoms separate two layers $10\times$ closer than the depth of field (d, e); free modes under the same likelihood predict unseen patterns best (a--c). (a)--(c)~Projected phase of the real 25\,mrad WSe$_2$ bilayer on the held-out sub-scan (shared scale, ramp removed, low-passed at $2\alpha$) with held-out $D_H$ per pixel (\protect\cref{tab:heldout}): (a)~our free-mode pixel fit, shown alone on the left half and with our fitted W (circles) and Se (diamonds) atoms overlaid on the right half; (b)~\code{PtyRAD} and (c)~\code{fold\_slice} at published settings. (d)~On the simulated bilayer, our fit separates both layers. (e)~Depth of field $\lambda/\alpha^2=67$\,\AA\ vs.\ the 6.6\,\AA\ interlayer spacing. (f)~Depth error $\delta z$ of each fitted atom: rms 50/49\,pm (W/Se, solid), vs.\ 84/127\,pm for published 3D peaks (dashed) and 5.3/5.5\,pm for prior-refined coordinates (band; \protect\cref{tab:accsim}).}
  \label{fig:recon}
\end{figure*}

\vspace*{-6pt}
\label{sec:depth}
The two layers lie 6.6\,\AA\ apart, $10\times$ closer than the depth of field $\lambda/\alpha^2=67$\,\AA. Pixel reconstructions blur them along the beam, and 3D phase-peak depths are off by 84\,pm (W) and 127\,pm (Se) on simulated data (\cref{fig:recon}). An atom model replaces that volume with three coordinates per atom. Multiple scattering across both layers shapes the dark field and constrains them. We start from atoms read off the published reconstruction, one plane per layer, and fit every depth freely without structural priors. In-plane, our atoms beat the published accuracy even after its priors (\cref{tab:accsim}). With depth polished to the likelihood minimum, they end at 50\,pm (W) and 49\,pm (Se); prior-refined coordinates remain best (5.3 and 5.5\,pm). The FI bounds single-atom depth at 8.0\,pm rms ($1.2\times10^{-3}\,\lambda/\alpha^2$; 0.34\,pm in-plane), close to the 6.4\,pm rms of flat tilted layers, and the likelihood of one atom's depth is indeed 5--9\,pm wide. The likelihood is thus sharp in depth, but mismatch with the simulator displaces its minimum from truth (\S S6); the priors of Huang \etal\ supply the rest.\vspace*{-6pt}

\subsection{Precision under the model}\vspace*{-6pt}
\label{sec:bounds}
When the model is well specified, the bars are calibrated: on Poisson redraws of the fitted model, truth falls within $1\sigma$ for 69\% of position coordinates and 68\% of widths (nominal 68\%), and within $2\sigma$ for 95\% of position coordinates (nominal 95\%).
Across an $8\times$ dose range, position errors scale as $\rho^{-0.48}$ vs.\ the Fisher prediction $\rho^{-0.44}$, and the bars stay calibrated (\S S5).

With the instrument profiled, species widths on the 25\,mrad bilayer are $\sqrt U$ = 7.35 $\pm$ 0.03\,pm (0.044\,$\lambda/\alpha$) for W and 8.98 $\pm$ 0.04\,pm for Se, with sandwich bars (robust bars from per-pattern score scatter) of 0.15 and 0.13\,pm. These effective widths exceed the $\approx$5.5\,pm molecular-dynamics value because static displacements, vacancies, and drift add to thermal motion. Zhang \etal\ read per-atom vibration amplitudes from phase reconstructions of these data and traced them to moir\'e phasons~\cite{zhang2025phasons}. Against an independent multislice simulation, our model underestimates widths by $\sim$1\,pm (8--12\%) due to a 3--7\% dark-field intensity deficit (\S S4). Per atom, median width bars are 0.68\,pm (W) and 1.36\,pm (Se) across the 236 interior atoms. Every width is nearly uncorrelated with each instrument parameter ($|r|\le0.07$), and profiling the instrument leaves the species-width bars essentially unchanged (\S S4). Untying the two Se atoms in each column splits them laterally by a median of 0.10\,\AA\ at an earlier fit and improves held-out prediction. The independent simulation recovers planted splits of this size (7 of 7 from 0.05\,\AA) without generating false splits from misfit, bonding charge, or vacancies (\S S4).

On real data and the independent simulator, model mismatch sets position error at about 10$\times$ the Fisher bars (\S S15). The bars there state shot-noise precision; \cref{sec:analysis} budgets the gap.
Per atom, sandwich bars are 1.6 (W) and 1.4 (Se) times the Fisher bars in median (\S S12).
Per-atom widths absorb part of this misfit: refitting with one width per atom lowers $D_H$ by at least $(1.21\pm0.18)\times10^{-3}$ per pixel (paired over held-out patterns). At the species-width fit, the FI determines about 200 of 235 interior widths (\S S3).

\vspace*{-6pt}
\subsection{Prediction and speed}
\label{sec:heldout}\vspace*{-6pt}
With atoms in place of pixels and free modes, the atom fit beats every baseline on unseen bilayer patterns with the smallest train--test gap; a 1\% detector ellipse, calibrated from the bright-field rim, widens its lead to 0.024 per pixel (\cref{tab:heldout}, \S S14). Pixels and eight free probe modes under the same likelihood predict best: 0.029 per pixel ahead of the calibrated atom fit, 0.053 below \code{fold\_slice} and 0.071--0.079 below \code{PtyRAD}. On the low-count bilayer the likelihood drives this margin: on the same model, data, and optimizer, $L_1$ and amplitude least squares lose 0.058 and 0.054 per pixel (\S S2), and \code{fold\_slice} and \code{PtyRAD} optimize amplitude losses.

On thick PrScO$_3$ and PZT, our free-mode fit leads every code by 0.07--0.53 per pixel, on PrScO$_3$ mostly through the fitted detector blur; Poisson pays at low counts (\cref{tab:heldout}; \S\S S2, S8, S10).

The fast mode cuts the wait for a reconstruction from a day to minutes. On one A6000, the free-mode fit matches every bilayer baseline after 5.3\,min ($D_H$ 0.453 vs.\ 0.455), against 22\,min for \code{PtyRAD} and 20.5\,h for \code{fold\_slice} (A5000, published settings). It reaches 0.4025 after 53\,min. On PrScO$_3$ it passes the best \code{torchslice} score in 95\,s vs.\ 4.7\,h (\S\S S9, S11). The sub-scan takes about 1\,s to record, so reconstruction time over acquisition time falls from $7\times10^4$ to $3\times10^2$; near one, the fit can steer the next scan.

\section{Identifiability and model error}
\label{sec:analysis}\vspace*{-6pt}
\paragraph{Exact reference.} Atom-by-atom multiple scattering via Foldy--Lax~\cite{foldy1945multiple,lax1951multiple} with partial-wave T-matrices~\cite{salvat2005elsepa} is our reference, exact for static atoms of a given potential. With slices at atomic planes, multislice matches dark-field amplitudes of the bilayer's mean-field columns within 1\%, while the tabulated potential~\cite{peng1996robust} differs from partial waves by $+1.6\%$ to $-4.8\%$ (\S S18).

\vspace*{-6pt}
\paragraph{Identifiability.} The FI answers two questions before an experiment: which parameters the counts determine, and what dose a target precision needs. Stacking whitened Jacobians $\operatorname{diag}(h_j)^{-1/2}\mathbf J_j$ ($j\in T$) into $\mathbf A$ yields $\mathcal I=\mathbf A^{\!\top}\mathbf A$; a rank-revealing (column-pivoted) QR factorization gives
\begin{equation*}
  \mathbf A\boldsymbol\Pi=\mathbf Q\mathbf R,\qquad \mathcal I^{-1}=\boldsymbol\Pi\,\mathbf R^{-1}\mathbf R^{-\top}\boldsymbol\Pi^{\top}.
\end{equation*}
In natural units ($\lambda/\alpha$ in-plane, $\lambda/\alpha^2$ in depth, $\log U$ for widths), $R_{kk}^2$ is the information on pivot $k$ with earlier pivots profiled out and later ones fixed; counts determine the pivot when $|R_{kk}|\ge1$ (one shot-noise unit). With background scaling with dose, $\mathbf A\propto\mathrm{dose}^{1/2}$, so one factorization covers all doses: pivot $k$ reaches conditional standard deviation $1/c$ at $D_k=D_0c^2/R_{kk}^2$ for measured dose $D_0$. On local patches of the 25\,mrad bilayer with per-atom widths, 95--96\% of parameters are determined at the dataset dose and 88\% at $0.01\times$ dose.
With all other patch parameters profiled, $0.1\,\lambda/\alpha$ in-plane precision requires 196--212\,e/\AA$^2$ (W) and 731--1{,}090\,e/\AA$^2$ (Se); 10\% per-atom width precision 0.1--2.5 times the dataset dose, and 1\,\AA\ depth $2\times10^3$--$2\times10^4$\,e/\AA$^2$ at an earlier fit (\S S16).

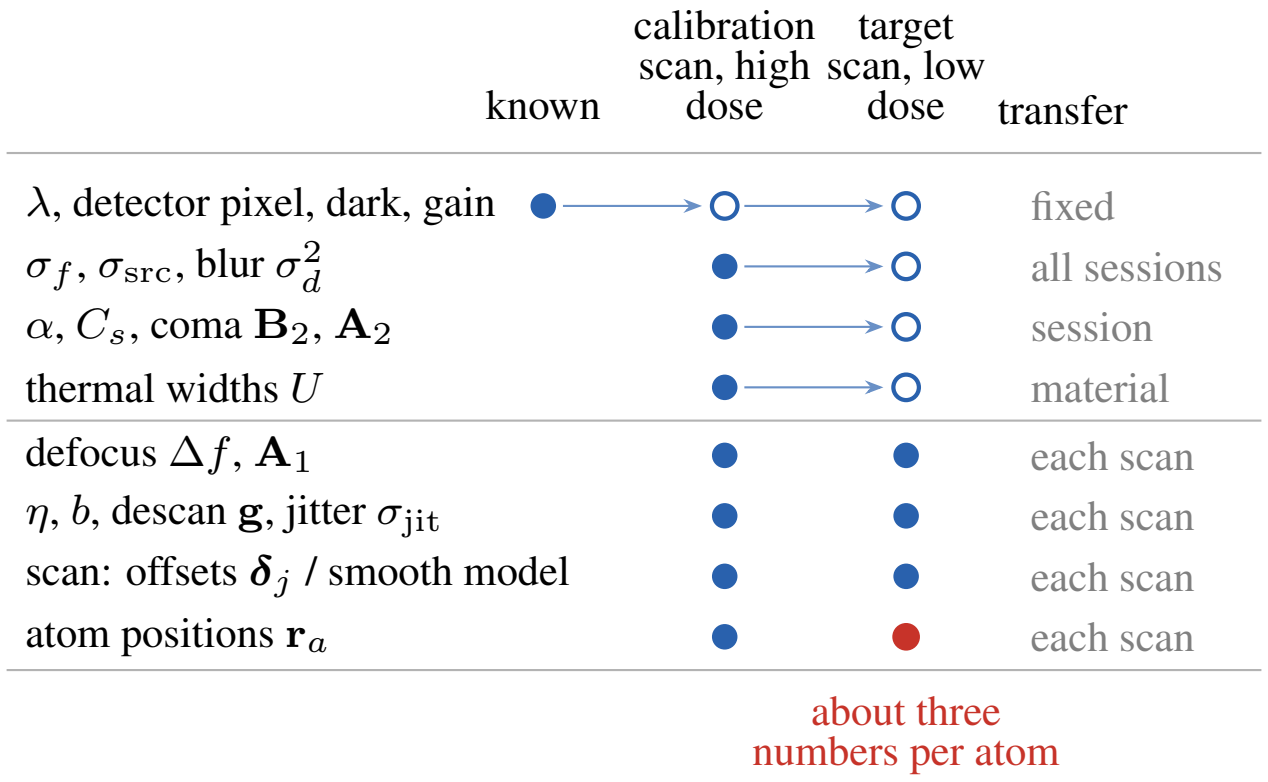
\begin{figure}[hbt]
  \centering
  \begin{tikzpicture}[font=\footnotesize, >={Stealth[length=3.5pt]}]
  \definecolor{phys}{rgb}{0.16,0.38,0.68}
  \definecolor{free}{rgb}{0.78,0.20,0.16}
  \def\cA{3.55}\def\cB{4.75}\def\cC{5.95}\def\cT{6.65}
  \node[align=center,anchor=south] at (\cA,0.25) {known};
  \node[align=center,anchor=south] at (\cB,0.25) {calibration\\[-2pt]scan, high\\[-2pt]dose};
  \node[align=center,anchor=south] at (\cC,0.25) {target\\[-2pt]scan, low\\[-2pt]dose};
  \node[anchor=south west,inner sep=0pt] at (\cT-0.1,0.33) {transfer};
  \draw[gray!60] (0,0.15) -- (8.3,0.15);
  \newcommand{\fit}[2]{\fill[phys] (#1,#2) circle (2.4pt);}
  \newcommand{\held}[2]{\draw[phys,thick,fill=white] (#1,#2) circle (2.4pt);}
  \newcommand{\carry}[3]{\draw[->,phys!70] (#1+0.13,#3) -- (#2-0.15,#3);}
  \node[anchor=west] at (0,-0.2) {$\lambda$, detector pixel, dark, gain};
  \fit{\cA}{-0.2}\held{\cB}{-0.2}\held{\cC}{-0.2}\carry{\cA}{\cB}{-0.2}\carry{\cB}{\cC}{-0.2}
  \node[anchor=west,gray] at (\cT,-0.2) {fixed};
  \node[anchor=west] at (0,-0.6) {$\sigma_f$, $\sigma_{\mathrm{src}}$, blur $\sigma_d^2$};
  \fit{\cB}{-0.6}\held{\cC}{-0.6}\carry{\cB}{\cC}{-0.6}
  \node[anchor=west,gray] at (\cT,-0.6) {all sessions};
  \node[anchor=west] at (0,-1.0) {$\alpha$, $C_s$, coma $\mathbf B_2$, $\mathbf A_2$};
  \fit{\cB}{-1.0}\held{\cC}{-1.0}\carry{\cB}{\cC}{-1.0}
  \node[anchor=west,gray] at (\cT,-1.0) {session};
  \node[anchor=west] at (0,-1.4) {thermal widths $U$};
  \fit{\cB}{-1.4}\held{\cC}{-1.4}\carry{\cB}{\cC}{-1.4}
  \node[anchor=west,gray] at (\cT,-1.4) {material};
  \draw[gray!60] (0,-1.62) -- (8.3,-1.62);
  \node[anchor=west] at (0,-1.85) {defocus $\Delta f$, $\mathbf A_1$};
  \fit{\cB}{-1.85}\fit{\cC}{-1.85}
  \node[anchor=west,gray] at (\cT,-1.85) {each scan};
  \node[anchor=west] at (0,-2.25) {$\eta$, $b$, descan $\mathbf g$, jitter $\sigma_{\mathrm{jit}}$};
  \fit{\cB}{-2.25}\fit{\cC}{-2.25}
  \node[anchor=west,gray] at (\cT,-2.25) {each scan};
  \node[anchor=west] at (0,-2.65) {scan: offsets $\boldsymbol\delta_j$ / smooth model};
  \fit{\cB}{-2.65}\fit{\cC}{-2.65}
  \node[anchor=west,gray] at (\cT,-2.65) {each scan};
  \node[anchor=west] at (0,-3.05) {atom positions $\mathbf r_a$};
  \fit{\cB}{-3.05}\draw[free,fill=free] (\cC,-3.05) circle (2.4pt);
  \node[anchor=west,gray] at (\cT,-3.05) {each scan};
  \draw[gray!60] (0,-3.27) -- (8.3,-3.27);
  \node[anchor=north,align=center,free] at (\cC,-3.32) {about three\\[-2pt]numbers per atom};
\end{tikzpicture}
  \caption{Separate readings fix different parameters. A high-dose calibration scan -- adjacent to the target or on a radiation-hard specimen -- fixes the instrument and thermal widths; the low-dose target scan fits only atom positions, a smooth scan model, drifting terms, and one position jitter that catches the scan's own blur. Filled: fitted; open: fixed from the left reading; right: transfer classes of Tab.~S1; symbols as in \protect\cref{sec:model}. Repeating the calibration bounds the drift (\S S17(f)).}
  \label{fig:calib}
\end{figure}

Separate readings enable low-dose imaging (\cref{fig:calib}). At 100\,e/\AA$^2$, a dose for beam-sensitive specimens, fixing the calibrated instrument, coma included, and the widths from a high-dose scan, and refitting positions, a smooth scan model, defocus, two-fold astigmatism, and one jitter, estimates $3\times$ fewer parameters and tightens position bounds by 24\% (W) and 6\% (Se), if coma holds to well within 30\,nm, as between scans minutes apart (\S S16).

\vspace*{-6pt}
\paragraph{Error budget.} The FI converts any model error into a shift in error-bar units. We project ten omitted or truncated terms and our three cost-saving approximations through the FI of a reference fit (\S S12).
Each term shifts a median width by 0.2--0.9\,sd.
A $\sim$20\,mrad local tilt or a frozen-phonon diffuse-scattering error reaches the measured per-atom misfit, which names the next terms to model.
\vspace*{-6pt}
\section{Conclusion}
\label{sec:conclusion}
An atom with an error bar turns atomic imaging into hypothesis testing: whether an atom moved by 5\,pm, or a width changed with temperature, is a test against a stated bar (\cref{sec:bounds}), once model error is budgeted (\cref{sec:analysis}) or cancels between scans on one instrument. The FI states what each electron tells about each parameter, the limit any reconstruction faces, and so makes the experiment designable. It gives the dose each parameter needs before the scan, and after one high-dose calibration a low-dose scan fits $3\times$ fewer parameters (\cref{fig:calib}). That is the route to beam-sensitive materials and, with known molecular fragments as primitives, to frozen biological specimens. As fit time approaches acquisition time (now minutes for a 1\,s scan, from 20.5\,h), the fit can steer the next scan and follow a process, such as a defect forming under the beam, while it happens. Two steps remain: forward models accurate enough that counts alone limit each atom, and event-driven detectors, now reaching the market, that timestamp each electron~\cite{jannis2022event}, as SPADs do photons~\cite{wei2023passive}, so the likelihood absorbs drift and dynamics.

\paragraph{Limitations.} Each bar is a model-conditional bound, which per-atom misfit on the bilayer exceeds (\cref{sec:analysis}) until diffuse scattering and coherence are better modeled. At published doses, counts constrain single-atom depth no better than layer geometry (\cref{sec:depth}); on real data we fit atoms in their atomic planes, so we do not resolve two atoms of one layer in depth.
The fit neither adds nor removes atoms. In specimens tens of nm thick, every method, ours included, is challenged, since a column spans tens of atoms. We fit pixels there while building a joint column model (\S S10).

{
    \small
    \bibliographystyle{ieeenat_fullname}
    \bibliography{main}
}

\clearpage
\twocolumn[{\centering\Large\bfseries Supplementary Material\par\vspace{1.5em}}]
\setcounter{figure}{0}\setcounter{table}{0}\setcounter{equation}{0}
\renewcommand{\thefigure}{S\arabic{figure}}
\renewcommand{\thetable}{S\arabic{table}}
\renewcommand{\theequation}{S\arabic{equation}}

\section*{S0. Physical picture}
\textbf{Source.} A field-emission gun emits electrons, accelerated through 80--300\,kV. At 80\,kV an electron moves at half the speed of light and has a wavelength $\lambda=4.18$\,pm. The source has a finite size and the electrons a small spread of energies. The first blurs the probe across the beam; the second, through the lens's chromatic aberration, blurs its focus. These are the two partial-coherence parameters, $\sigma_{\mathrm{src}}$ and $\sigma_f$.

\textbf{Probe.} Condenser lenses and an aberration corrector focus the beam, and an aperture limits its angles to a semi-angle $\alpha$. The probe at the specimen is about $\lambda/\alpha$ across (1.7\,\AA\ for the bilayer). Its shape is set by the residual aberrations: defocus, astigmatism, coma and spherical aberration.

\textbf{One electron at a time.} At picoampere currents electrons are nanoseconds, meters, apart. They do not interact. Each one propagates as a wave, interferes with itself, and is detected once, as a particle, at one pixel. The diffraction pattern is a histogram of these detections. Its expected value is the wave's intensity, and its fluctuations are counting noise.

\textbf{Specimen, layer by layer.} Each atom's electrostatic potential (a positive nucleus screened by its electron cloud) slows the electron and shifts the phase of its wave. A thin layer acts as a phase screen. Between layers the wave spreads by Fresnel diffraction. In a thin bilayer the electron is scattered a few times. In a crystal many depths of field thick, it is scattered many times and channels along atomic columns. Multislice alternates the two steps, transmit and propagate (\cref{eq:multislice}).

\textbf{Heat.} At room temperature the atoms vibrate by a few picometers, with periods of about a picosecond. An electron crosses the specimen in about $10^{-17}$\,s, so it sees the atoms frozen in one configuration. Successive electrons see different configurations. The recorded pattern is therefore an average of intensities over thermal configurations (\cref{eq:configs}). That average weakens the sharp diffraction and spreads intensity to high angles (thermal diffuse scattering). A few electrons also lose energy to the specimen (inelastic scattering), which redistributes them by small angles.

\textbf{Detector.} After the specimen, the pattern spreads over the camera length onto a pixelated detector. On the EMPAD, each electron deposits its energy in silicon. The charge spreads sideways, by an amount that grows with beam energy, into neighboring pixels, and each pixel integrates its charge over its area and the frame time. The readout converts charge to a number through a gain (151\,ADU per electron at 80\,kV), and a dark frame is subtracted. The recorded value is thus a blurred, integrated, scaled count with a little read noise (\cref{eq:detector}).

\textbf{Dose.} Dose is the number of electrons per unit area, in electrons per \AA$^2$. Beam-sensitive specimens are damaged above a threshold dose, so they are imaged at low dose. There most dark-field detector pixels record zero or one electron (median 0.35 on the bilayer). Poisson counting noise then sets what can be learned. At high dose the counts per pixel are large, the noise is nearly Gaussian, and errors in the model, not counting noise, set the limit (\S S2). Because electrons arrive independently, the count in each pixel is a Poisson variable whose mean is the model's expected count. The natural loss is therefore the Poisson negative log-likelihood of the counts, which weights each pixel by its noise, rather than least squares or L1 on intensities (\S S2).

\textbf{The scan.} The probe steps across the specimen on a grid finer than its width, so neighboring illuminated regions overlap. One pattern is recorded at each position. The overlap makes ptychography possible. The forward model of \cref{sec:model} follows this path in order: source and probe, multislice through vibrating atoms, the detector, and the Poisson likelihood of the counts.

\section*{S1. Model components}
We test the object (atoms, pixels or a hybrid), the thermal model (frozen phonons with $K$ = 1 to 32, mean field, absorptive potential), partial coherence and focal spread, aberration order, slicing, the depth-placement kernel, and the exact Foldy--Lax reference.
\begin{itemize}
  \item Exact reference: atom-by-atom Foldy--Lax multiple scattering with partial-wave T-matrices; back-scattering below $10^{-14}$ for the tabulated potential, forward-only equals full to $3\times10^{-13}$ (\S S18).
  \item Forward-only Foldy--Lax is multislice with one atom per slice, which links the reference to the fitted model (\S S18).
  \item Partial waves agree with \code{ELSEPA} to $10^{-4}$; the tabulated potential~\cite{peng1996robust} differs from it by $+1.6\%$ to $-4.8\%$ (80--300\,kV).
  \item 1\,\AA\ slice grid: tungsten (W) dark-field amplitude error 3.3\% (selenium, Se, 1.0\%); plane-aligned or $\le0.25$\,\AA\ slices: $\le1\%$; with the eikonal-exact phase: $\le0.3\%$. These hold for mean-field atoms at the reference sampling. Static atoms are in \S S18.
  \item Probe window: agreement with \code{abTEM} 1.1\% at 50\,\AA\ (dark field 4.8\%), against 4.3\% at 25\,\AA.
  \item With point-sampled detector pixels, mean field and frozen phonons ($K=32$) predict the held-out patterns equally well within noise. Each is fitted from two starts and scored on the 128 held-out patterns with a freshly computed probe-mode basis on common configurations. Paired per pattern, the frozen-phonon $D_H$ differs from the mean field's by $(+4.1\pm3.1)\times10^{-4}$ per pixel from one start and $(-4.4\pm3.6)\times10^{-4}$ from the other. The two starts fit the same counts, so they are not pooled as independent samples. The standard errors treat the 128 patterns as independent. Neighboring held-out patterns are 1.2\,\AA\ apart and share atom misfit, so the standard errors are lower bounds.
The score is converged in $K$ ($K=64$ changes it by $10^{-5}$).
  \item A third thermal model keeps the mean field and adds the absorptive phase $\tfrac12\sigma^2\operatorname{Var}_{\boldsymbol\epsilon}[v]$, the second cumulant of the frozen-phonon phase~\cite{hall1965effect}, in closed form for Gaussian terms.
  \item The absorptive potential is worse by 0.0012 per pixel, 4.9 times its paired standard error, because it removes flux that the detector records as thermal diffuse scattering.
  \item The two thermal models differ in their widths. Frozen phonons give $\sqrt{U_a}$ = 7.35\,pm for W and 8.97--8.98\,pm for Se (two starts). These effective widths exceed the $\approx$5.5\,pm molecular-dynamics value because static displacements, vacancies and drift add to thermal motion (\cref{sec:bounds}).
The propagation distance $\Delta z$ between the two W slices of the six-plane model (one slice per atomic plane, the Se planes fixed at $\pm1.67$\,\AA\ from their W plane) is 6.74\,\AA\ from both starts. $\Delta z$ is a parameter of that model, not the interlayer spacing. Its Fisher standard deviation is 0.013\,\AA. Its offset from the nominal 6.5\,\AA\ spacing is 0.24\,\AA, only 0.0036 depths of field. With depth-resolved atoms and no priors, the W--W spacing is 6.51\,\AA\ on the published 6$^\circ$ bilayer and 6.62\,\AA\ on the simulation (truth 6.60\,\AA; \S S15). In that comparison the mean field reaches the same prediction only with $\sqrt{U_a}$ = 11.3--11.4\,pm for W, twice the molecular-dynamics value, a collapsed 1.4\,pm for Se, and $\Delta z$ = 6.40--6.43\,\AA.
  The mean field suffices to predict a thin bilayer. The thermal widths require frozen phonons. Image-based thermometry on twisted-bilayer molybdenum disulfide (MoS$_2$) under the same acquisition conditions (80\,kV, 25\,mrad, EMPAD) finds room-temperature rms displacements of about 9.6\,pm for Mo and 11.7\,pm for S, with 1--2\,pm precision per atom~\cite{bae2026thermal}; W and Se are heavier. The bounds of \cref{sec:experiments} are 0.7--1.4\,pm per atom. On the counts before the dark-reference correction (\S S14), the widths are converged in $K$ to 0.1\,pm: fits with $K=8$ and $K=32$ give W 5.17 against 5.21\,pm and Se 5.07 against 5.08\,pm from one start, and 5.39 against 5.49\,pm and 4.96 against 4.97\,pm from the other.
  \item The atom model trails a free-pixel fit by 0.047 per pixel (\cref{tab:heldout}), far more than the phonon effect, so the thermal model does not set the remaining deficit.
The thick specimens (PdH, PrScO$_3$) are fitted with a pixel object only (\S S10), so the thermal model is untested there.
  \item Representation size. An atom carries four parameters (three in a layered specimen), so the 25\,mrad bilayer needs 2.6 per \AA$^2$, against 26 real values per \AA$^2$ for our one-slice pixel fit and 208 for a four-slice complex object (\cref{sec:model}). Over its scanned field, 1{,}985 atoms carry 7{,}940 parameters. For the thick PdH particle a pixel object of 14 slices holds $1.0\times10^7$ complex values, plus free probe modes.
  \item Depth derivatives need a placement kernel of fixed physical width. Placing an atom on the two nearest depth nodes (linear weights) gives a correct forward model. It leaves a sawtooth artifact in the depth derivative that does not shrink as $\Delta z\to0$:
    \begin{itemize}
      \item single W atom, forward-mode derivative against finite differences: artifact / true tangent 16, 28 and 51 at $\Delta z$ = 0.5, 0.25 and 0.125\,\AA;
      \item a cubic B-spline gives 1.1--3.4 and grows as $1/\Delta z$;
      \item a Gaussian profile of fixed width $\sigma_z$ = 0.5\,\AA\ gives $6\times10^{-3}$ and $9\times10^{-4}$ at $\Delta z$ = 0.5 and 0.25\,\AA.
    \end{itemize}
    Each atom is therefore a 3D object of depth extent $\sigma_z\ll\lambda/\vartheta_{\max}^2$ that moves rigidly. Its derivatives agree with finite differences to $10^{-8}$ in FP64.
  \item Slicing on real data: one slice per plane equals one per layer; misplacing atoms in depth costs 0.004; finite projection and the eikonal phase change the score by $\le3\times10^{-4}$.
  \item 50\,\AA\ window: the same 60-epoch difference ($D_H$ 0.42493 for frozen phonons, 0.42601 for the mean field) and the same widths as at 25\,\AA.
\end{itemize}

\paragraph{Aberrations and parameters.}
The full aberration function of \cref{sec:model}, extending \cref{eq:chi}, is
\begin{multline}
  \chi(\mathbf k)=\tfrac{2\pi}{\lambda}\Big[-\tfrac12\Delta f\,|\boldsymbol\omega|^2+\tfrac12\mathbf A_1\!\cdot\!(\omega_x^2-\omega_y^2,2\omega_x\omega_y)\\
  +\tfrac13\mathbf A_2\!\cdot\!(\omega_x^3-3\omega_x\omega_y^2,\,3\omega_x^2\omega_y-\omega_y^3)\\
  +|\boldsymbol\omega|^2\,\mathbf B_2\!\cdot\!\boldsymbol\omega+\tfrac14C_s|\boldsymbol\omega|^4\Big],
  \label{eq:chifull}
\end{multline}
with $\boldsymbol\omega=\lambda\mathbf k$.
\begin{table}[h]
  \centering\footnotesize
  \setlength{\tabcolsep}{3pt}
  \caption{Sixteen fitted numbers describe the optics and detector of the 25\,mrad WSe$_2$ bilayer, vs.\ 475 for the specimen and 1{,}804 for the scan (counts in parentheses; fit$^{p}$: weak Gaussian prior). The last column shows how fixed quantities are known or how far fitted ones transfer: each scan; a session (between corrector retunes); all sessions; or the material (\protect\cref{fig:calib}). $^{j}$Only in the transferred-calibration design (\S S17(f)); in the fitted 25\,mrad model $\sigma_{\mathrm{src}}$ covers it, and it is not in the counts above.}
  \label{tab:params}
  \resizebox{\columnwidth}{!}{\begin{tabular}{@{}lll@{}}
    \toprule
    quantity (fitted numbers) & status & source / transfer \\
    \midrule
    \multicolumn{3}{@{}l}{\emph{Specimen} $\boldsymbol\theta_A$: 1{,}702 atoms, 236 free (78 W, 158 Se)}\\
    in-plane positions $\mathbf r_a$ of free atoms (472) & fit$^{p}$ & each scan \\
    positions of the other 1{,}466 atoms & fixed & earlier fit, lattice \\
    thermal widths $U$, one per species (2) & fit & material \\
    interlayer distance (1) & fit & each scan \\
    species, occupancy 1, Se planes $\pm1.67$\,\AA & fixed & lattice \\
    \midrule
    \multicolumn{3}{@{}l}{\emph{Instrument} $\boldsymbol\theta_\nu$: beam and optics}\\
    energy 80\,kV, $\lambda=4.18$\,pm & fixed & high tension \\
    convergence angle $\alpha$ (1) & fit & session \\
    aperture edge width, 0.3 pixel & fixed & construction \\
    defocus $\Delta f$ (1) & fit & each scan \\
    $\mathbf A_1$ (2) & fit$^{p}$ & each scan \\
    $\mathbf A_2$, $\mathbf B_2$ (4) & fit$^{p}$ & session \\
    spherical aberration $C_s$ (1) & fit$^{p}$ & session \\
    coherence widths $\sigma_f$, $\sigma_{\mathrm{src}}$ (2) & fit$^{p}$ & session \\
    probe modes, $1-10^{-6}$ of the trace & derived & rows above \\
    \midrule
    \multicolumn{3}{@{}l}{\emph{Instrument} $\boldsymbol\theta_\nu$: scan, $32\times32$ positions, 896 training, 128 held out}\\
    step 0.429\,\AA\ ($0.26\,\lambda/\alpha$) & fixed & scan calibration \\
    polynomial distortion, degree 2 (10) & fit$^{p}$ & each scan \\
    training offsets $\boldsymbol\delta_j$ (1{,}792) & fit & each scan \\
    specimen-to-scan shift (2) & fit & each scan \\
    position jitter $\sigma_{\mathrm{jit}}$, one Gaussian (1)$^{j}$ & fit & each scan \\
    mean offset, polynomial--offset split (12) & gauge & removed exactly \\
    \midrule
    \multicolumn{3}{@{}l}{\emph{Instrument} $\boldsymbol\theta_\nu$: detector, $128\times128$ pixels}\\
    pixel 1.666\,mrad & fixed & diffraction standard \\
    pixel box, orientation & fixed & construction, \S S8 \\
    projector ellipse $\epsilon$, axis (2) & rim fit & camera length, aperture (\S S14) \\
    dark frame, gain 151\,ADU/e, $N_e$ (mean count per pattern, 94{,}571\,e; its scale is absorbed by the dose scale $\eta$, fitted on training patterns) & fixed & measured~\cite{tate2016empad} \\
    effective detector blur $\sigma_d^2\ge0$ (1) & fit & all sessions \\
    dose scale $\eta$, background $b$, descan $\mathbf g$ (4) & fit & each scan \\
    \bottomrule
  \end{tabular}}
\end{table}

\section*{S2. Loss}
We compare the Poisson likelihood with least squares on counts and on amplitudes, for the same model, data and held-out set. Each fit is scored by held-out Poisson deviance and by the metric of each loss. The WSe$_2$ numbers of this section use point-sampled detector pixels, as do the \code{torchslice} rows of \S S9. \Cref{tab:heldout} integrates over each pixel.
\begin{itemize}
  \item Free-mode pixel fit (8 probe modes, one slice, 2{,}500 iterations): held-out $D_H$ 0.4077 (Poisson), 0.4618 (least squares on amplitudes), 0.4733 (least squares on counts).
  \item Held-out squared amplitude error is 0.258 (Poisson), 0.231 (amplitudes) and 0.308 (counts); squared count error is 14.47, 14.52 and 15.18, lowest for the Poisson fit. Least squares on counts over-fits most (train 11.2 against test 15.2).
  \item The amplitude loss gives up 0.054 in $D_H$, and least squares on counts 0.066.
  \item Atom model (mean field, one width per species) on the counts before the dark-reference correction (\S S14), 300 L-BFGS iterations, held-out $D_H$ scored with a freshly computed probe-mode basis: from the shared-width solution (warm) 0.4222 (Poisson), 0.4364 (least squares on amplitudes), 0.4660 (least squares on counts); from a cold start through Adam, 0.4224, 0.4419 and 0.4669. The ordering is that of the pixel fit on the same counts, with a smaller penalty (0.014--0.045 against 0.034--0.080). Under the two least-squares losses the W width drifts to 8--12\,pm.
\end{itemize}

\paragraph{Loss ablation across datasets.} \Cref{tab:lossabl} trains our pixel solver, unchanged, under four losses on three datasets, with only the loss swapped (Poisson path bit-identical). It also trains the released \code{torchslice} code~\cite{diederichs2024exact} under its L1 and under our Poisson loss (\S S9). The datasets differ in how many electrons reach a dark-field pixel: median 0.35 on the bilayer, 2.3 on PrScO$_3$ and 3.1 on PZT (training patterns, $|\mathbf k|>1.3$ bright-field radii, inside the scoring disc).
\begin{itemize}
  \item The likelihood's advantage grows as the counts fall. L1 costs our solver 0.058 per pixel on the bilayer and 0.004 (4 modes; 0.003 with 8) on PrScO$_3$, and gains 0.016 on PZT. Least squares on counts costs 0.066 and 0.035 where it was run. \code{torchslice} with our loss gains 0.51 on the bilayer at 50 epochs (0.14 at 500) and nothing on PrScO$_3$.
  \item On PZT, L1 wins. The residual dispersion $\phi=\sum(n-\lambda)^2/\lambda/N$ at the fitted $\lambda$ is about 0.77 on held-out dark-field pixels but 1.6 in the bright-field disc, where $\lambda\ge10$. We read the excess there as model error rather than counting noise; a robust loss limits its pull. Two checks rule out the noise model. A refit with counts at an effective gain of 270\,ADU/e, which brings the dark-field $\phi$ to about 1, keeps L1 ahead (0.583 against 0.596 per pixel at the 378 scale, three seeds each). A quasi-Poisson loss weighted by $1/\phi(\lambda)$, with $\phi(\lambda)$ fitted on training residuals, does worse (0.610 $\pm$ 0.009).
  \item All three detectors are sub-Poisson at the nominal gain ($\phi$ column), so the Poisson bars are conservative (\S S12). The PrScO$_3$ and PZT values imply effective gains 25--45\% below the nominal ones.
\end{itemize}
\begin{table}[hbt]\centering\footnotesize
\caption{Loss ablation: held-out $D_H$ (mean $\pm$ sd over three seeds; gap in parentheses) of our pixel solver under four losses, and of \code{torchslice} under its L1 and our Poisson loss, with identical counts and scoring per dataset. e/px: median electrons per dark-field pixel; $\phi$: Pearson dispersion of held-out dark-field pixels at the Poisson fit (1 for Poisson counts). Ours: 8 modes (WSe$_2$), 4 modes (PrScO$_3$, PZT). WSe$_2$: point-sampled detector pixels (\S S9). $^{*}$Their object learning rate $\times10$, tuned by us; $3c$: Poisson weight three times the calibrated $c$ (\S S9).}
\label{tab:lossabl}
\setlength{\tabcolsep}{2.5pt}
\resizebox{\columnwidth}{!}{%
\begin{tabular}{@{}lccc@{}}
\toprule
 & WSe$_2$ bilayer & PrScO$_3$ & PZT \\
\midrule
e/px, dark field & 0.35 & 2.3 & 3.1 \\
$\phi$ & 0.88 & 0.56 & 0.77 \\
\midrule
ours, Poisson & \textbf{0.4077} (0.028) & \textbf{0.4224}$\pm$0.0006 (0.131) & 0.5981$\pm$0.0017 (0.151) \\
ours, L1 & 0.4657$\pm$0.0005 (0.023) & 0.4265$\pm$0.0002 (0.127) & \textbf{0.5826}$\pm$0.0006 (0.137) \\
ours, amp. & 0.4618 & 0.4237$\pm$0.0001 (0.131) & -- \\
ours, counts & 0.4733 & 0.4570$\pm$0.0003 (0.122) & -- \\
\midrule
\code{torchslice}, L1, 50 ep. & 1.105 (0.007) & 0.745 (0.094) & 1.106 (0.078) \\
\quad Poisson, 50 ep. & 0.594 (0.032) & 0.754 (0.089) & 1.035 (0.083) \\
\quad Poisson, $3c$, 50 ep. & -- & 0.724 (0.101) & -- \\
\code{torchslice}, L1, 100 ep. & -- & 0.627 (0.101) & -- \\
\quad Poisson, 100 ep. & -- & 0.668 (0.097) & -- \\
\code{torchslice}, L1, 500 ep. & 0.688 (0.058) & -- & -- \\
\quad Poisson, 500 ep. & 0.548 (0.043) & -- & -- \\
\code{torchslice}, L1, 1{,}000 ep. & 0.660 (0.061) & -- & -- \\
\code{torchslice}, L1, lr$\times$10$^{*}$, 500 ep. & 0.639 (0.053) & -- & -- \\
\quad Poisson, lr$\times$10$^{*}$, 500 ep. & 0.473 (0.012) & -- & -- \\
\quad Poisson, lr$\times$10$^{*}$, 1{,}000 ep. & 0.458 (0.009) & -- & -- \\
\code{torchslice}, L1, lr$\times$10$^{*}$, 50 ep. & -- & 0.905 (0.094) & -- \\
\bottomrule
\end{tabular}}
\end{table}

\section*{S3. Parameterization of per-atom widths}
Per-atom widths replace the two species widths of \cref{tab:params} by one $\log U_a$ per atom. Every other row of that table is unchanged. We compare unscaled and Fisher-scaled $\log U$ at a species-width fit. With unit scaling, the per-atom $\log U$ directions are about $10^4$ times flatter than the position directions, and L-BFGS leaves 98\% of the 1{,}702 widths within 1\% of their start after 400 iterations. With each $\log U$ scaled by its Fisher standard deviation, 66\% of the interior atoms move within 40 iterations and the held-out deviance falls.

The scaled refit starts from the converged species-width fit. After 40 L-BFGS iterations it lowers the held-out deviance by $(1.21\pm0.18)\times10^{-3}$ per pixel, paired over the 128 held-out patterns on common phonon configurations (\cref{sec:bounds}), and 79\% of them improve. The bright field gains $(5.8\pm0.5)\times10^{-3}$ per pixel and the dark field $(0.58\pm0.16)\times10^{-3}$. Later iterates score slightly lower still, so this gain is a lower bound. The standard errors treat the patterns as independent. Neighbors 1.2\,\AA\ apart share atom misfit, so the significance is lower than the ratio suggests.

\begin{figure}[hbt]\centering\includegraphics[width=\columnwidth]{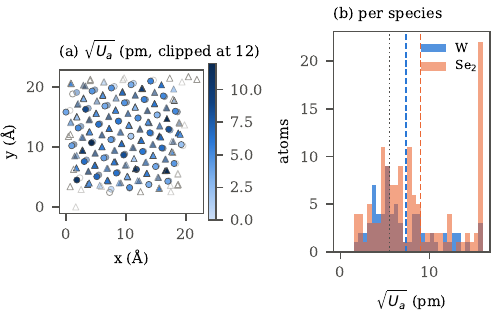}\caption{The Fisher information determines 200 of the 235 unclamped interior widths of the 25\,mrad bilayer. (a)~After 40 iterations of the per-atom refit, $\sqrt{U_a}$ for the 236 interior atoms (W circles, Se$_2$ triangles, second layer faded; gray outlines: the 36 widths the Fisher information does not determine). (b)~Distribution of the 200 determined widths per species; dashed lines mark the one-width-per-species values, the dotted line the molecular-dynamics value.}\label{fig:widthmap}\end{figure}

The per-atom Fisher information at the species-width fit determines 200 of the 235 interior widths with an unclamped $\log U_a$ (standard deviation of $\log U_a$ at most 1): 71 of 78 W and 129 of 157 Se$_2$. One Se$_2$ width sits at its upper bound. The determined widths have a median standard deviation of 0.65\,pm in $\sqrt{U_a}$ for W and 1.11\,pm for Se$_2$. The refit is stopped after 40 iterations. Single widths still move between iterations 20 and 40 (rms 1.9\,sd), so \cref{fig:widthmap} shows the population of widths and we quote no single-atom value. \Cref{fig:widthmap} maps them at iteration 40.

In twisted bilayers, ultrasoft phason modes near 1\,cm$^{-1}$ appear at any nonzero twist, and uneven interlayer coupling mixes in-plane with out-of-plane vibration~\cite{maity2020twistnonics,zhang2025phasons}. Per-atom widths with an out-of-plane term could measure this.

\section*{S4. Error-bar coverage}
This test checks the estimator and its bars under a correct model. It draws Poisson counts from the fitted model at the measured dose (a redraw) and refits each from the generating point with the same objective. Model error is tested in \S\S S6 and S15. $\zeta=(\hat\theta_k-\theta_k^{\mathrm{true}})/\mathrm{sd}$ uses the Fisher standard deviation at the truth. We score 42 redraws at full dose and 47 at half dose (\S S5 adds 1/4 and 1/8). Refits that fail the convergence test are dropped, which can raise the coverage. There are 236 free atoms (472 position coordinates, 153 informed $\log U$; the other 83 $\log U$, with Fisher standard deviation above 1, are held at the truth). With the redraw as the unit, a fraction $f$ over $N$ redraws has the binomial standard error $\sqrt{f(1-f)/N}$: 0.07 for the fractions within 1\,sd below, and 0.03--0.05 for those within 2\,sd.
\begin{center}\small
\resizebox{\columnwidth}{!}{\begin{tabular}{lcccc}
\toprule
group, dose & within 1\,sd & within 2\,sd & sd$(\zeta)$ & mean $\zeta$ \\
\midrule
positions, $\rho=1$ & 0.687 & 0.954 & 1.00 & $+0.06$ \\
positions, $\rho=1/2$ & 0.686 & 0.952 & 1.00 & $+0.05$ \\
$\log U$, Fisher sd $<0.3$, $\rho=1$ & 0.671 & 0.940 & 1.15 & $-0.17$ \\
$\log U$, Fisher sd 0.3--1, $\rho=1$ & 0.687 & 0.900 & 1.57 & $-0.37$ \\
$\log U$, Fisher sd 0.3--1, $\rho=1/2$ & 0.681 & 0.865 & 1.67 & $-0.52$ \\
\midrule
nominal & 0.683 & 0.954 & 1 & 0 \\
\bottomrule
\end{tabular}}
\end{center}
\begin{itemize}
  \item Positions are calibrated at both levels and both doses. This holds for redraws from the fitted model; model error is treated in \S\S S6 and S15. The widths are calibrated at one standard deviation. At two, the interval covers 90\% where the Fisher standard deviation of $\log U$ exceeds 0.3 (widths known to worse than 15\%), because the refits scatter 1.6 times the Fisher value there (sd of $\zeta$ in the table above) and the low tail is heavier (estimates narrower than the truth).
  \item The interval for $\sqrt U$ in picometers inherits the $\log U$ coverage exactly (monotone map): median per-atom standard deviation 0.56\,pm for the well-determined widths and 1.19\,pm for the rest, at a true width of 5.1--5.2\,pm.
\end{itemize}
\subsection*{Atom-level bounds}
\textit{Synthetic} (twisted bilayer at the conditions of the 25\,mrad data; instrument fitted jointly). The model has 1{,}985 atoms with position and in-plane covariance (six parameters each) and 16 instrument parameters ($\Delta f$, $\mathbf A_1$, $\mathbf A_2$, $\mathbf B_2$, $C_s$, $\sigma_d$, $\eta$, a four-term scan affine map, $\sigma_f$, $\sigma_{\mathrm{src}}$), 11{,}926 in all. The Fisher matrix costs about 1{,}500 CPU core-hours. It is assembled in blocks of nearby positions truncated at 6\,\AA: 2--3\% of the diagonal information of a tungsten (W) atom lies beyond, and the truncation changes those bounds by about 1\%. Doses are $5.2\times10^5$\,e/\AA$^2$, with $2.4\times10^5$ in brackets.
\begin{itemize}
  \item Single atom, sd($\sqrt{U_a}$): W 0.14\,pm [0.21]; Se 0.43\,pm [0.65].
  \item Region means of $\sqrt{U_a}$: sd 0.006--0.014\,pm. AA and AB stacking are resolved at 65\,sd [44].
  \item Domain-wall phason ellipse (1.3:1), single W atom: detected at 4.5\,sd, orientation to 6$^\circ$ [9$^\circ$].
  \item Identifiability: $|r|\le0.042$ with every instrument parameter. Fitting the instrument jointly widens the bounds by $<0.3\%$.
  \item Recovery on synthetic refits: empirical over predicted spread 0.88--0.99; $\chi^2$ 0.93--1.12 (1 dof).
  \item The bounds are 20--50$\times$ below published image-fitting errors.
\end{itemize}
\textit{Real data} (Laplace bounds at the fitted optimum, conditional on the model).
\begin{itemize}
  \item Public data, 10.3\,mrad, atom model:
    \begin{itemize}
      \item Columns resolved from the other layer by more than 1\,\AA\ (0.41\,$\lambda/\alpha$) have median $U$ = 0.0062\,\AA$^2$ ($\sqrt U$ = 7.9\,pm) for both W and Se$_2$.
      \item Pairs closer than 0.4\,\AA\ (108 of 282 interior sites) merge into one broad column and are flagged.
      \item The W and Se$_2$ sublattices are not distinguished at this aperture, so per-site $U$ is an effective column width.
    \end{itemize}
  \item Main dataset, 25\,mrad~\cite{zhang2025phasons}, the main real-data result. Frozen phonons, $K=32$:
    \begin{itemize}
      \item Species widths with the instrument profiled: $\sqrt{U}$ = 7.35 $\pm$ 0.03\,pm (W) and 8.98 $\pm$ 0.04\,pm (Se) at the converged fit (Laplace bars, dense $K=8$ Fisher information over the 502 non-offset parameters of the species-width model, with the offsets eliminated; the per-atom plan of \S\S S12 and S16 has 1{,}636 columns). With the instrument known the bars are 0.033 and 0.039\,pm. From the data term alone, without the priors, they are 0.034 and 0.041\,pm. The cluster-by-pattern sandwich bars are 0.15 and 0.13\,pm, 4.5 and 3.2 times the Laplace bars. The species-width model is therefore misspecified at this level; the per-atom misfit of \S S3 is the likely carrier. The Pearson dispersion of the training counts at this fit is 0.92.
      \item Per-atom widths for the 236 interior atoms: median sd($\sqrt{U_a}$) 0.68\,pm (W) and 1.36\,pm (Se), from the per-atom Fisher information at the converged species-width fit.
      \item All Fisher bars on the 25\,mrad data use $K=8$ of the fit's 32 phonon configurations. Their medians hold to about 10\%; the bar of a single atom depends on its four antithetic pairs of draws. The per-atom position bars have medians of 0.5\,pm (W) and 0.8\,pm (Se) (\cref{fig:teaser}).
      \item Region means (78 W, 158 Se): sd 0.23 and 0.76\,pm. %
      \item Correlation with every instrument parameter: $|r|\le0.07$ (W; 0.02 for Se). %
      \item The real per-atom bounds are 3--5$\times$ the synthetic ones (0.68 against 0.14\,pm for W, 1.36 against 0.43\,pm for Se). Most of the gap is accounted for (\S S12). Thermal diffuse scattering leaves frozen-phonon intensities with 3.9 times less diagonal width information for W (1.3 for Se) than the static potential of the synthetic test. The 71\,mrad scoring disc keeps 58\% of the detector's information, and marginalizing the neighbors costs a further factor of 1.3--1.4. The rest is the per-atom misfit, measured at the budget fit (\S S12) at 3.5\,sd (W) and 3.4\,sd (Se) against the 2.0 and 1.3\,sd that the budgeted terms explain there.
    \end{itemize}
    The width map after 40 iterations of the per-atom refit is \cref{fig:widthmap} (\S S3).
  \item Held-out comparison on the same data: the 128 held-out patterns of one $32\times32$ sub-scan, identical counts and scoring (exact detector-pixel integration, \cref{tab:heldout}).
    \begin{itemize}
      \item Our free-mode pixel fit (8 modes, one slice): $D_H$ 0.4025, gap 0.027.
      \item \code{PtyRAD} at its published settings, retrained without the held-out patterns: 0.4817 (full field, gap 0.014); 0.4735 trained on the sub-scan alone (gap 0.012).
      \item Our atom model (frozen phonons, $K=32$, one width per species): 0.4496, gap 0.005; 0.4312 with the calibrated detector ellipse (\S S14).
      \item Our pixel fit leads \code{PtyRAD} by 0.071--0.079 per pixel. The atom model is 0.047 per pixel behind (0.029 with the ellipse), returns physical parameters with error bars, and has the smallest train--test gap.
      \item \code{fold\_slice} at its published schedule (400 + 125 iterations), retrained without the held-out patterns: 0.4552, gap 0.007.
    \end{itemize}
\end{itemize}
\paragraph{Start independence.} The species-width fit is run from two starts. In projection, the tungsten atoms of the two layers can trade places where their columns coincide, and so can the two selenium atoms of one column. Such a relabeling barely changes the counts, so two starts can differ by a few of these discrete swaps. An exchange screen and a relaxed exchange test remove them (\S S17), after which the two starts agree inside the scan box. Of the 236 free atoms, 232 agree to better than 0.1\,\AA\ (median 0.04\,pm; 0.05\,pm rms for the 114 inside the box), and the species widths agree to 0.001 and 0.010\,pm (0.04 and 0.24\,sd). The other four are selenium atoms in the margin, 1.6--4.9\,\AA\ outside the scan box. Their Fisher standard deviations, 0.017--0.17\,\AA, are 3--26 times the median inside the box (0.0066\,\AA). Yet the two starts place them 0.28--3.4\,\AA\ apart. They sit in different local arrangements, which a local bound does not describe. The rearrangement is worth at most 28 nats of the training objective. It does not move the reported widths.
\paragraph{Planted-truth test.} To test whether the fit recovers or creates the Se-pair splits, we simulate data from known structures with an independent multislice code (\code{abTEM}: Lobato potentials, 0.25\,\AA\ slices, 3D frozen phonons, exact pixel integration). The simulation uses the fitted instrument, dose, scan and held-out split of the real data, on the central $16\times16$ positions. Our pipeline fits it from the tied start.
\begin{itemize}
  \item Coincident Se pairs: release gives splits of median 0.017\,\AA\ (0.010\,$\lambda/\alpha$), a gain of 8 nats against 9 expected by chance, and no held-out improvement. The result is the same with an anisotropic thermal ellipsoid and a $\pm$10\% width at each site (0.017\,\AA), and with DFT bonding charge in the potential (0.013\,\AA).
  \item Planted splits of 0.05--0.20\,\AA\ in random directions are all detected (7 of 7, no false detection among the null pairs), with median vector error 0.016\,\AA\ and median angle error 9$^\circ$. The release gains 58 nats per split pair and lowers the held-out deviance by $2.7\times10^{-4}$ per pixel. The Fisher bars of the split components understate the scatter by a factor of 1.5--2.
  \item Outer-plane Se vacancies leave a different signature. The remaining atom of the column moves by about 0.6\,\AA, worth about 2{,}000 nats.
  \item At an earlier fit, the real data give in-box splits of median 0.10\,\AA\ (at most 0.28\,\AA), about 63 nats per pair and a held-out improvement of $8\times10^{-4}$ per pixel. This matches the planted case and none of the confounds. Not tested: dislocations, Se buckling along the beam, larger site anisotropy, interstitials.
  \item The same test bounds the accuracy of the species widths. Against a sampling-converged simulation the fitted $\sqrt U$ are low by 0.9\,pm (W) and 0.75\,pm (Se), 8--12\%, because our forward model under-predicts dark-field intensity by 3--7\% (static cores without the Coulomb cusp, the 2.55\,\AA$^{-1}$ band limit, plane slicing). The fit compensates with a smaller $U$ and a higher background. On data from our model the pipeline recovers the widths to within 1.4 error bars. Coherence, vibration along the beam, the scattering-factor table, bonding and the number of phonon configurations are each excluded as causes.
\end{itemize}

\section*{S5. Dose}
The same synthetic bilayer is redrawn at fractions $\rho$ = 1, 1/2, 1/4 and 1/8 of the measured dose. As in \S S4, only refits that reach the convergence test are scored (42, 47, 44 and 23 draws; at $\rho=1/8$ the rest stop on the iteration budget or the FP32 line-search floor). With the redraw as the unit, the binomial standard error of the coverage is 0.03--0.05 for the positions and 0.04, 0.05, 0.05 and 0.08 for the widths.
\begin{center}\small
\resizebox{\columnwidth}{!}{\begin{tabular}{lcccc}
\toprule
$\rho$ & 1 & 1/2 & 1/4 & 1/8 \\
\midrule
positions, rms sd (\AA): Fisher / meas. & 0.011 / 0.011 & 0.015 / 0.015 & 0.021 / 0.021 & 0.028 / 0.029 \\
positions, within 2\,sd & 0.954 & 0.952 & 0.948 & 0.944 \\
$\log U$ informed, rms sd: Fisher / meas. & 0.50 / 0.83 & 0.71 / 1.19 & 1.00 / 1.56 & 1.42 / 1.91 \\
$\log U$ informed, within 2\,sd & 0.912 & 0.883 & 0.854 & 0.817 \\
\bottomrule
\end{tabular}}
\end{center}
\begin{itemize}
  \item Positions: measured scatter follows $\rho^{-0.48}$, the Fisher standard deviation $\rho^{-0.44}$ (below 1/2 because the weak position prior matters more at low dose). The bars stay calibrated at every dose.
  \item Widths: the scatter exceeds the Fisher value at every dose, and the two-standard-deviation coverage falls as the dose drops. The widths that stay well determined (Fisher sd of $\log U$ below 0.3 at full dose) follow $\rho^{-0.52}$ against $\rho^{-0.5}$.
\end{itemize}

\section*{S6. Depth}
\begin{figure*}[hbt]\centering\includegraphics[width=\textwidth]{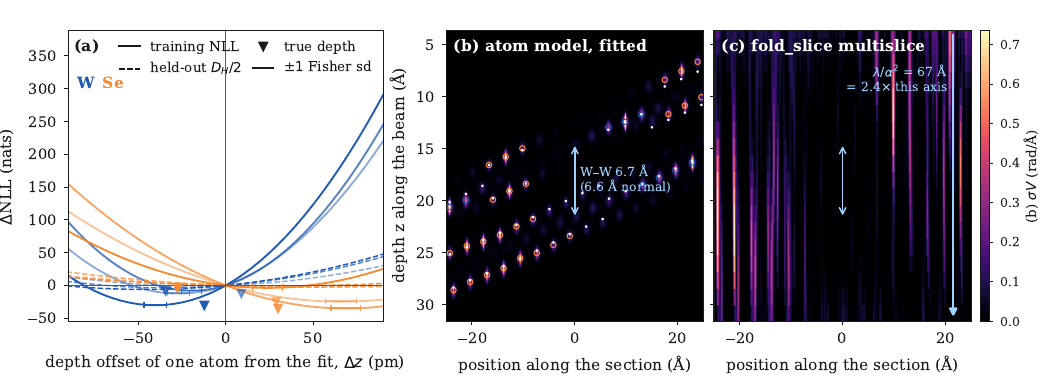}\caption{A single atom's depth is determined at the Fisher bound, and an atom model resolves the layers that a pixel reconstruction blurs along the beam (simulated bilayer; full fit from the published reconstruction, before the polish of \S S6). (a)~Change in training negative log-likelihood (solid) and half the held-out deviance (dashed) as one atom's depth moves, all else fixed; three W (blue) and three Se (orange) atoms. Ticks: $\pm1$ Fisher sd; triangles: true depth. The polish moves the atoms toward these minima. (b)~$xz$ section of the fitted atomic potential through both layers (circles: fitted atoms; dots: truth). (c)~The same section of the published \code{fold\_slice} reconstruction.}\label{fig:depthprofile}\end{figure*}

The fit separates the two layers and places each atomic plane without structural priors: the W--W spacing comes out at 6.62\,\AA\ against the true 6.60\,\AA\ (\S S15). From the published reconstruction, with depth polished to the likelihood minimum, depth error is 49\,pm rms (W 50, Se 49\,pm), against 114\,pm for the 3D phase peaks. Freeing in-plane positions as well leaves it at 51\,pm. The Fisher bound on single-atom depth is 8\,pm, about what flat tilted layers give, and the likelihood of one atom's depth has that width (\cref{fig:depthprofile}). Mismatch between our model and the simulator (0.02--0.035 per pixel in $D_H$) places these minima tens of picometers from truth. The items below give the evidence.
\begin{itemize}
  \item Task: recover $(x,y,z,U)$ for every atom of the simulated bilayer, with no bond-length or interlayer prior; a weak Gaussian prior (0.5\,\AA) acts on in-plane position increments only, none on depth. Two starts:
    \begin{itemize}
      \item (i) From the published reconstruction (main text): in-plane, the 3-D phase peaks; in depth, one plane fitted to the peak depths of each layer. Truth enters only through one global similarity (four numbers), the depth gauge and atom identity. Start error: 4.9\,pm in-plane, 47\,pm in depth.
      \item (ii) With the true layer geometry: flat layers at the true tilt and depth, so no per-atom depth information, and in-plane offsets of about 14\,pm rms from truth. Start error in depth: 11--12\,pm.
    \end{itemize}
  \item Instrument: fitted jointly. The simulation is coherent, so the fitted focal spread and source size measure how far our model departs from the simulator.
  \item Fit floor at the true atoms: with the instrument fitted, $D_H=0.384$, against a Poisson floor of 0.35--0.36 for our model.
    \begin{itemize}
      \item The difference, 0.02--0.035 per pixel, is the model mismatch between their simulator and ours (potential parameterization, slicing, detector binning).
    \end{itemize}
  \item References for depth error (rms over atoms):
    \begin{itemize}
      \item peaks of the published multislice reconstruction, 84\,pm (W) and 127\,pm (Se);
      \item the same after bond-length and interlayer priors, 5.3\,pm~\cite{huang2026moire3d};
      \item flat tilted planes fitted to the truth, which carry no per-atom depth information, 6.3--6.5\,pm per plane.
    \end{itemize}
    Per-atom depth resolution therefore requires an error below about 6.4\,pm, about $10^{-3}$ of the depth of field.
  \item Block test: $48\times48$ positions (20.6\,\AA\ = 12\,$\lambda/\alpha$ square; 266 atoms inside the window, all free):
    \begin{itemize}
      \item Start: flat layers at the true tilt; in-plane offsets of 13.5\,pm rms; depth offsets of 10.9\,pm rms.
      \item Held-out $D_H$ is 0.3842, with gap $D_H-D_T$ = 0.0008. Both are within 0.001 of the fit at the true atoms.
      \item Fit time 2.8\,h on one consumer GPU.
    \end{itemize}
  \item In-plane positions on this block are recovered to 3.75\,pm rms ($2.2\times10^{-2}\,\lambda/\alpha$; W 3.65\,pm, Se 3.80\,pm), with no structural priors (3.2\,pm over the full field below). Huang \etal\ reach 4.4\,pm using bond-length and interlayer priors~\cite{huang2026moire3d}.
  \item At convergence, counting statistics and the model set depth:
    \begin{itemize}
      \item Fisher bound on per-atom depth, instrument profiled: 8.0\,pm rms ($1.2\times10^{-3}\,\lambda/\alpha^2$; W 6.7\,pm, Se 8.6\,pm). This is about the 6.4\,pm that flat tilted planes give. At $6\times10^5$ electrons per resolution cell, the data alone cannot locate single atoms in depth much better than the layer geometry does. The 5.3\,pm of Huang \etal\ is below this bound because their bond-length and interlayer priors supply the missing information.
      \item The joint fit stays at 11.2\,pm, 1.44 bound-sd per atom, with 85\% of atoms inside 2\,sd.
      \item A depth-only search with the instrument and in-plane positions held at the joint fit lands at the same point from either start, the fit or the true depths. That point is 36.6\,pm rms from the truth, at a lower objective ($D_H$ 0.38387 against 0.38418). The true depths are not a stationary point of our objective.
      \item Rescaling depth steps by 10$\times$ in the joint fit gives 17.1\,pm at a similar objective ($D_H$ 0.38393).
    \end{itemize}
    The next steps are structural priors and a potential matched to the simulator.
  \item In-plane, the same mismatch dominates: the bound is 0.33\,pm, against the block error of 3.75\,pm.
  \item Widths: the 25\,mrad fits use $\log U$, so $U>0$ by construction (\cref{sec:model}). The simulated truth has $U=0$, on the boundary, so these fits use $U$ directly, bounded below at $-3$\,pm$^2$, to test for bias at zero.
  \item Width at zero truth: W $U=-1.5\pm4.9$\,pm$^2$ (bound 1.8\,pm$^2$), unbiased. Se is biased by $+12.5$\,pm$^2$ (bound 4.7\,pm$^2$), consistent with the Se$_2$ column absorbing the depth and potential mismatch.
  \item Full field from start (i): $128\times128$ positions, all 1{,}928 atoms in the window free; 14\,h on one RTX A6000.
    \begin{itemize}
      \item $D_H$ 0.38403.
      \item In-plane: 3.2\,pm rms (W 2.8\,pm, Se 3.4\,pm), without priors, from 4.9\,pm at the start.
      \item Depth: 35.8\,pm rms (W 29.0, Se 38.8\,pm) at the full fit, 49.0\,pm (W 49.6, Se 48.8\,pm) with depth polished to the likelihood minimum, and 50.9\,pm (W 51.3, Se 50.7\,pm) after a joint polish of $x$, $y$ and $z$ (below). The block's depth-only searches end at a similar distance, 36.6\,pm. The fit from start (ii) has held-out $D_H$ lower by $2\times10^{-4}$, so the counts separate the two depth configurations only weakly.
    \end{itemize}
  \item Full field from start (ii), the same flags; 14\,h on one RTX A5000. In-plane agrees with start (i) to 0.05\,pm. Depth starts at 12.1\,pm and ends at 16.7\,pm. This fit has no depth polish, so its depth stays near the start.
    \begin{itemize}
      \item $D_H$ 0.38384, gap 0.0003.
      \item In-plane: 3.20\,pm rms ($1.9\times10^{-2}\,\lambda/\alpha$; W 2.9\,pm, Se 3.4\,pm), without priors.
      \item Depth: 16.7\,pm rms. Whole atomic planes shift by $-13.5$ to $+8.5$\,pm, so the mismatch acts on the layer geometry as much as on single atoms.
      \item Widths: W sits near the lower bound of $U$ ($-3$\,pm$^2$; mean 0.5\,pm$^2$), and Se is biased by $+47$\,pm$^2$. The same mismatch causes both.
      \item Fisher bounds on the full field (9{,}495 parameters, instrument profiled): depth 8.2\,pm rms (W 6.6, Se 8.9\,pm), the same as on the block; in-plane 0.34\,pm. The errors are 2.2 bound-sd in depth and 10 in-plane, with whole planes shifted by $-13.5$ to $+8.5$\,pm. Width bounds: 1.8\,pm$^2$ (W) and 5.2\,pm$^2$ (Se).
    \end{itemize}
  \item Sensitivity to the depth profile: $\sigma_z$ = 0.25, 0.4 and 0.6\,\AA\ give the same depth error (11.2, 11.2 and 11.1\,pm) and in-plane error (3.67, 3.75 and 3.98\,pm). $D_H$ is 0.3835, 0.3842 and 0.3851, so the narrowest profile fits best.
  \item Freezing focal spread and source size near 0 on these coherent data leaves depth at 11.1\,pm and costs 0.017 per pixel in $D_H$. The fitted partial coherence absorbs the simulator mismatch; it does not limit depth.
  \item \Cref{fig:depthfull} shows the per-atom depth errors of the full field against their bounds.
  \item One-atom depth profiles at the fit from start (i), all other parameters fixed (three W and three Se atoms, interior; \cref{fig:depthprofile}a). The training likelihood is 4.8--5.3\,pm wide for W and 7.5--9.0\,pm for Se, against Fisher standard deviations of 5.8--6.5 and 8.4--8.8\,pm, so the counts locate a single atom in depth at the bound. At the full fit each minimum lies 20--70\,pm away, 4--35 nats lower, and the held-out deviance falls in the same direction. A depth-only L-BFGS polish of all in-window atoms, run to a relative decrease below $10^{-7}$, moves them toward the minima by 56\,pm rms: $D_H$ falls from 0.38403 to 0.38364, depth error is 49.6\,/\,48.8\,pm (W\,/\,Se), and in-plane is unchanged. A joint polish of $x$, $y$ and $z$ from there lowers $D_H$ to 0.38360 and leaves depth at 51.3\,/\,50.7\,pm and in-plane at 3.2\,pm. The minima lie 14--51\,pm from truth, the model mismatch above.
  \item An $xz$ section through both layers (\cref{fig:depthprofile}b,c): the fitted atoms show the two layers and their Se planes; the published multislice reconstruction of the same data shows streaks along the beam, its depth of field $\lambda/\alpha^2=67$\,\AA\ being $10\times$ the layer spacing.
\end{itemize}

\begin{figure*}[hbt]\centering\includegraphics[width=\textwidth]{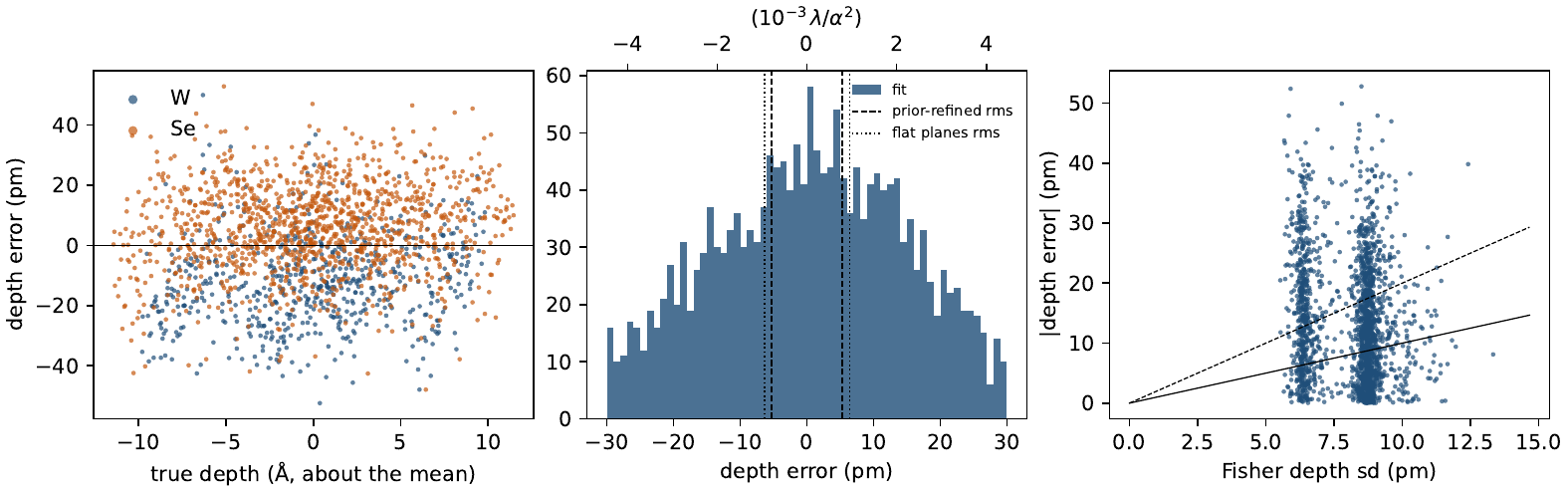}\caption{Depth on the simulated bilayer, full field from start (ii) of \S S6 (1{,}928 atoms, no structural priors). Left: depth error against true depth (W blue, Se orange). Middle: its distribution against the rms of the prior-refined published coordinates (dashed) and of flat tilted planes (dotted). Right: each atom's depth error against its Fisher standard deviation; the solid line is one bound-sd, the dashed line two. The errors cluster by species at 6.6 and 8.9\,pm of bound and exceed it by 2.2 on average, so model mismatch sets the depth error.}\label{fig:depthfull}\end{figure*}

\section*{S7. Held-out protocol}
One pattern in eight is held out by the rule $(j_1+3j_2)\bmod 8=0$ on the scan grid. Held-out positions come from their scan neighbors (neighbor rule). Deviance is scored on the 2/3 disc. A floor of $10^{-9}$ in the model intensity changes scores by $10^{-4}$. Baselines train only on the remaining seven eighths, in every stage, and are scored on the same 128 (sub-scan) or 2{,}048 (full field) patterns through our likelihood.

\section*{S8. Calibration sensitivity}
On the 25\,mrad bilayer the center-of-mass curl test fixes the detector orientation (a flip, no residual rotation), and the scan calibration fixes the scan step. The convergence angle is fitted. For the uncorrected 200\,kV instrument the rotation prior comes from the camera-length table ($+1.1\pm0.5^\circ$) and the detector scale and $\alpha$ from fits of a silicon calibration specimen. Each calibration error is rendered and projected through the per-atom Fisher information of the fit of \S S12, as there. The table gives the median / 90th-percentile width bias over the 236 interior atoms, in units of each atom's standard deviation. It also gives $\Delta D_H$ per pixel before and after the refit absorbs the error.
\begin{center}\small
\resizebox{\columnwidth}{!}{\begin{tabular}{lcccc}
\toprule
error & $|\zeta|$ W & $|\zeta|$ Se & $\Delta D_H$ ($10^{-3}$) & absorbed \\
\midrule
scan rotation $1^\circ$ ($0.07\,\lambda/\alpha$ at the scan edge) & 0.12/0.33 & 0.12/0.25 & 2.86 / 0.04 & 98.6\% \\
scan step $+2\%$ & 0.16/0.44 & 0.14/0.37 & 3.67 / 0.07 & 98.2\% \\
convergence angle $+2\%$ & 0.88/2.11 & 0.71/1.72 & 242 / 22.6 & 90.6\% \\
\bottomrule
\end{tabular}}
\end{center}
\begin{itemize}
  \item A scan rotation or scale error is a linear map of the probe positions, which the fitted per-pattern offsets and drift absorb. At sizes larger than the calibration uncertainty the widths move by at most 0.4\,sd for one atom in ten, and the held-out deviance by $<10^{-4}$ after the refit. Projected on the data, the residuals favor $0.18\pm0.05^\circ$ of rotation and $0.3\pm0.1\%$ of step, both inside the calibration.
  \item The convergence angle is fitted and pinned by the bright-field edge. A 2\% error is 1{,}360 of its standard deviations, \ie\ the data fix $\alpha$ to $1.5\times10^{-5}$ relative. The row is an upper bound for an error the fit cannot make. At the size the data allow, the widths move by less than $10^{-3}$\,sd.
\end{itemize}

\paragraph{Detector blur on PrScO$_3$.} Our pixel solver fits a Gaussian detector blur $\sigma$ (clamped to [0.05, 3]\,px). Freezing it measures its effect (4 modes, three seeds, full depth, same held-out rule, 1{,}336\,s each on one RTX A4000):
\begin{center}\small
\begin{tabular}{@{}lccc@{}}\toprule
blur & $D_H\!\downarrow$ & gap & $\sigma$ (px)\\\midrule
fitted & 0.4224\,$\pm$\,0.0006 & 0.131 & 0.862\\
pinned at the hardware value & 0.4766\,$\pm$\,0.0004 & 0.105 & 1.34\\
off & 0.4927\,$\pm$\,0.0023 & 0.153 & 0.05\\\bottomrule
\end{tabular}
\end{center}
\begin{itemize}
  \item The fitted blur is worth 0.070 per pixel. Without it, our fit trails \code{PtyRAD} (0.4877 at 1{,}000 and 0.4846 at 4{,}000 iterations) by 0.005--0.008. \code{PtyRAD} also applies a blur kernel, so the margin comes from fitting the blur width with the specimen.
  \item The data reject the per-axis charge spread measured for the EMPAD-G2 at 300\,keV (201\,$\mu$m~\cite{philipp2022empadg2}, 1.34\,px at 150\,$\mu$m pixels): pinning it costs 0.054. The fitted 0.86\,px is an effective blur after the free modes take their share (below).
  \item One run suggests that free modes absorb detector blur. On the 7-slice first stage (seed 1), removing the blur costs 0.151 with four modes and 0.092 with eight. The fitted blur falls from 0.81 to 0.66\,px (0.86 to 0.79\,px at full depth; 0.91 to 0.70\,px on PZT). The difference, 0.059, is twenty times the spread over seeds and GPUs. This is consistent with \cref{fig:subspace}. A full-depth check with eight modes needs a card with more than 16\,GB.
\end{itemize}

\section*{S9. Baselines at their own settings}
\begin{center}\small
\begin{tabular}{@{}lll@{}}
\toprule
method & object & role\\
\midrule
\code{fold\_slice}~\cite{wakonig2020ptychoshelves,chen2021latticevib} & pixels & held-out baseline\\
\code{PtyRAD}~\cite{lee2025ptyrad} & pixels & held-out baseline\\
published reconstruction~\cite{huang2026moire3d} & pixels & accuracy (\S S15)\\
\code{PtyRANNOSAUR}~\cite{loehr2026ptyrannosaur} & pixels & learned warm start\\
ours, pixel object & pixels & held-out, ablation (\S S1)\\
ours, atom model & atoms & with Fisher bars\\
\bottomrule
\end{tabular}
\end{center}
Every baseline is retrained without the held-out patterns and scored through our likelihood (\S S7). Our pixel fit of \cref{tab:heldout} shares the fixed rows of \cref{tab:params}, the scan and $\sigma_d^2$. Its specimen is one slice of real phase, its probe eight free modes in place of the optics rows, and it fits no $\eta$, $b$ or $\mathbf g$.
\code{fold\_slice} is run with the published settings for this instrument (multislice LSQ-ML, 19 slices of 1.535\,\AA, 8 probe modes, 400 + 125 iterations) and scored at every saved iteration. Its held-out $D_H$ on the sub-scan, with point-sampled pixels, is 0.4651, 0.4611, 0.4629, 0.4653 at 100--400 iterations of stage 1 and 0.4721, 0.4696, 0.4672, 0.4648, 0.4602 at 25--125 iterations of stage 2. \code{PtyRAD} v1.0.0 is run with its shipped twisted-bilayer tungsten diselenide (WSe$_2$) parameters at 200, 400 and 3{,}000 iterations. Run to 4{,}000 iterations, as in its paper, it reaches 0.485 on PrScO$_3$ (0.488 at 1{,}000) and 0.477 / 0.487 on the WSe$_2$ sub-scan / full field (point-sampled; 0.473 / 0.488 at the published iterations), flat after 2{,}000--2{,}500 iterations. Its run-to-run spread is 0.003--0.005. The PrScO$_3$ run takes 16.8\,h on one A6000 without compilation.
\code{PtyRANNOSAUR} on the WSe$_2$ data:
\begin{itemize}
  \item 25\,mrad data, at the conditions of the network's 80\,kV model, $D_H$: free modes 0.409; ours, 4 slices, 0.414; learned 0.419 (scaled) and 0.421 (raw), 0.417 and 0.421 with rigid stitching of the patches.
  \item Started from the learned output, our fit reaches the final score of a fit from a blank object 3.1$\times$ sooner.
  \item 18.4\,mrad data, outside the network's range:
    \begin{itemize}
      \item position correction rejects at least 15{,}362 of 15{,}376 patches;
      \item the phase is about 12$\times$ too weak, with no amplitude.
    \end{itemize}
\end{itemize}
\paragraph{\code{PtyRANNOSAUR} as a warm start.} \code{PtyRANNOSAUR} is a learned reconstructor. Convolutional networks trained on multislice simulations of Materials Project structures map patches of diffraction patterns to a real projected phase~\cite{loehr2026ptyrannosaur}. Three properties keep it out of \cref{tab:heldout}.
\begin{itemize}
  \item Its models hold only at their training conditions: 80\,kV, 25\,mrad and $2\pm0.6$\,nm; and 300\,kV, 25\,mrad, $10\pm2$ or $20\pm2$\,nm, with defocus of 5--7.5\,nm and a 0.5\,\AA\ step. Its authors note that accuracy decreases outside these regimes. Our bilayer is at the conditions of the first. PrScO$_3$ (21.4\,mrad, 20\,nm defocus), PZT (200\,kV) and PdH (30\,mrad, 28\,nm) are not, and on our 18.4\,mrad data the network fails (above). A fair comparison there needs a model retrained at each dataset's conditions, about 64 A100-hours each by its authors' account, with training choices that would be ours, not theirs.
  \item It returns a real projected phase, not a complex multislice object with a probe, scan and coherence, so it cannot predict a diffraction pattern by itself. We score it through our fitted instrument, with and without a fitted global scale.
  \item Its paper validates by SSIM against simulated truth and by comparison with an iterative reconstruction. No held-out score is reported on experimental data.
\end{itemize}
Where it applies, it is fast and close: 0.419 against 0.409 for our free-mode fit, from 5\,s of inference. As a warm start it makes our fit 3.1$\times$ faster. Our fit adds the instrument and the error bars. We therefore treat it as a learned warm start that our method refines. Its speed suits further roles, not tested here:
\begin{itemize}
  \item live view during acquisition, with the FI evaluated at its output to choose the next scan positions and dose (\S S16);
  \item an initial atom list, by peak finding on its phase, for the atom model;
  \item training once over the range of instrument states a calibration scan reports (probe, coherence and detector blur), which would remove the retraining per condition;
  \item triage of large fields: regions where its output and a fit disagree, or its held-out deviance is high, get a full fit. The same score flags structure imported from training data.
\end{itemize}

\paragraph{\code{torchslice} at its released settings.} Diederichs \etal\ released their code (\code{torchslice}) with a pixel-object notebook for the PrScO$_3$ data of \cref{tab:heldout}~\cite{diederichs2024exact}; their PZT atom pipeline is not released. We convert the notebook to a script and change only two things: the held-out patterns leave the epoch loop, and the snapshots are saved. The intensity scale (mean total per pattern) uses training patterns only.
\begin{itemize}
  \item Model as released: 40 complex phase gratings of 5\,\AA\ on a $512^2$ grid at 0.093\,\AA\ ($0.10\,\lambda/\alpha$, half our pixel), 13 spatial and 1 temporal coherence states with fitted weights and offsets, five aberrations, per-pattern positions; about $2.1\times10^7$ parameters (ours with 4 modes: $7.6\times10^6$). Adam with their four learning rates, warm-up and switch-on epochs and halving on a worse epoch, one step per pattern, 50 epochs, L1 on intensities over the whole $256^2$ frame, sparse-frequency regularizer.
  \item Our loss: only $\mathrm{L1}\cdot10^6$ is replaced by $c\sum[\lambda-n+n\log(n/\lambda)]$ with $\lambda$ their fixed dose times the model intensity. Adam ignores the constant $c$ except against their regularizer, so $c$ matches the L1 data term at the L1 run's epoch-10 snapshot (training patterns). Calibrated at the empty model, it over-weights the regularizer 5--10 times late in training. The regularizer's share of the loss still ends 1.5 times that of the L1 run on PrScO$_3$. With $3c$ the Poisson run reaches 0.724 at 50 epochs, against 0.754 at $c$ and 0.745 for L1, so on PrScO$_3$ the effect of the loss on \code{torchslice} is the size of the effect of the regularizer weight.
  \item WSe$_2$ and PZT: the same pixel-object setup at each dataset's geometry, with slices and sampling at the notebook's non-dimensional values (2 slices of 6.5\,\AA\ and 41 of 5\,\AA; 0.12 and 0.11\,$\lambda/\alpha$); detector orientation and defocus for PZT screened on training loss. A one-row roll of the WSe$_2$ frame aligns its pattern center, which the code cannot fit.
  \item Scoring: their forward model, reloaded from each snapshot, is cut to our 128-pixel window and scored by our likelihood. Held-out positions take the mean learned offset of their four neighbors. Gates: the reloaded forward reproduces the training run's to 0 (bitwise) on four training patterns, and their input frame equals our counts bitwise, with the same held-out set.
  \item Run to plateau (WSe$_2$, one A6000): $D_H$ of the L1 run falls from 1.105 at 50 epochs to 0.688 at 500 and 0.660 at 1{,}000, flat to $6\times10^{-4}$ over its last 50 epochs. The Poisson run reaches 0.548, flat to $7\times10^{-4}$ over its last 50 epochs. With ten times their object learning rate (tuned by us), L1 reaches 0.639 at 500 epochs and Poisson 0.458 at 1{,}000, flat to $3\times10^{-4}$ over its last 50 epochs and still 0.051 above our solver. On PrScO$_3$ the same tuning makes L1 worse, 0.905 at 50 epochs against 0.745 as released, so the released rate stands there. Our solver: 0.4077 in 7\,min. On PrScO$_3$ both arms still fall at 50 epochs (0.016 per 5 epochs). At 100 epochs L1 reaches 0.627 and Poisson 0.668, both still falling (L1 by 0.013 over the last 10 epochs), so our margin shrinks from 0.32--0.34 at the released 50 epochs to 0.21--0.25 (epochs 51--100 on an A4000, so no single-GPU time).
  \item Time to accuracy on one A6000 (PrScO$_3$): \code{torchslice} needs 4.7\,h for 50 epochs (337\,s each) under either loss. Our 4-mode fit reaches the best \code{torchslice} $D_H$, 0.745, after 95\,s, and its own 0.4224 after 12.8\,min. On WSe$_2$, 50 epochs take 249\,s; ours takes 413\,s. On PZT, on one RTX A4000, \code{torchslice} takes 9.5\,h for 50 epochs under either loss and our 4-mode fit 26\,min, 21.6 times less. Repeats of our fits on another GPU change $D_H$ by at most $10^{-4}$ (Poisson) and $5\times10^{-4}$ (L1), below the spread over seeds.
\end{itemize}

\paragraph{Reported evidence of fit.} None of the methods we compare reports how well its fit predicts diffraction patterns it did not see. Their papers judge a reconstruction by an information limit, a Fourier ring correlation against a known model or a repeated scan, SSIM, a training loss, or the scatter of positions over repeated unit cells (\cref{tab:reported}). A small gap is not a good fit: \code{torchslice} at 50 epochs and \code{fold\_slice} on PrScO$_3$ have gaps of 0.007 and 0.051 at $D_H$ 1.105 and 0.736, both under-fitted.
\begin{table*}[hbt]
  \centering\footnotesize
  \caption{What each method's paper reports as evidence of fit, beside what we measure for it on withheld patterns. $D_T$, $D_H$: half the Poisson deviance per detector pixel on training and held-out patterns, identical counts and likelihood within a block; gap $=D_H-D_T$. WSe$_2$: $^\dagger$exact detector-pixel integration (\protect\cref{tab:heldout}); other rows point-sampled pixels. Baselines retrained without the held-out patterns; \code{torchslice} with Poisson loss is our loss swap in their code. PrScO$_3$ gaps include about 0.09\,\AA\ of neighbor-mean position error for every method (\S S10).}
  \label{tab:reported}
  \setlength{\tabcolsep}{3pt}
  \begin{tabular}{@{}l p{7.4cm} c c c c@{}}
    \toprule
    method & what the paper reports & held out? & $D_T\!\downarrow$ & $D_H\!\downarrow$ & gap \\
    \midrule
    \multicolumn{6}{@{}l}{\emph{WSe$_2$ twisted bilayer, 80\,kV, 25\,mrad}}\\
    \code{fold\_slice}, 400 + 125 it.$^\dagger$ & info.\ limit 0.31\,\AA; rms against simulated truth 4.5\,pm in-plane~\cite{huang2026moire3d}; no data-fit value & no & 0.4479 & 0.4552 & 0.007 \\
    \code{PtyRAD}, sub-scan & info.\ limit 30\,pm (other data); SSIM; data error rejected as a metric~\cite{lee2025ptyrad} & no & 0.4615 & 0.4735 & 0.012 \\
    \code{PtyRAD}, full field & as above & no & 0.4680 & 0.4817 & 0.014 \\
    \code{torchslice}, L1, 50 ep. & L1 training loss; 2--4\,pm s.d.\ over 25 unit cells (PZT)~\cite{diederichs2024exact} & no & 1.098 & 1.105 & 0.007 \\
    \code{torchslice}, L1, 500 ep. & as above & no & 0.631 & 0.688 & 0.058 \\
    \code{torchslice}, Poisson, 500 ep. & (our loss swap) & -- & 0.505 & 0.548 & 0.043 \\
    \code{torchslice}, L1, 1{,}000 ep. & as above & no & 0.598 & 0.660 & 0.061 \\
    \code{torchslice}, Poisson, lr$\times$10, 1{,}000 ep. & (our loss swap, tuned by us) & -- & 0.449 & 0.458 & 0.009 \\
    ours, pixels, 8 modes$^\dagger$ & $D_H$, gap; Fisher bars with coverage & yes & 0.3758 & \textbf{0.4025} & 0.027 \\
    ours, atoms$^\dagger$ & $D_H$, gap; per-atom Fisher bars & yes & 0.4447 & 0.4496 & \textbf{0.005} \\
    \midrule
    \multicolumn{6}{@{}l}{\emph{PrScO$_3$ [001], 300\,kV, 21\,nm}}\\
    \code{fold\_slice}, 200 + 200 it. & info.\ limit 0.23\,\AA; 0.7\,pm s.d.\ of repeated Pr--Pr distances; data error against trial thickness (training)~\cite{chen2021latticevib} & no & 0.685 & 0.736 & 0.051 \\
    \code{PtyRAD}, 1{,}000 it. & info.\ limit 23\,pm~\cite{lee2025ptyrad} & no & 0.394 & 0.488 & 0.094 \\
    \code{PtyRAD}, 4{,}000 it. & as above & no & 0.389 & 0.485 & 0.096 \\
    \code{torchslice}, L1, 50 ep. & released notebook (``a demonstration, not fine-tuned'') & no & 0.651 & 0.745 & 0.094 \\
    \code{torchslice}, Poisson, 50 ep. & (our loss swap) & -- & 0.664 & 0.754 & 0.089 \\
    \code{torchslice}, L1, 100 ep. & (run on by us) & no & 0.526 & 0.627 & 0.101 \\
    \code{torchslice}, Poisson, 100 ep. & (our loss swap) & -- & 0.572 & 0.668 & 0.097 \\
    ours, pixels, 8 modes & $D_H$, gap & yes & 0.287 & \textbf{0.416} & 0.129 \\
    ours, pixels, 4 modes & $D_H$, gap & yes & 0.291 & 0.422 & 0.131 \\
    \midrule
    \multicolumn{6}{@{}l}{\emph{PZT film, 200\,kV, 20\,nm}}\\
    \code{torchslice}, L1, 50 ep. & 2--4\,pm s.d.\ over 25 unit cells, from their atom pipeline~\cite{diederichs2024exact} & no & 1.028 & 1.106 & 0.078 \\
    \code{torchslice}, Poisson, 50 ep. & (our loss swap) & -- & 0.951 & 1.035 & 0.083 \\
    ours, pixels, 8 modes & $D_H$, gap & yes & 0.430 & \textbf{0.575} & 0.146 \\
    \bottomrule
  \end{tabular}
\end{table*}

\paragraph{Atom fits in the literature.} Diederichs \etal\ fit atoms to experimental 4D-STEM data of a 20\,nm PbZr$_{0.2}$Ti$_{0.8}$O$_3$ film, about five depths of field. Their differentiable multislice includes frozen phonons, five aberrations, a fitted partial-coherence ensemble, crystal tilt and scan positions, under an L1 loss on intensities. They report a spread of 2--4\,pm over 25 unit cells, not a per-atom bar~\cite{diederichs2024exact}. Herdegen \etal\ recover temperature in such inversions on simulated data~\cite{herdegen2024thermal}. Singh \etal\ fit isotropic Gaussian atoms to ptychographic phase projections for tomography~\cite{singh2025gaussian}, and Mani \etal\ train a network on randomized Gaussian atoms to clean tomograms~\cite{mani2025anyatom}. Both work on reconstructed images, not detector counts. Crystallographic refinement fits positions and displacement parameters to diffraction data by maximum likelihood~\cite{murshudov1997refinement}, and to cryo-EM maps in real space~\cite{afonine2018real}.

\section*{S10. More datasets}
\Cref{tab:quantitiesfull} gives the characteristic numbers of every dataset, and \cref{tab:quantities} repeats the main ones. Lengths are in $\lambda/\alpha$ across the beam and in $\lambda/\alpha^2$ along it.
\begin{table*}[t]
  \centering\small
  \caption{Characteristic quantities of the datasets of tungsten diselenide (WSe$_2$), palladium hydride (PdH) and praseodymium scandium oxide (PrScO$_3$), computed from the counts and the instrument parameters. Doses are the electrons recorded inside the detector disc; $\sigma V$ is the projected phase band-limited to the probe's transfer disc, $2\alpha$. $^{a}$Defocus of the probe of the published reconstruction ($-238$\,\AA\ $=-0.45\sqrt{C_s\lambda}$); our physics-built fits do not determine it (the estimate follows its start). $^{b}$Fitted, 5.7\,nm underfocus; 7\,nm underfocus nominal at acquisition. $^{c}$Nominal, 20\,nm overfocus~\cite{chen2021latticevib}.}
  \label{tab:quantitiesfull}
  \resizebox{\textwidth}{!}{%
\begin{tabular}{lrrrrrrrrrr}
\toprule
dataset & $\lambda/\alpha$ (\AA) & $\lambda/\alpha^2$ (\AA) & step & pixel & $t$ & $\Delta f$ & dose (e/\AA$^2$) & e per $(\lambda/\alpha)^2$ & $\sqrt{C_s\lambda}$ (\AA) & $\sigma V$ W / Se$_2$ \\
 & & & \multicolumn{2}{c}{in $\lambda/\alpha$} & \multicolumn{2}{c}{in $\lambda/\alpha^2$} & & & & (rad) \\
\midrule
WSe$_2$ bilayer, Nguyen 2024, 7.5 mrad & 3.34 & 446 & 0.16 & 0.039 & 0.01 & -- & 4.1e+05 & 4.6e+06 & 525 & 0.28 / 0.34 \\
WSe$_2$ bilayer, Nguyen 2024, 10.3 mrad & 2.43 & 236 & 0.22 & 0.053 & 0.03 & $-1.01^{a}$ & 4.2e+05 & 2.5e+06 & 525 & 0.42 / 0.50 \\
WSe$_2$ bilayer, Nguyen 2024, 22.5 mrad & 1.11 & 50 & 0.36 & 0.187 & 0.13 & -- & 7.1e+05 & 8.8e+05 & 525 & 0.95 / 1.06 \\
WSe$_2$ twisted bilayer~\cite{zhang2025phasons}, 25 mrad & 1.68 & 67 & 0.26 & 0.078 & 0.15 & $0.91^{b}$ & 5.1e+05 & 1.4e+06 & -- & 0.89 / 1.02 \\
WSe$_2$ monolayer (unpublished), 18.4 mrad & 2.26 & 123 & 0.13 & -- & 0.05 & -- & 9.5e+05 & 4.9e+06 & -- & 0.64 / 0.74 \\
WSe$_2$ twisted bilayer, simulated (Zenodo), 25 mrad & 1.68 & 67 & 0.26 & 0.078 & 0.15 & 1.30 & 2.2e+05 & 6.3e+05 & -- & 0.89 / 1.02 \\
WSe$_2$ twisted bilayer, published 6$^\circ$ (Zenodo), 25 mrad & 1.68 & 67 & 0.26 & 0.078 & 0.15 & -- & 6.5e+05 & 1.8e+06 & -- & 0.89 / 1.02 \\
PdH nanoparticle (PtychoHub), 30 mrad & 0.66 & 22 & 0.64 & 0.234 & 12.80 & -- & 5.4e+05 & 2.3e+05 & -- & -- \\
PrScO$_3$ crystal (PARADIM), 21.4 mrad & 0.92 & 43 & 0.45 & 0.203 & 4.88 & $-4.65^{c}$ & 1.2e+06 & 1.0e+06 & -- & -- \\
\bottomrule
\end{tabular}}

\end{table*}
\paragraph{Datasets.}
\begin{itemize}
  \item \textbf{Thin, public}~\cite{nguyen2024uncorrected}: twisted bilayer WSe$_2$, uncorrected 200\,kV Talos ($C_s$ 1.1\,mm, the vendor value; $\sqrt{C_s\lambda}=525$\,\AA), EMPAD, $128\times128$ scan, apertures 7.5, 10.3 and 22.5\,mrad; dose $4.1$--$7.1\times10^5$\,e/\AA$^2$.
  \item \textbf{Thin, experimental}: the twisted bilayer WSe$_2$ data of Zhang \etal~\cite{zhang2025phasons}, 80\,kV (not calibrated), aberration-corrected (retuned each session; $C_s$ not measured), Schottky source with 0.8--1\,eV spread, EMPAD; 25\,mrad (the main real-data set, nominal 7\,nm underfocus). From the same microscope, an unpublished 18.4\,mrad scan of a monolayer region tests the learned baseline outside its training range (\S S9). Counts are dark-subtracted raw values over 151\,ADU per electron (\S S14). $\sigma V$ = 0.89\,rad for W and 1.02\,rad for an Se$_2$ column, so multiple scattering matters even in a bilayer. At $5.1\times10^5$\,e/\AA$^2$ about 5\% of outer-plane Se sites are vacant after imaging~\cite{huang2026moire3d}, so the dose is at the onset of damage.
  \item \textbf{Simulated truth}~\cite{huang2026moire3d}: the bilayer with known coordinates (1{,}928 atoms in the window) at 80\,kV, 24.92\,mrad, step 0.42900\,\AA\ = 0.256\,$\lambda/\alpha$, tilt 12.7$^\circ$, coherent probe, static atoms (thermal width zero), Poisson counts at $2.4\times10^5$\,e/\AA$^2$ nominal ($2.2\times10^5$ in the disc).
  \item \textbf{Published 6$^\circ$ bilayer}~\cite{huang2026moire3d}: untilted $6^\circ$ bilayer under the conditions of the main dataset, $6.5\times10^5$\,e/\AA$^2$. The published tilted bilayer (12.5$^\circ$ tilt) is not fitted.
  \item \textbf{Thick:} PdH nanoparticle (PtychoHub~\cite{ptychohub,shi2025pdh}), 300\,kV, 30\,mrad, 280\,\AA; PrScO$_3$ [001]~\cite{chen2021latticevib,paradim_ssmm2j11}, 300\,kV, 21.4\,mrad, 210\,\AA, whose public release is a $64\times64$ part of the published scan, without a reconstruction; PbZr$_{0.2}$Ti$_{0.8}$O$_3$ (PZT) film of Diederichs \etal~\cite{diederichs2024exact} (CC BY 4.0), 200\,kV, 24.6\,mrad, $\lambda/\alpha=1.02$\,\AA, 52 cells of 3.95\,\AA\ = 205\,\AA\ = 5.0 depths of field, step 0.35\,\AA\ = 0.34\,$\lambda/\alpha$, $70\times70$ scan, EMPAD, $1.1\times10^6$\,e/\AA$^2$ ($1.2\times10^6$ per resolution cell). PZT holds dark-subtracted ADU without a gain. We estimate 378\,ADU/e from the single-electron peak and EMPAD calibrations scaled to 200\,keV; a refit at 270 keeps every ranking (\S S2). Its lattice ($a_x=3.95$, $a_y=4.15$\,\AA\ against 3.953 and 4.148\,\AA) confirms step and detector calibration to better than 1\%. We report no atom positions for it.
\end{itemize}
Every fit uses the Poisson likelihood and held-out rule of \cref{sec:inference}. We report $D_H$ (lower is better) and the gap $D_H-D_T$. Depth nodes are every 0.25\,\AA\ ($3.7\times10^{-3}\,\lambda/\alpha^2$), and each atom is spread in depth by a Gaussian of $\sigma_z=0.4$\,\AA\ (\S S1; sensitivity in \S S6).

\paragraph{Public bilayer.} On the sub-scan, the physics-built instrument halves the train--test gap at matched accuracy: free modes 0.299--0.307 (gap 0.025), physics-built 0.324 (0.012), physics-built + 2 modes 0.305 (0.012). On the full field (2{,}048 held-out patterns) the three give 0.2906 (gap 0.010), 0.3235 (0.006) and 0.2913 (0.008). The information limit is the largest frequency at which a lattice spot exceeds the 99th percentile of the noise maximum in its window. It is 0.41--0.49\,\AA\ = 0.17--0.20\,$\lambda/\alpha$ for free modes over three seeds and 0.45\,\AA\ with the physics-built instrument, against 0.46\,\AA\ for the published reconstruction. Requiring two symmetry-equivalent spots gives 0.49--0.52 and 0.46\,\AA, so the two are comparable, not ranked.
The free-mode fit takes 381\,s on one V100; the published run took 16.7\,h by its log, 158$\times$ longer (19--113$\times$ against the 2--12\,h the article quotes). The published code, retrained without the held-out patterns, reaches 0.345 near 1{,}000 iterations and then over-fits (0.352 at 9{,}000), with transmission above 1 on 57\% and then 94\% of the field. Our physics-built fits reproduce across seeds to $2\times10^{-4}$.

\paragraph{Instrument on the public data.} The fits do not determine $C_s$ or focal spread at 7.5 and 10.3\,mrad ($C_s$ about 0 and 0.15\,mm against 1.1\,mm), and at 22.5\,mrad bound only $C_s>0.6$\,mm. The counts carry the information (Cram\'er--Rao bound 2.5--8\,\textmu m, under 1\% of 1.1\,mm; \S S16). Model error coupled to the free modes leaves $C_s$ undetermined.

\paragraph{Thick specimens} (pixel object, free modes, same held-out rule for every method). Atoms are not fitted here: a column holds tens of atoms, about ten per depth of field, so it needs a joint parameterization, \eg\ a lattice with displacements. This joint column model is in development.
\begin{itemize}
  \item \textbf{PdH} (12.8 depths of field): $D_H$ 0.401 (thick region) and 0.350 (mid), against 0.419 and 0.385 for \code{PtyRAD} retrained without the held-out patterns~\cite{lee2025ptyrad}; full field ($256\times256$, 8{,}192 held-out) 0.367 against 0.390 (one seed; thick corner 0.414, mid 0.369). The physics-built instrument gives 0.486 and 0.458, 0.085 and 0.108 above free modes. Adding two free modes (0.429, 0.359) lets them absorb what the instrument model leaves out.
  \item \textbf{PrScO$_3$} (4.9 depths of field, 512 held-out; \cref{tab:heldout_prsco3}): free modes 0.4157 with 8 modes (seeds within $6\times10^{-4}$, gap 0.129), 0.4224 with 4 (0.131); \code{PtyRAD} at its shipped settings (21 slices of 10\,\AA, 4 modes, 200 iterations)~\cite{lee2025ptyrad} 0.559 (gap 0.070, still falling), 0.488 (0.094) at 1{,}000 iterations after 2.8\,h; and \code{fold\_slice} at its published script 0.736 (0.051), best saved state 0.717. Our fit passes \code{PtyRAD}'s 200-iteration score after 110\,s, 19 times sooner on the same GPU. Scored at the nominal raster instead of the neighbor rule, our $D_H$ rises from 0.423 to 0.795. The offsets jitter along the fast scan (rms 0.17\,\AA), and the neighbor mean misses them by 0.09\,\AA\ ($0.10\,\lambda/\alpha$) rms, so part of every method's $D_H$ here is position error.
  \item \textbf{PZT}: 613 held-out, depth schedule as for PrScO$_3$ (7 slices of 30\,\AA, then 21 of 9.8\,\AA), orientation and defocus ($-100$\,\AA) screened on a $32\times32$ sub-scan; $D_H$ 0.575 (8 modes, 30\,min on an A6000) and 0.598 (4 modes, 15\,min), gaps 0.146 and 0.151 (\cref{tab:heldout}).
\end{itemize}

\paragraph{6$^\circ$ bilayer, full field.} After screening registry and stacking by held-out deviance, the atom fit (all in-window atoms free, 34\,h on one RTX A6000) reaches $D_H$ 0.549, gap $8\times10^{-5}$. The published \code{fold\_slice} reconstruction, which used every pattern, scores 0.456 through our likelihood, so here atoms trail pixels by 0.09 per pixel. The layer-mean W--W spacing is accurate. Per-atom bounds would need about 6{,}800 A100-hours (\S S16).

\section*{S11. Timing and hardware}

\paragraph{Reconstruction against acquisition.} The 1{,}024-pattern sub-scan is recorded in about 1\,s (1\,ms per pattern; the full $128\times128$ scan takes 16\,s). \code{fold\_slice} reconstructs it in 20.5\,h and \code{PtyRAD} in 22\,min, $7\times10^4$ and $1.3\times10^3$ times the acquisition time (RTX A5000, published settings). Our free-mode fit passes both scores after 5.3\,min on one RTX A6000, $3\times10^2$ times the acquisition time; from a learned warm start it converges 3.1$\times$ sooner (\S S9). A ratio near one would let the fit steer the next scan.

\paragraph{Time to accuracy, WSe$_2$.} Our free-mode pixel fit with exact detector-pixel integration on one RTX A6000 reaches held-out $D_H$ 0.453 after 250 iterations (5.3\,min), already below \code{fold\_slice} (0.455, 20.5\,h) and \code{PtyRAD} (0.474, 22\,min). It reaches 0.412 after 500 iterations (10.6\,min) and 0.4025 after 2{,}500 (53\,min). With exact integration on a $2\times$ sub-sampled pixel instead of $3\times$, it reaches 0.4023 in 26\,min. Each budget is a separate run from the same start.

\paragraph{Time to accuracy, simulated bilayer.} The atom fit of \cref{tab:accsim}, started from the published peaks with the same flags, is run with shorter budgets on one RTX A6000 (each a separate run; last column: held-out $D_H$):
\begin{center}\small
\resizebox{\columnwidth}{!}{\begin{tabular}{@{}lccc@{}}\toprule
budget & in-plane (pm)\,$\downarrow$ & depth W\,/\,Se (pm)\,$\downarrow$ & $D_H$\,$\downarrow$\\\midrule
start & 4.9 & 49\,/\,46 & 0.513\\
2 Adam epochs, 0.31\,h & 4.0 & 47\,/\,45 & 0.392\\
5 Adam epochs, 0.93\,h & 3.5 & 45\,/\,44 & 0.387\\
10 Adam epochs, 1.96\,h & 3.3 & 41\,/\,43 & 0.386\\
full fit (30 epochs + 40 L-BFGS), 14\,h & 3.2 & 29\,/\,39 & 0.384\\
+ depth-only polish, 17.8\,h & 3.2 & 50\,/\,49 & 0.384\\
+ joint $x,y,z$ polish, 20.6\,h & 3.2 & 51\,/\,51 & 0.384\\
published pipeline~\cite{huang2026moire3d}, 1.9\,h & 4.8 & 84\,/\,127 & --\\\bottomrule
\end{tabular}}
\end{center}
In-plane accuracy passes the published value within 20\,min and is within 0.3\,pm of the full fit after 1\,h. Depth moves slowly. An hour leaves it near the start; the polish moves it to the likelihood minimum, 50\,/\,49\,pm. A block of 5 L-BFGS iterations after 5 epochs changes neither number. A joint polish of $x$, $y$ and $z$ after the depth-only polish lowers $D_H$ by $4\times10^{-5}$ and ends at 51\,/\,51\,pm in depth, with in-plane unchanged. Freeing all three coordinates together does not improve depth over the depth-only polish.
\paragraph{Other timings.} Free-mode pixel fit of the public 10.3\,mrad data (full $128\times128$ scan, 10 modes): 381\,s on one V100. The published \code{fold\_slice} run of the same data: 16.7\,h on a V100 by its log (9{,}000 iterations at 6.7\,s). \code{fold\_slice} on the 25\,mrad data at its published settings: 140\,s per iteration on an RTX A5000 (15.6\,h for stage 1, 4.9\,h for stage 2). Atom model with frozen phonons ($K=32$ configurations, 81 coherence modes, no free modes, 1{,}702 atoms, 896 training patterns): exact pixel integration at bin 2 costs 850\,s per gradient pass on two RTX A6000 (211\,s with sinc pixels). The final two stages of the fit of \cref{tab:heldout} take 42\,h wall on 2--5 RTX A6000 (118 GPU-hours), and the refit with the calibrated detector ellipse 24\,h on six. The 300-iteration fits take 4.4\,min per iteration on an RTX A6000.
Fisher-vector product on one A100: 1.7\,s against 1.2\,s for the likelihood and its gradient over 4{,}096 patterns. Depth block of the simulated bilayer: 2.8\,h on an RTX 2080\,Ti. The full field: 14\,h on an RTX A6000 (start (i); A5000 for start (ii)). The 6$^\circ$ bilayer full field: 34\,h on an RTX A6000. Free-mode pixel fit of PrScO$_3$ (full $64\times64$ scan, 21 slices): 403\,s with 4 modes and 783\,s with 8 modes on one A100 (6.5 and 12.7\,s per pass over the 3{,}584 training patterns). \code{PtyRAD} takes 34\,min for 200 iterations (10.2\,s each), and \code{fold\_slice} 95\,min for 200 + 200 iterations. The published reconstruction of the full $128\times128$ scan (12 modes, 42 slices, 256-pixel patterns) took about 45\,min per iteration for fewer than 100 iterations on a Quadro P5000~\cite{chen2021latticevib}. These A100 times are the 7 and 13\,min of \cref{tab:heldout}. On one RTX A6000, the card of the \code{torchslice} comparison: our PrScO$_3$ fit takes 12.8\,min with 4 modes and 25.1\,min with 8 (L1 in place of the likelihood: the same), and \code{torchslice} 4.7\,h for its 50 epochs under either loss (337\,s per epoch over 3{,}584 patterns). On WSe$_2$ the times are 413\,s against 249\,s for its 50 epochs, and on PZT 15 and 30\,min.

\section*{S12. Error budget}
The Fisher bounds of \S S4 are dose bounds that assume the model is right. This section projects measurement errors through the Fisher information of one reference fit of the 25\,mrad bilayer, the \emph{budget fit}, and compares them with the per-atom width misfit measured there. There the misfit, one Gauss--Newton step from the data, has a median of 3.5 (W) and 3.4 (Se) Fisher standard deviations (p90 14 and 8.9; last row of \cref{tab:errbudget}), where shot noise alone gives 1. The real per-atom bounds are 3--5 times the synthetic ones. The error budget identifies the measurement errors that account for this, and those below the shot-noise floor.

\paragraph{Conditions.} 80\,kV, $\alpha=24.92$\,mrad, $\lambda/\alpha=1.676$\,\AA, depth of field $\lambda/\alpha^2=67.2$\,\AA; EMPAD, $128\times128$ scan, step 0.429\,\AA\ $=0.256\,\lambda/\alpha$; $9.6\times10^4$\,e per pattern, $5.2\times10^5$\,e/\AA$^2$ ($1.46\times10^6$\,e per $(\lambda/\alpha)^2$) on the uncorrected counts. The bilayer spans 11.6\,\AA\ $=0.17\,\lambda/\alpha^2$ in the fitted model of this section and 10.08\,\AA\ $=0.150\,\lambda/\alpha^2$ at the converged fit. \Cref{tab:quantities} uses the nominal 9.84\,\AA\ ($t=0.15$). The three approximations that save cost are $K=32$ phonon configurations, slices at the atomic planes and a lateral sampling of $0.078\,\lambda/\alpha$.
The budget fit is a frozen-phonon fit with one width per species ($\sqrt U$ = 5.18\,pm W, 5.08\,pm Se), on the counts before the dark-reference correction (\S S14), evaluated with $K=32$ configurations. Its 236 interior atoms carry a free width.

\paragraph{Method.} For each source we switch it on at a plausible size in the fitted model, or switch a modeled term off, and compute the change $\Delta h_j$ in the expected counts on the 71\,mrad disc of all 1024 patterns. A fit that lacks the source is biased by one Gauss--Newton step,
\[ \Delta\boldsymbol\theta = \mathcal I^{-1}\textstyle\sum_j\mathbf J_j^{\!\top}\operatorname{diag}(1/h_j)\,\Delta h_j, \]
with $\mathcal I^{-1}$ the gauge-fixed Laplace covariance of \S S4 (1636 global and per-atom columns, 1792 per-pattern offsets). We report the induced width and position shifts in units of their Fisher sd, the change in held-out deviance per pixel $D_H$ before and after the linear refit, and how much $D_H$ the measured data would gain if the source were fitted at its best amplitude. Checks: a log-width step of one atom projects onto that atom to 0.8\% (the curvature of a 2\% step) with every other parameter below $2\times10^{-5}$\,sd; a 0.002\,\AA\ position step projects to $10^{-6}$; the data residual reproduces the stored per-pattern scores to $4\times10^{-6}$.

\begin{table*}[t]
\centering\scriptsize
\resizebox{\textwidth}{!}{%
\setlength{\tabcolsep}{3pt}
\begin{tabular}{lllcccccc}
\toprule
Source & Size & Model & $|\zeta|$ W & $|\zeta|$ Se & $|\zeta_{xy}|$ & $\overline{\Delta\sqrt U}$ W/Se & $\Delta D_H$ & data gain \\
 & & & med/p90 & med/p90 & p90 & pm & no refit/refit & \\
\midrule
\multicolumn{9}{l}{\textit{1. Shot noise}} \\
Poisson counts & $9.6\times10^4$\,e/pattern & includes & 1 & 1 & 1 & sd 1.09/2.05 & -- & -- \\
\multicolumn{9}{l}{\textit{2--4. Detector, scan, specimen}} \\
Gain & 151$\to$140\,ADU/e ($-7.3\%$) & fits (dose) & $<$0.01/$<$0.01 & $<$0.01/$<$0.01 & $<$0.01 & $<$0.01/$<$0.01 & 49.9/$<$0.01 & -- \\
Flat background & $+0.05$\,e/px & fits & $<$0.01/$<$0.01 & $<$0.01/$<$0.01 & $<$0.01 & $<$0.01/$<$0.01 & 2.14/$<$0.01 & -- \\
Read noise, clipped at 0 & $\sigma_r=0.33$\,e (uncorrected); bias 0.13\,e/px dark & ignores & 0.66/1.45 & 0.43/1.23 & 1.37 & 0.94/1.10 & 5.17/0.08 & 0.15 \\
PSF & 0.05$\to$0.174\,px & pinned at floor & 0.20/0.55 & 0.20/0.44 & 0.74 & -0.28/0.50 & 2.28/1.64 & 2.28 \\
Scan affine & $0.057^\circ$, steps $\pm0.35\%$ & fits (offsets, drift) & $<$0.01/0.02 & $<$0.01/0.02 & 0.02 & $<$0.01/$<$0.01 & 0.12/$<$0.01 & $<$0.01 \\
Specimen drift & 1\,\AA/min: 0.068\,\AA\ = 0.04\,$\lambda/\alpha$ & fits (drift, offsets) & $<$0.01/0.02 & $<$0.01/0.01 & 0.02 & $<$0.01/$<$0.01 & 0.11/$<$0.01 & $<$0.01 \\
Se vacancies (static) & 5\% of outer Se & ignores & 0.29/1.31 & 0.30/1.43 & 1.93 & -0.04/0.98 & 2.40/1.09 & $<$0.01 \\
Saturation & max 162\,e/px/frame & ignores & \multicolumn{6}{l}{$<10^{-3}$ of the EMPAD range; no effect} \\
Scan flyback & 100--200\,\textmu s/line & ignores & \multicolumn{6}{l}{sub-scan columns 84--115, far from the line start} \\
Se loss during the scan & 0.6--1.1\% per scan & ignores & \multicolumn{6}{l}{occupancy deficit $\le$0.006 per site while sampled: $\le$0.006$\times$ the bias of one vacancy} \\
\multicolumn{9}{l}{\textit{5--6. Partial coherence and aberrations beyond $C_s$}} \\
Focal-spread rule & 9 vs 19 nodes & includes & $<$0.01/$<$0.01 & $<$0.01/$<$0.01 & $<$0.01 & $<$0.01/$<$0.01 & $<$0.01/$<$0.01 & 0.87 \\
Focal-spread shape & skewness $-0.38$ & ignores & 0.19/0.41 & 0.13/0.30 & 0.35 & -0.14/0.02 & 0.46/0.18 & 0.02 \\
$A_3$ & $\pi/4$ at $\alpha$ (5.4\,\textmu m) & ignores & 0.77/2.12 & 0.68/1.61 & 2.26 & 0.36/0.52 & 5.74/5.38 & $<$0.01 \\
$S_3$ & $\pi/4$ (1.35\,\textmu m) & ignores & 0.30/0.63 & 0.25/0.60 & 1.14 & 0.27/0.52 & 6.33/1.01 & 0.14 \\
$D_4$ & $\pi/4$ & ignores & 0.44/1.05 & 0.30/0.71 & 1.02 & 0.63/0.55 & 5.60/0.79 & 0.34 \\
$B_4$ & $\pi/4$ & ignores & 0.10/0.27 & 0.08/0.17 & 0.76 & 0.12/0.10 & 2.32/0.14 & 0.04 \\
$A_4$ & $\pi/4$ (0.27\,mm) & ignores & 0.74/1.70 & 0.52/1.33 & 2.27 & 0.60/0.40 & 4.09/3.72 & 0.08 \\
$C_5$ & $\pi/4$ (13\,mm) & ignores & 0.05/0.11 & 0.05/0.09 & 0.19 & 0.02/0.09 & 1.30/0.05 & 0.03 \\
\multicolumn{9}{l}{\textit{7--8. Thermal scattering and geometry}} \\
Frozen-phonon sampling & $K=32$ & includes & 0.39/1.43 & 0.20/0.68 & 0.26 & 0.04/$<$0.01 & 0.05/0.02 & 0.01 \\
Fit at $K=8$ vs $K=32$ & $K$ 8$\to$32 & includes & 0.61/2.53 & 0.43/1.20 & 0.46 & 0.10/0.03 & 0.15/0.07 & $<$0.01 \\
TDS (mean field, same $U$) & switch off & includes & 4.76/10.8 & 0.88/2.72 & 2.07 & -6.55/-1.82 & 6.17/0.62 & 0.34 \\
Absorptive instead of TDS & switch & includes & 1.22/3.21 & 0.43/1.27 & 0.78 & -1.34/-0.76 & 1.66/0.09 & 0.26 \\
In-plane anisotropy & 1.3:1 & ignores & 0.24/0.97 & 0.12/0.40 & 0.28 & $<$0.01/-0.07 & 0.02/$<$0.01 & 0.35 \\
Local tilt & 10\,mrad (0.12\,\AA\ over 11.6\,\AA) & ignores & 0.92/2.86 & 0.38/1.13 & 7.70 & 1.20/0.76 & 4.21/0.09 & 0.03 \\
Local tilt & 20\,mrad & ignores & 3.29/9.76 & 1.14/4.08 & 13.6 & 4.36/2.74 & 15.1/1.04 & 0.02 \\
Moir\'e corrugation & height $\approx0.15$\,\AA\ (est.) $=0.002\,\lambda/\alpha^2$ & ignores & \multicolumn{6}{l}{interlayer change across the 13.7\,\AA\ window; below the depth of field by 450$\times$} \\
\multicolumn{9}{l}{\textit{9. Forward-model discretization and potential}} \\
Lateral sampling & $\Delta x$ 0.078$\to$0.058\,$\lambda/\alpha$ & includes & 0.10/0.35 & 0.08/0.21 & 0.20 & 0.05/0.04 & 0.01/$<$0.01 & 0.04 \\
Probe window & 25$\to$50\,\AA & includes & 0.02/0.05 & 0.02/0.04 & 0.14 & $<$0.01/-0.01 & 0.08/0.08 & 0.37 \\
Slicing & planes$\to$0.5\,\AA\ finite projection & includes & 0.67/2.26 & 0.41/1.13 & 0.85 & -1.10/0.03 & 0.65/0.16 & 0.72 \\
Peng vs Dirac--Fock atom & DF $+1.6\%$ (W), $-1.1\%$ (Se) & ignores & \multicolumn{3}{l}{common to a species: $-0.27$/$+0.19$\,pm (band center $2\alpha$)} & \multicolumn{3}{l}{1.2/0.5 sd of the species mean} \\
Schr\"odinger vs Dirac & DF $-0.24\%$/$+0.29\%$ & ignores & \multicolumn{6}{l}{0.2$\times$ the Peng row} \\
\midrule
\textbf{Measured misfit} & one GN step from the data & & 3.5/14 & 3.4/8.9 & 8.7 & & & \\
\bottomrule
\end{tabular}}
\caption{Error budget of the 25\,mrad bilayer at the budget fit (counts before the dark-reference correction). $|\zeta|$: induced per-atom bias in units of its Fisher sd over the 236 interior atoms (78 W, 158 Se). $\overline{\Delta\sqrt U}$: mean width shift. $\Delta D_H$ in $10^{-3}$ per pixel (held-out before refit, training after the linear refit). Data gain: decrease of $D_H$ ($10^{-3}$) if the source were fitted to the data at its best amplitude. ``--'': absorbed by fitted parameters to $10^{-4}$ (unresolved). Aberrations are set to $\pi/4$ of phase at the aperture edge, the corrector tuning tolerance; the bias scales linearly below it.}
\label{tab:errbudget}
\end{table*}

\paragraph{Below the shot-noise floor.} Everything the model fits is absorbed: a 7\% gain error goes into the dose scale, a background offset into the background, the scan affine between two scans of the session (0.057$^\circ$, steps differing by 0.7\%; 0.030\,\AA\ $=0.018\,\lambda/\alpha$ across the sub-scan) and 1\,\AA/min of specimen drift into the scan polynomial and per-pattern offsets. Each moves widths by less than 0.01\,sd. So do the focal-spread quadrature (9 against 19 nodes), finer lateral sampling ($\Delta x$ 0.078$\to$0.058\,$\lambda/\alpha$), a 50\,\AA\ probe window in place of 25\,\AA, $B_4$ and $C_5$ at the tuning tolerance, saturation and flyback.

\paragraph{Comparable to the floor.} At plausible sizes, ten sources each move the median width by 0.2--0.9\,sd and the worst tenth by up to 2.9\,sd: clipped read noise (0.33\,e, sized to the 31--39\% negative pixels of the uncorrected counts; \S S14 measures 0.02--0.04\,e on the corrected counts),%
 the detector PSF held at its floor, $A_3$, $S_3$, $D_4$ and $A_4$ at $\pi/4$, finite-projection slicing, a 10\,mrad local tilt, the frozen-phonon sampling at $K=32$, and the difference between the $K=8$ fit and the $K=32$ evaluation. Static Se vacancies (5\% of outer sites) move only their neighbors (p90 1.4\,sd). Combined in quadrature these give a median of 2.0\,sd (p90 6.0) for W and 1.3\,sd (p90 3.7) for Se, against the measured 3.5 (p90 14) and 3.4 (p90 8.9). At the budget fit they account for about a third of the W misfit variance and a sixth of the Se.

\paragraph{Closing the gap.} Two terms reach the measured scale. Switching thermal diffuse scattering off (mean field at the same $U$) biases W by 4.8\,sd median and $-6.6$\,pm on average. The model includes TDS, so only a fraction of this, the error of independent-atom frozen phonons, is a candidate. A uniform 20\,mrad (1.1$^\circ$) tilt gives 3.3/9.8\,sd for W, because the Se pair and the two layers shear by the tilt times their separation along the beam. With it the quadrature sum reaches 3.7/11 for W, matching the measurement, but only 1.7/5.4 for Se. The data do not favor a uniform tilt along the fast axis (best amplitude $-0.6\pm0.03$ of 10\,mrad in training, not significant on held-out patterns). Fitting a tilt along the slow scan axis and one tilt per moir\'e region as nuisance parameters is the next step. It adds two to six parameters and repeats the per-atom Fisher pipeline (about 390 CPU node-hours for this sub-scan; \S S16). It is not run here.

\paragraph{Held-out deviance.} At the budget fit, fitting every source above at its best amplitude would lower $D_H$ by at most $6.1\times10^{-3}$ per pixel in total. That is small against the 0.047 per pixel that separates the converged atom model (0.4496) from our free-mode pixel fit (0.4025; \cref{tab:heldout}), so most of that gap lies outside this list. The absolute $D_H$ lies below the Poisson expectation at the fitted counts (0.511 per pixel; 0.453 with clipped read noise). The counts are sub-Poisson (Pearson dispersion 0.92 at the converged atom fit), consistent with a gain of 140 rather than 151\,ADU/e plus clipping. $D_H$ is therefore compared only between models.

\paragraph{Real against synthetic bounds.} The real bounds are 3--5 times the synthetic ones. Three factors are measured at the budget fit. (i) Thermal diffuse scattering: at the same $U$ the static potential carries 3.9 times the diagonal width information of frozen phonons for W (1.3 for Se). Diffuse scattering returns to the detector much of what the Debye--Waller factor removes from the Bragg beams, and the synthetic bounds use the static potential. (ii) Detector disc: the scorer's 71\,mrad disc holds 58\% of the width information of the full 106\,mrad detector for both species. (iii) Marginalizing neighbors, stacked partners and the instrument widens the conditional sd by 1.40 (W) and 1.29 (Se). Together they give 3.7 of the 4.9 for W and 1.9 of the 3.2 for Se. We attribute the remaining factor of 1.3 (W) and 1.7 (Se) to the synthetic test: its truth widths (36--45\,pm$^2$ against the fitted 26--27\,pm$^2$) and its 13-slice static model carry more width information than the real data at the same dose. Closing it needs the synthetic coverage test rerun with frozen phonons at the fitted widths (about 60 GPU-hours), which is not done here. The cluster-by-pattern sandwich bounds are 1.6 (W) and 1.4 (Se) times the Fisher bounds (per-atom medians): the same misfit, seen from the score side. The ratio grows with the size of the pattern clusters, which share misfit, and its scores use $K=8$ phonon configurations against the fit's $K=32$ (\S S4). It indicates the misfit; it does not calibrate the bars.

\paragraph{Dose.} The dose dependence of the bars and of the measured scatter is in \S S5.

\paragraph{Counting statistics.} The counts are slightly sub-Poisson (Pearson dispersion 0.88 on held-out patterns, flat in the expected count). The Fisher bars on the experimental data are therefore conservative by about 6\%. The dispersion $\phi=\sum(n-\lambda)^2/\lambda/N$ is measured at our point-sampled pixel fit (\S S2; iteration 2{,}500, $D_H=0.4077$) on the 5{,}721 pixels of the scoring disc. For variance $\phi\lambda$ the covariance of the Poisson-likelihood estimate is $\phi\,\mathcal I^{-1}$, so the true standard deviations are $\sqrt\phi\approx0.94$ of the Fisher ones. Clipping at zero leaves 26.5\% of the disc pixels at $n=0$ (25.1\% of the dark-subtracted values were negative). At the atom model the larger held-out misfit is model error, not noise.
\begin{center}\footnotesize
\begin{tabular}{@{}lccccc@{}}
\toprule
expected count $\lambda$ (e) & $<0.5$ & 0.5--2 & 2--10 & $\ge10$ & all \\
\midrule
share of disc pixels & 41\% & 32\% & 13\% & 13\% & \\
$\phi$, held-out (128 patterns) & 0.885 & 0.859 & 0.887 & 0.890 & 0.877 \\
$\phi$, training (896) & 0.728 & 0.806 & 0.843 & 0.823 & 0.781 \\
\bottomrule
\end{tabular}
\end{center}

\section*{S13. Information per detector pixel}
\begin{figure*}[hbt]\centering\includegraphics[width=\textwidth]{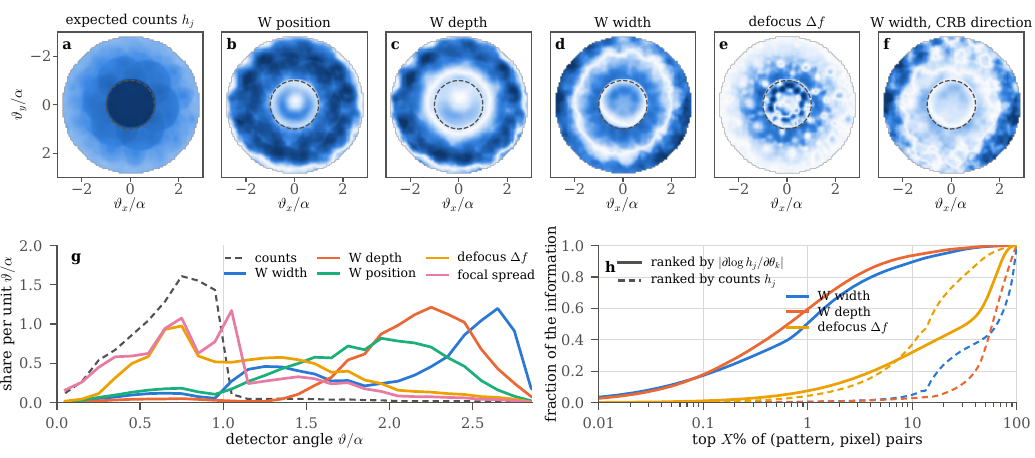}\caption{Per-pixel Fisher information of the 25\,mrad bilayer at the budget fit (\S S12). (a--f)~Probe on a W atom: expected counts $h_j$ (log scale), the Fisher density $(\partial h_j/\partial\theta_k)^2/h_j$ for the W atom's in-plane position, depth and width and for defocus, and the per-pixel share $(\mathbf J_j\mathbf v)^2/h_j$ of the Cram\'er--Rao direction $\mathbf v=\mathcal I^{-1}\mathbf e_k$ of the W width with the instrument profiled out; densities on a square-root scale, each to its own 99.5th percentile; dashed circle $\vartheta=\alpha$; the disc is the 71\,mrad scoring disc ($2.86\alpha$). (g)~Share of the information per unit $\vartheta/\alpha$, summed over the 896 training patterns, against the share of the counts. (h)~Fraction of the information carried by the top $X\%$ of (pattern, pixel) pairs, ranked by relative sensitivity $|\partial\log h_j/\partial\theta_k|$ (solid) or by counts (dashed).}\label{fig:pixfisher}\end{figure*}

A detector pixel with expected count $h$ adds $(\partial h/\partial\theta_k)^2/h$ to the information on $\theta_k$ (\cref{eq:fisher}). We evaluate it at the budget fit (\S S12) from the forward-mode Jacobian columns, which match central differences to $6\times10^{-5}$ and sum to the diagonal of the Fisher matrix to rounding. Depth is not a parameter of the fit. It is added for one W and one Se atom by moving each onto its own sub-slice, which reproduces the interlayer and defocus columns to $2\times10^{-5}$ (\cref{fig:pixfisher}).
The atoms draw 76--98\% of their information from beyond $1.1\alpha$, with medians near $2\alpha$ for width and depth. The brightest 10\% of the pattern pixels carry 1--18\% of it, and the 10\% with the largest relative sensitivity 76--94\%.
The instrument uses the bright field more. Defocus and the two coherence widths take 41--54\% of their information from inside $0.9\alpha$, and it grows with the counts (rank correlation 0.6, against at most 0.2 for the atoms).
The Cram\'er--Rao direction of one W width lies 96\% beyond $1.1\alpha$, and profiling the instrument changes its bound by 0.5\%.
At the budget fit, with the instrument profiled, the Fisher information bounds an atom's depth on the real bilayer to 12.9\,pm for W and 14.6\,pm for Se, about $2\times10^{-3}$ of the depth of field $\lambda/\alpha^2$. Freeing the depth widens the W width bound by 15\%.

\section*{S14. Detector check}
The 25\,mrad frames are raw detector values (\cref{sec:data}; dark reference subtracted in the detector orientation). We check the outer detector, beyond 3.7$\alpha$, where almost every frame records no electron:
\begin{itemize}
  \item The dark reference leaves a static residual of 2.6\,ADU rms per pixel (0.017 electrons at 151\,ADU per electron).
  \item The zero-electron peak has a standard deviation of 4\,ADU, of which 3\,ADU (0.02 electrons) is temporal read noise per pixel and frame. With the frame-to-frame common offset left in, as the fits see the counts, it is about 6\,ADU (0.04 electrons) around a zero level of $-5$\,ADU. Electrons that share their charge between pixels add a tail of partial deposits above the peak. Second differences over the scan and the variance--mean intercept include this tail and overstate the read noise by a factor of about seven.
\end{itemize}
Clipping the counts at zero (\cref{sec:inference}) turns this noise into a positive pedestal. With $o$ the zero level and $s$ the read noise, both known before the fit, the expected count $\lambda$ of \cref{eq:detector} becomes the expected clipped count
\begin{gather*}
  h=e^{-\lambda}G(o)+\big(1-e^{-\lambda}\big)\,G\Big(o+\frac{\lambda}{1-e^{-\lambda}}\Big),\\
  G(m)=s\,\phi(m/s)+m\,\Phi(m/s),
\end{gather*}
\ie\ no electron, or at least one at its conditional mean. This is within 1.6\% of the exact Poisson sum for $\lambda$ from $10^{-3}$ to 10. With the measured $o$ and $s$ the atom model ($K=32$, 20 L-BFGS iterations from the fit of \cref{tab:heldout}) changes $D_H$ by $-1\times10^{-5}$ per pixel, below the seed-to-seed spread of $2\times10^{-4}$. A read noise of 30\,ADU raises it by $4\times10^{-4}$. In the free-mode pixel fit every read-noise setting widens the train--test gap. The detector of \cref{eq:detector}, with its fitted flat background, is therefore used for every fit.

A vacuum scan from an earlier session on the same detector calibrates the detector blur. With the pixel box resolved by $16\times16$ sub-samples per pixel, the bright-field disc has a hard edge seen through a Gaussian blur of $\sigma_d=0.173$\,px ($\sigma_d^2=0.0298\pm0.0001$\,px$^2$) at $\alpha=18.65$\,mrad. At coarser sub-sampling the blur and the edge width trade off, with deviances equal to $10^{-5}$ per pixel. The atom model fixes the aperture edge at 0.3\,px. This soft edge stands in for the blur at the disc edge, so its fitted $\sigma_d^2$ goes to 0.

\paragraph{Projector ellipse.} The bright-field rim is an ellipse: rim fits give semi-axes $R(1\pm\epsilon)$ with $\epsilon=0.0097$ and the major axis at $55.6^\circ$ from the detector $x$ axis, a distortion of the whole diffraction plane (it improves the dark field as well as the bright field, below). This is $1-b/a=1.9\%$. Bright-field disks on other aberration-corrected microscopes are elliptical by 1--4\% in the same measure. The projector and condenser settings set the ellipse. On one microscope, for each condenser aperture, its axis holds to within a few degrees over all camera lengths beyond the last projector-mode change, and its size to about a percentage point. It is therefore calibrated once per camera length and aperture. On the six scans of our session (about 90\,min, two magnifications, five defoci), the rim gives $\epsilon=0.0095$ to within $10^{-4}$, the axis at 55.0--55.2$^\circ$, and the same rim radius to 0.004 pixel. The value above, from the summed training patterns of the fitted sub-scan (896 of 1{,}024; held-out patterns do not enter), lies within the $1.3\times10^{-4}$ scatter of single sub-scans.
The atom model can carry it as a fixed, area-preserving stretch of detector frequencies, calibrated from the rim and never fitted: aperture, aberrations, propagator and band limits see the stretched frequencies, the potential is synthesized at the matching sheared frequencies, and the scoring disc and detector pixels stay unchanged, with no resampling; at $\epsilon=0$ every array is unchanged. Switching it on for the converged atom fit lowers $D_H$ from 0.4496 to 0.4318 without a refit (bright field 0.599 to 0.567, dark field 0.429 to 0.413); a refit gains another $6\times10^{-4}$ and leaves the train--test gap at 0.005 (\cref{tab:heldout}). The refit lowers the species widths by 0.08 (W) and 0.03\,pm (Se), within their sandwich bars, the rms atom displacement from the lattice from 0.41 to 0.32\,\AA, and the interlayer parameter from 6.74 to 6.69\,\AA. The free-mode pixel fit does not improve with the same correction, since its free modes absorb the distortion. The widths and bars quoted in the main text are those of the fit without the ellipse.

\section*{S15. Accuracy against published results}
We read atom positions from the published reconstructions as a user would, with no structural priors. We use Gaussian fits in 2-D to the projected phase or to the max-phase image, and Huang \etal's 3-D peak retrieval (voxel maxima of the phase refined to sub-voxel by a spline)~\cite{huang2026moire3d}. A 2-D peak cannot separate atoms closer than 0.35\,\AA\ in projection, so 2-D peaks are compared with the $\sigma V$-weighted centroids of such columns. Every set is aligned to its reference by one similarity transform (rotation and isotropic scale). The published simulated reconstruction used a scan step 0.9\% long. With translation-only alignment it scores 19--23\,pm against our 3.2\,pm; the difference is the step calibration, not the method. Values are rms, with ours on the same atoms or columns in brackets.

\begin{table}[hbt]\centering\small
\caption{Without priors, our atoms have the lowest in-plane error on the simulated WSe$_2$ bilayer from either start; prior-refined coordinates keep the lowest depth error. Rms error against truth (1{,}928 atoms).}
\label{tab:accsimS}
\resizebox{\columnwidth}{!}{\begin{tabular}{@{}llcccc@{}}\toprule
method & priors & in-plane (pm)\,$\downarrow$ & W\,$\downarrow$ & Se\,$\downarrow$ & depth W\,/\,Se (pm)\,$\downarrow$\\\midrule
published, 2-D, projected phase & none & 7.0 [3.0] & 7.1 & 7.3 & --\\
published, 2-D, max-phase image & none & 5.5 [3.0] & 5.9 & 5.3 & --\\
published, 3-D peaks & none & 4.8 [3.1] & 3.9 & 5.2 & 84\,/\,127\\
refined coordinates & bond, interlayer & 4.3 [2.9] & 3.7 & 4.5 & 5.3\,/\,5.5\\
ours, from published peaks & none & 3.2 & 2.8 & 3.4 & 50\,/\,49\\
ours, true layer geometry (\S S6) & none & 3.1 & 2.8 & 3.3 & 18.8\,/\,15.5\\\bottomrule
\end{tabular}}
\end{table}
\begin{itemize}
  \item With no priors on either side, our in-plane error is 3.1--3.2\,pm against 4.8\,pm ($1.9\times10^{-2}$ against $2.9\times10^{-2}\,\lambda/\alpha$), from either start. The advantage holds for both species, on overlapping columns, and in the outer 3\,\AA\ of the field (3.5 against 6.2--7.9\,pm).
  \item The priors move the published in-plane error only from 4.8 to 4.4\,pm~\cite{huang2026moire3d}. They act on depth, 84\,/\,127 to 5.3\,/\,5.5\,pm. From the published reconstruction our depth error is about 3 times below the raw published peaks and 5--7 times above the refined coordinates (\S S6).
  \item Our in-plane error is 10 times our Fisher standard deviation (0.34\,pm), and only 8\% of atoms lie within one. The published peak-fit standard deviation understates its error by 3.5 times. Both methods are limited by model or readout error, not by the counts.
\end{itemize}

\begin{table}[hbt]\centering\small
\caption{Real data: our atoms against peaks of the published phase, similarity-aligned (pm rms). $n$: matched columns. Interior: away from the edge of the fitted field. Last column: rms of the difference over the combined standard deviation per axis (our Fisher bound and the peak-fit error).}
\label{tab:accreal}
\resizebox{\columnwidth}{!}{\begin{tabular}{@{}lccccccc@{}}\toprule
dataset & $n$ & all\,$\downarrow$ & $\lambda/\alpha$\,$\downarrow$ & W\,$\downarrow$ & Se$_2$\,$\downarrow$ & interior\,$\downarrow$ & diff.\,/\,sd\\\midrule
main, 25\,mrad & 127 & 13.7 & 0.082 & 14.1 & 12.2 & 9.0 & 10.4\\
public, 10.3\,mrad & 445 & 19.2 & 0.079 & 17.6 & 19.3 & 14.0 & --\\
6$^\circ$ bilayer & 766 & 31.9 & 0.19 & 33.3 & 32.6 & 31.6 & --\\
6$^\circ$ bilayer, ideal-lattice start & 1{,}033 & 17.9 & 0.11 & & & 17.1 & --\\\bottomrule
\end{tabular}}
\end{table}
\begin{itemize}
  \item \textbf{25\,mrad.} Agreement is 13.7\,pm overall and 9.0\,pm on the 16 columns farthest from the edge of the free-atom box (111 of the 127 lie within 3\,\AA\ of it in a 13.3\,\AA\ field). The difference is 10 times the combined standard deviation. The two starts of our fit agree to 0.05\,pm rms on the in-box atoms, so the difference from the published peaks is model or readout error. The position bars at one optimum are therefore not a credible interval for positions on these data.
The width bars stand as in \S\S S3--S4.
  \item \textbf{Public.} Agreement is 14\,pm in the interior and 26.7\,pm in the outer 3\,\AA, where our fitted sub-scan ends.
  \item \textbf{6$^\circ$ bilayer.} Our atoms agree with the published phase worse than the ideal two-lattice start does (32 against 18\,pm). The fit moves the atoms by 42--49\,pm in random directions, its coherence parameters sit at their bounds ($\sigma_f$ = 120\,\AA, $\sigma_{\mathrm{src}}$ = 0.5\,\AA), and it trails the published reconstruction by 0.09 per pixel in $D_H$ (\S S10). The positions absorb instrument mismatch and are not accurate.
  \item \textbf{Where the two differ.} At the edges of the fitted field, at overlapping columns (25\,mrad: 17.1 against 12.3\,pm on isolated columns), and at low-amplitude Se$_2$ columns, 5--6\% of them, about the fraction of Se vacancies in~\cite{huang2026moire3d}. Our full-occupancy model does not widen those sites ($\sqrt{U_a}$ of Se 6.1 against 6.6\,pm elsewhere on the 6$^\circ$ bilayer). These sites need an occupancy parameter.
\end{itemize}
On real data and an independent simulator, model error, not counting noise, sets both methods' position error: 14--19\,pm on real data ($0.08\,\lambda/\alpha$), about 10 times our Fisher bars.

\paragraph{Spacing of the W planes.} We rotate each layer so its mean W plane is level, interpolate both W surfaces on a 0.5\,\AA\ grid, and take their difference. Published depths and interlayer distances come from refinement under bond-length constraints~\cite{huang2026moire3d}, so they are a constrained reference, not a measurement.
\begin{center}\small
\begin{tabular}{@{}lcc@{}}\toprule
set & W--W (\AA) & 5--95\% (\AA)\\\midrule
molecular dynamics, flat & 6.57 & 6.45--6.78\\
simulation, truth & 6.60 & 6.45--6.84\\
simulation, refined coordinates (priors) & 6.58 & 6.37--6.83\\
simulation, ours (no priors) & 6.62 & 6.28--6.88\\
6$^\circ$ bilayer, refined coordinates (priors) & 6.57 & 6.28--6.83\\
6$^\circ$ bilayer, ours (no priors) & 6.51 & 6.13--6.89\\\bottomrule
\end{tabular}
\end{center}
Without priors, the mean W--W spacing of our depth-resolved fit matches the simulation and agrees with the prior-refined coordinates on the 6$^\circ$ bilayer (the paper reports 6.2--6.9\,\AA\ between stacking regions~\cite{huang2026moire3d}). The Se--W spacings do not match. On the simulation one Se plane per layer shifts by about 17\,pm (1.49--1.86\,\AA\ against 1.64\,\AA), and on the 6$^\circ$ bilayer they stretch to 1.82--1.86\,\AA.

The 6.74\,\AA\ of \S S1 is a different quantity: the propagation distance $\Delta z$ between the two W slices of the six-plane model used for the 25\,mrad fits, in which each atomic plane is one thin slice and the Se planes sit at $\pm1.67$\,\AA\ from their W plane. It depends on the model: 6.738 and 6.735\,\AA\ for frozen phonons from two starts, 6.87\,\AA\ at iteration 40 of the per-atom-width refit, and 6.40--6.43\,\AA\ for the mean field at an earlier fit.
Its Fisher standard deviation is 0.013\,\AA\ ($2\times10^{-4}\,\lambda/\alpha^2$). Its offset from the nominal 6.5\,\AA\ spacing of the W planes is 0.24\,\AA, 0.0036 depths of field, a direction in which the counts are weakly informative. The fitted defocus of the 25\,mrad set is $\Delta f$ = $+57$\,\AA\ (5.7\,nm underfocus, \ie\ $C_1=-5.7$\,nm in the microscope's sign), against 7\,nm underfocus at acquisition.

\paragraph{Image level.} Our free-mode phase of the public data and the published phase, registered by rotation, scale and sub-pixel shift, correlate at 0.80 (0.79 after a 2\,\AA\ high-pass), and their Fourier ring correlation crosses 0.143 at 0.46\,\AA\ ($0.19\,\lambda/\alpha$). Both come from the same counts, so this measures agreement, not resolution. The two agree to about the information limit of either (\S S10).

\paragraph{Thermal widths.} An atom in a published phase image is 15--21\,pm wide (short axis), three to four times the thermal width, so the reconstruction sets the image width. On the 6$^\circ$ bilayer it is 20.7\,pm, no larger than on the static simulation (21.1\,pm), so the thermal width cannot be read from one published image. Image-based thermometry separates the two only through a temperature series~\cite{bae2026thermal}. Our $\sqrt{U_a}$ is fitted directly, with its bar.

\section*{S16. Identifiability by rank-revealing QR}
\paragraph{Setup.} The whitened Jacobian $\mathbf A$ stacks $\operatorname{diag}(h_j)^{-1/2}\mathbf J_j$ over the training positions, so $\mathcal I=\mathbf A^{\!\top}\mathbf A$ (\cref{eq:fisher}) with unit noise per count. Columns are forward-mode tangents with every parameter in its natural unit: positions and scan offsets in $\lambda/\alpha$, depths in $\lambda/\alpha^2$, $\log U$, aberrations in radians of phase at the aperture edge, and $\log\alpha$, $\log\sigma_f$, $\log\sigma_{\mathrm{src}}$. A patch centered on a W site, Se site or hollow takes the 372--375 training patterns within 5\,\AA\ and all parameters they see (about 1{,}260 with per-atom widths, 1{,}100 with one width per species), $2.1\times10^6$ rows. The specimen-to-scan shift (2) trades exactly with a common shift of a patch's atoms. Fixing the fit's 12 gauges (mean offset 2, scan polynomial 10; \cref{sec:inference}) and this shift removes 14 exact null directions and cuts the equilibrated $\kappa(\mathbf A)$ from $2.1\times10^6$ to 10--12.

\paragraph{Rule.} $R_{kk}^2$ is the information on pivot $k$ with the earlier pivots profiled and the later ones fixed; adding it gains about $R_{kk}^2/2$ in expected log-likelihood. The test $|R_{kk}|\ge c$ asks whether the data support a standard deviation of $1/c$ natural units ($c=1$ is of AIC type). Since $\mathbf A(D)=(D/D_0)^{1/2}\mathbf A(D_0)$ exactly, pivot $k$ is determined at the dose $D_k=D_0c^2/R_{kk}^2$.

\paragraph{Rank and dose.} With per-atom widths, at $D_0=5.1\times10^5$\,e/\AA$^2$, 1{,}200--1{,}213 of 1{,}246--1{,}271 parameters (95--96\%) pass $|R_{kk}|\ge1$; the widths of outer, partly seen atoms fail. At $0.1D_0$ more widths fail (91--92\%), and at $0.01D_0$ still only widths fail (88\%). These scalings assume the fitted background (0.024 electrons per pixel, 7\% of the median dark-field count) scales with dose. A fixed detector pedestal of that size would dominate dark-field pixels below about $0.1D_0$ and raise the low-dose figures. On all three patches the instrument pivots first and the widths last. With one width per species, as the paper fits, every parameter passes at all three doses.
\begin{center}\small
\resizebox{\columnwidth}{!}{\begin{tabular}{@{}lccc@{}}\toprule
parameter & $D_k$, $c=1$ (e/\AA$^2$) & target & $D_k$ at target (e/\AA$^2$)\\\midrule
W position & 0.6--0.9 & $0.1\,\lambda/\alpha$ & 61--94\\
Se position & 4.4--9.8 & $0.1\,\lambda/\alpha$ & 440--980\\
W depth, earlier fit & 0.5--1.3 & 1\,\AA & $2$--$6\times10^3$\\
Se depth, earlier fit & 2--4 & 1\,\AA & $1$--$2\times10^4$\\
W width, per atom ($\log U$) & $2.5\times10^3$--$5.3\times10^4$ & 10\% in $\sqrt U$ & $6\times10^4$--$1.3\times10^6$\\
Se width, per atom ($\log U$) & $3.1\times10^3$--$1.8\times10^4$ & 10\% in $\sqrt U$ & $8\times10^4$--$4.5\times10^5$\\
width, per species ($\log U$) & 100--140 & 10\% in $\sqrt U$ & $2.5$--$3.6\times10^3$\\\bottomrule
\end{tabular}}
\end{center}
These pivot-conditional doses, for a detector that counts every electron once, are lower bounds. With every other parameter profiled, the core positions reach $0.1\,\lambda/\alpha$ at 196--212\,e/\AA$^2$ (W) and 731--1{,}090\,e/\AA$^2$ (Se), above the 10--100\,e/\AA$^2$ that organic specimens tolerate. The two Se atoms of the core column lie 0.15\,\AA\ apart in the plane and trade positions, which raises the Se doses.
Per-atom widths to 10\% need 0.1--2.5 times the dataset dose, and one width per species about $3\times10^3$\,e/\AA$^2$. The depth rows come from an earlier fit, at which depth to 1\,\AA\ needs $2\times10^3$--$2\times10^4$\,e/\AA$^2$.

\paragraph{QR against Cholesky.} Worst relative error of $\operatorname{diag}\mathcal I^{-1}$ against refined double-precision QR, on the patches of an earlier fit:
\begin{center}\small
\resizebox{\columnwidth}{!}{\begin{tabular}{@{}lccccc@{}}\toprule
problem & $\kappa(\mathbf A)$ & Cholesky, FP64 & Cholesky, FP32 & QR, FP32 & Cholesky + 1 refinement\\\midrule
patch, gauges fixed & 12.4 & $4\times10^{-14}$ & $3$--$5\times10^{-6}$ & $2$--$4\times10^{-5}$ & $1$--$3\times10^{-11}$\\
full field, gauge held by the prior & $1.8\times10^3$ & $2\times10^{-10}$ & 3.9\% & $6\times10^{-5}$ & --\\
patch, gauges left in & $2.4\times10^6$ & $1.1\times10^{-3}$ & breakdown & 98\% & breakdown\\\bottomrule
\end{tabular}}
\end{center}
With the gauges fixed, every method works. The bad conditioning is all in the near-gauge of scan polynomial, offsets and global shift. Held only by a weak prior, FP32 Cholesky errs by $\epsilon_{32}\kappa^2$ and QR by $\epsilon_{32}\kappa$. With gauges left in, only the pivoted QR truncated at $|R_{kk}|\ge1$ degrades gracefully: it flags the gauges and gives finite bars conditional on them. The bars fix the mean-offset gauge exactly; the scan-polynomial gauge is held by its prior (bars without priors agree within 1.2\%). $\mathcal I$ is accumulated in FP64 from the FP32 Jacobian, Jacobi-scaled, and inverted by eigendecomposition on the gauge null space, with no eigenvalue truncated. Its condition number of $3.4\times10^6$ costs about $4\times10^{-10}$ relative accuracy.
A randomized interpolative decomposition~\cite{martinsson2011randomized,halko2011finding} finds the same skeleton as the strong rank-revealing QR~\cite{gu1996efficient} but costs 7.0--7.2 times its tangent passes at $D_0$, so it does not pay here.

\paragraph{Cost of the Fisher matrix.} Every column over every training pattern of the 25\,mrad sub-scan (1{,}636 columns, 896 patterns) costs about 390 CPU node-hours or 75 A100-hours (0.18\,s per column and pattern). A field the size of the 6$^\circ$ bilayer (14{,}336 patterns, 9{,}458 parameters) would cost about 6{,}800 A100-hours.
Local assembly fails. Width columns of fixed atoms outside the 13.3\,\AA\ scan draw information from the whole sub-scan through the 25\,\AA\ probe window. Truncating them changes the width bars by a median of 33\% (p90 183\%) at a 5\,\AA\ footprint, 3.4\% (44\%) at 6\,\AA\ and 1.1\% (23\%) at 8\,\AA. The dense matrix is therefore formed. Its factorization takes seconds on a GPU.

\paragraph{Instrument across datasets.} Ordering local columns first and shared ones ($C_s$, $\log\sigma_f$, $\log\sigma_{\mathrm{src}}$) last, the trailing $3\times3$ block $\mathbf R_{ss}^{(e)}$ of each dataset is the Schur complement (\cref{eq:schur}), and the joint factor is the QR of the stacked blocks. On the full 25\,mrad field the bound is 0.12\,\textmu m for $C_s$ (0.018\,rad at the aperture edge); on the $32\times32$ sub-scan the converged fit gives 0.20\% for $\sigma_f$ ($70.62$\,\AA) and 0.13\% for $\sigma_{\mathrm{src}}$ ($0.3203$\,\AA);
 at 18.4\,mrad 0.25--0.26\,\textmu m; for an uncorrected 200\,kV lens with $C_s=1.1$\,mm, 8.3\,\textmu m (7.5\,mrad) and 2.5\,\textmu m (10.3\,mrad). Pooling $N$ datasets narrows the coherence bounds only as $N^{-1/2}$. These are lower bounds: on real data model error, not information, limits the instrument.

\paragraph{Low-dose design.}
About 70\% of a patch's columns are per-pattern scan offsets. A low-dose scan can replace them with a cubic scan model (18 columns), freeze calibrated parameters, and take the instrument, including the second-order aberrations $\mathbf A_2$ and $\mathbf B_2$, and the widths from a high-dose calibration. The target scan then refits positions, the interlayer distance, defocus $C_1$, two-fold astigmatism $\mathbf A_1$, dose scale, background and descan. One Gaussian position jitter, a free $\log\sigma_{\mathrm{src}}$, catches the scan's blur. Each row adds to the one above. The last row is the alternative in which coma is refit:
\begin{center}\small
\resizebox{\columnwidth}{!}{\begin{tabular}{@{}lccccc@{}}\toprule
design & $n$ & passes & rank, 100\,e/\AA$^2$ & $D$, W / Se (e/\AA$^2$) & variance, W / Se\\\midrule
all profiled (species widths) & 1{,}084--1{,}104 & 342--358 & 87--89\% & 193--206 / 657--1{,}076 & 1 / 1\\
scan model & 358--374 & 342--358 & 62--67\% & 173--185 / 622--1{,}058 & 0.90 / 0.97\\
+ calibration and jitter (\cref{fig:calib}) & 349--365 & 333--349 & 62--67\% & 110--122 / 540--995 & 0.58 / 0.88\\
+ $C_1$ and $\mathbf A_1$ also held & 342--358 & 326--342 & 61--66\% & 110--122 / 535--993 & 0.58 / 0.87\\\midrule
$\mathbf A_2$, $\mathbf B_2$ refit & 353--369 & 337--353 & 62--67\% & 173--184 / 616--1{,}057 & 0.90 / 0.97\\\bottomrule
\end{tabular}}
\end{center}
$D$ is the dose at which the core W or Se position reaches $0.1\,\lambda/\alpha$ with every other estimated parameter profiled; the pivot-conditional $D_k$ does not change (61--94 and 440--980\,e/\AA$^2$). Variance is relative to the profiled design (median over the core coordinates). The design of \cref{fig:calib} estimates 3.02--3.11 times fewer parameters. The passes barely change: the offsets take two passes in any design, and one width per species adds only two columns. At 100\,e/\AA$^2$ every core position stays determined, and the position bounds tighten by 24\% (W) and 6\% (Se). The jitter leaves them unchanged and is determined to 12\% (relative standard deviation of the total Gaussian spread, 0.32\,\AA). Holding $C_1$ and $\mathbf A_1$ as well changes little (24\% and 7\%).
The gain comes from the axial coma $\mathbf B_2$, which trades with in-plane position. Holding $\mathbf B_2$ alone gives the same 24\% and 6\%, and $\mathbf A_2$ does not matter. If coma must be refit, the bounds tighten by only 5\% and 2\%. Holding it is valid when coma drifts by much less than the target scan can resolve. The target scan alone pins $\mathbf B_2$ only to about 30\,nm at 100\,e/\AA$^2$ (0.4\,nm at the dataset dose), and a coma error of that size biases the positions by about half a standard deviation. Coma drift is negligible between scans taken minutes apart and reaches about 20\,nm over an hour, a bias of a third of a standard deviation. The calibration scan is therefore taken next to the target in time as well as in space. Defocus and two-fold astigmatism drift in minutes and are refit~\cite{barthel2013optical}.

A frozen group off by $\delta\boldsymbol\theta_\nu$ biases the estimate by $-\mathcal I_{AA}^{-1}\mathcal I_{A\nu}\delta\boldsymbol\theta_\nu$ at every dose while the standard deviation falls as $D^{-1/2}$. At $D_0$ the tolerable errors are about the bounds, so the design suits low dose only. The binding assumption is the scan model: the fitted offsets are 10.6\,pm rms and a cubic model leaves 8.8--9.0\,pm, five times their 1.8\,pm bound. In the design of \cref{fig:calib} this residual biases core W positions by 0.15\,sd at 100\,e/\AA$^2$, 1.1 at $0.01D_0$ and 11 at $D_0$; the bias reaches half a standard deviation at about $1\times10^3$\,e/\AA$^2$ and one at $4\times10^3$\,e/\AA$^2$. A smooth scan model therefore suits organic doses, not the dataset dose.

\paragraph{Transfer to later scans.} Two later scans of the same session are fitted with the same pipeline. One is taken 70\,min after the first at the same magnification and dose. The other is taken 83\,min after it at 10 instead of 14.5\,Mx and half the dose. We express each change of a calibrated parameter in units of its low-dose threshold, the Fisher standard deviation of the first scan scaled to 100\,e/\AA$^2$. The slow aberrations that the design holds fixed transfer on both scans within one threshold: $\mathbf B_2$ changes by at most 0.01, $\mathbf A_2$ by at most 0.92, $C_s$ by at most 0.11 and $\alpha$ by at most 0.30. The terms it refits move, as the design assumes: defocus by 2.0 and 2.1 thresholds, $\mathbf A_1$ by up to 4.5, and the descan by up to 6.4. The coherence widths stay inside one threshold at the same magnification, the focal spread changing by 0.52 and the source size by 0.10. On the 10\,Mx scan they move by 1.15 (71 to 83\,\AA) and 1.86 (0.32 to 0.38\,\AA). The design holds the focal spread fixed, and its jitter term adds to the source size. The species widths are 7.78 and 10.18\,pm at the same magnification, close to the 7.35 and 8.98\,pm of the first scan. The 10\,Mx scan fits 14.4 and 11.6\,pm, so part of its coherence change may be blur that the widths absorb. Within a session, the coherence widths therefore transfer to within one or two thresholds.

\section*{S17. Algorithms}
The steps below give the order of operations, with the cost and settings of each step. $N_p$ is the number of grid points of the probe window, $S$ the number of slices, $K_m$ the number of probe modes, $K$ the number of frozen-phonon configurations, $|T|$ the training positions and $n_\theta$ the parameters.

\paragraph{(a) Forward renderer} (one batch of positions $j$).
\begin{enumerate}
  \item Potential. For each slice $n$, evaluate $\hat v_Z(\mathbf k;U_a)$ (\cref{eq:atompot}) at every frequency of the oversampled grid and sum over the atoms of the slice, one matrix product per slice. Alternatively, evaluate the real-space Gaussians by the fast Gauss transform to a requested tolerance. The width enters the Gaussian variance $b_i^Z/(8\pi^2)+U_a$, and $\partial/\partial U_a$ follows from \cref{eq:heat}. Cost: $O(N_p\log N_p)$ per slice plus $O(\text{atoms})$. General depth spreads each atom over the slices by a Gaussian profile of width $\sigma_z=0.4$\,\AA.
  \item Frozen phonons. Draw $K=32$ fixed standard-normal displacement sets $\boldsymbol\epsilon_{a\kappa}$ in antithetic pairs and build $V_n^{(\kappa)}$ (\cref{eq:configs}); the mean field is $K=1$ with \cref{eq:atompot}. Cost: $K$ times steps 1, 3 and 4.
  \item Probe. Form $A\,e^{-i\chi}$ with the aberrations of \cref{eq:chifull}, the partially coherent copies of \cref{eq:coherence} on a product grid over $\pm3.5$ standard deviations, and compress them to the leading $K_m$ eigenmodes of their Gram matrix. This replaces hundreds of coherent copies by tens of modes: at least 32, holding all but $10^{-6}$ of the trace. The modes are refreshed every 50 steps during Adam and held fixed within each L-BFGS polish, step (b)3 (Fisher coherence columns: trace $\ge1-10^{-5}$, \S S16). Within a polish the eigenvector coefficients are frozen, so each mode is a fixed combination of the analytic copies. Gradients in $\chi$ are exact, since $\chi$ leaves the Gram matrix unchanged; gradients in $\alpha$, $\sigma_f$, $\sigma_{\mathrm{src}}$ are those of the frozen basis, corrected at each refresh. The modes have unit total intensity on the window. Shift each mode to $\mathbf r_j$ by a phase ramp.
  \item Multislice. For $n=1,\dots,S$: multiply by $O_n=e^{i\sigma V_n}$, band-limited to 2/3 of Nyquist, and propagate with $p_{\Delta z}$ (\cref{eq:multislice}). Modes are stacked along the batch axis. Cost: $2S$ FFTs of size $N_p$ per mode, position and configuration, $O(K K_m S N_p\log N_p)$ per position.
  \item Detector. Apply the descan ramp to $\psi_{j,S}$ ($\mathbf g$ in pixels times $1.666\,\text{mrad}/\lambda$) on a 50\,\AA\ window, twice the detector's conjugate cell, so its wrap at the window edge costs less than 0.02 in $\chi^2$ per pattern at the fitted $\mathbf g$. Take $|\mathcal F[\cdot]|^2$, sum the weighted modes and average the $K$ patterns, integrate over each detector pixel (exact integration at bin 2; bin 3 agrees to $3\times10^{-5}$), convolve with the Gaussian PSF $\hat d$ (variance fitted; 0 on the 25\,mrad data), scale by $(N_e-N_{\mathrm{px}}b)\eta$ and add $b$, held in $[0, N_e/(2N_{\mathrm{px}})]$ (\cref{eq:detector}). A smooth floor of $5\times10^{-4}$ electrons keeps $h_j>0$. Read noise is not in the fitted model (see \S\S S12 and S14). Cost: $O(N_p\log N_p)$ per position.
\end{enumerate}

\paragraph{(b) Fitting.}
\begin{enumerate}
  \item Initialize detector and scan parameters from a pixel fit with the same instrument model. Atoms start from the lattice, widths from a shared value per species.
  \item Adam on mini-batches of 32--512 training positions with a cosine learning-rate schedule (60 epochs by default), for fits from a coarse start; refits from a converged fit go straight to the polish. Loss: \cref{eq:likelihood} over $T$ and the detector pixels inside the band-limited disc, plus Gaussian priors (atom displacements 0.3\,\AA; scan polynomial; the aberrations $\mathbf A_1$, $\mathbf A_2$, $\mathbf B_2$ and $C_s$, with none on the defocus; $\log\sigma_f$, $\log\sigma_{\mathrm{src}}$), and offsets $|\boldsymbol\delta_j|\le1$\,\AA. Cost per epoch: one forward and one reverse pass over $T$, about twice step (a).
  \item Polish: full-batch L-BFGS with a strong-Wolfe line search (history 20), in blocks of 5 iterations and segments of 20. On real data every parameter is scaled by its Fisher standard deviation from a dense $K=8$ Fisher matrix computed once at an earlier nearby fit; the simulated-bilayer fits scale each parameter group by a fixed constant. Up to three polish rounds are run. The probe-mode basis is frozen within a round and refreshed between rounds, until a refresh changes the objective by less than 1 nat. Stop when the squared Newton decrement $\mathbf g^{\!\top}\mathcal I^{-1}\mathbf g$ is below 0.1 and every step is below 0.1 Fisher standard deviations, both on training patterns only. Here $\mathbf g$ includes the priors and $\mathcal I$ the prior precision. One start meets this rule (0.006, largest step 0.04\,sd). The other ends at 0.18, with its largest steps 0.38 and 0.34\,sd in $\log\sigma_{\mathrm{src}}$ and $\log\sigma_f$ and at most 0.04\,sd in any atom parameter. Its positions and widths are converged; its coherence widths are not.
  \item Label exchanges. Two atoms of one species whose projected positions coincide share a column, and only the weak depth contrast of the multislice decides which layer holds which. A continuous optimizer cannot swap them: the straight path passes through the coincident configuration. The exchange screen evaluates every such swap from the current point on the training patterns near the column (one frozen probe basis, the same frozen-phonon draws before and after) and accepts improving swaps with disjoint pattern sets. An accepted set is kept only if the full training objective drops by more than a threshold per swap, both at the fitting configurations and at a phonon seed never used for scoring. The relaxed exchange test imposes a candidate swap and refits from it, keeping the swap if the refit ends lower.
  \item Numerics: the slice loop is checkpointed, so memory is that of one slice per mode and position plus the stored slice boundaries. Waves are single-precision complex, and the likelihood is accumulated in double precision. Double precision throughout is an option, used for derivative checks. Fit state is checkpointed for restart.
\end{enumerate}

\paragraph{(c) Held-out protocol and score.}
\begin{enumerate}
  \item Split: $H=\{j:(j_1+3j_2)\bmod 8=0\}$, one position in eight, each with four training neighbors. $T$ is the rest. Baselines see only $T$.
  \item Neighbor rule: a held-out position takes the mean per-position parameters (offset $\boldsymbol\delta_j$) of its four scan neighbors. An edge position uses the opposite neighbor in place of the missing one.
  \item Score $D_H$ of \cref{eq:score} on the band-limited disc, with a model floor of $10^{-9}$. Report $D_H$ and $D_H-D_T$. Cost: one forward pass over $H$.
\end{enumerate}

\paragraph{(d) Fisher information, rank and error bars.}
\begin{enumerate}
  \item Tangents. One forward-mode pass per non-offset parameter gives a column $\mathbf J_{:,k}$ over all training patterns. Each pattern sees only its own offset $\boldsymbol\delta_j$, so all offset columns come from two combined passes, one per axis. This gives $p=n_\theta+2$ passes in all, where $n_\theta$ counts the non-offset parameters. A product $\mathcal I\mathbf v$ costs one tangent and one adjoint pass (1.7\,s against 1.2\,s for the gradient on 4{,}096 SrTiO$_3$ patterns, one A100).
  \item Whiten: $\mathbf A=[\operatorname{diag}(h_j)^{-1/2}\mathbf J_j]_{j\in T}$, every parameter in its natural unit (\S S16). Remove the exact gauge directions.
  \item Offsets: eliminate each pattern's two offsets exactly by a $2\times2$ Schur complement (\cref{eq:schur}), so the remaining matrix is that of atoms and instrument.
  \item Identifiability: factor $\mathbf A\boldsymbol\Pi=\mathbf Q\mathbf R$ by column-pivoted (strong rank-revealing) QR in double precision. Cost $O(mn^2)$, seconds on a GPU at $n\approx9{,}500$. The passes of step 1 dominate.
  \item Rank: pivot $k$ is determined when $|R_{kk}|\ge1$.
  \item Error bars: accumulate $\mathcal I=\mathbf A^{\!\top}\mathbf A$ in double precision from the single-precision columns, scale it by its diagonal, and invert it through its symmetric eigendecomposition on the null space of the gauge constraints. The Schur complement over the instrument gives the profiled bars, and $\mathrm{sd}(\sqrt{U_a})=\mathrm{sd}(U_a)/(2\sqrt{U_a})$.
  \item Misspecification: the cluster-by-pattern sandwich bars of the two species widths are 4.5 (W) and 3.2 (Se) times their Laplace bars (\S S4). The ratio measures how far the species-width model is from the data.
  \item Dose map: $\mathbf A$ scales as $\mathrm{dose}^{1/2}$, so pivot $k$ reaches sd $1/c$ at $D_k=D_0c^2/R_{kk}^2$ from the same factorization.
\end{enumerate}

\paragraph{(e) Coverage test.}
\begin{enumerate}
  \item Generate expected counts from the fitted model at the measured dose (and at fractions $\rho$ of it).
  \item Draw Poisson counts, and refit each redraw from the truth with the same objective (42 redraws at full dose, 47 at half, after dropping the refits that do not converge).
  \item Compute $\zeta=(\hat\theta_k-\theta_k^{\mathrm{true}})/\mathrm{sd}$ with the Fisher sd at the truth, and tabulate the fractions within 1 and 2\,sd, sd$(\zeta)$ and mean $\zeta$ (\S S4). Cost: one fit per redraw.
\end{enumerate}

\paragraph{(f) Separating calibration from imaging.} The parameters split into groups that a measurement determines at different doses (\S S16). The instrument (aberrations, coherence, aperture, detector) and the thermal widths need a high dose, the atom positions a low one.
\begin{enumerate}
  \item Calibrate: fit everything, steps (a)--(b), on a high-dose scan of an area next to the target, or of a radiation-hard specimen at the same imaging conditions. The gun (source size, energy spread), the detector, the aperture, $C_s$ and the second-order aberrations $\mathbf A_2$ and $\mathbf B_2$ transfer while the corrector is not retuned; on a second scan of the same session the coherence widths hold only to one or two low-dose thresholds (\S S16). Coma drifts negligibly over minutes and by up to about 20\,nm in an hour~\cite{barthel2013optical}. Defocus and two-fold astigmatism $\mathbf A_1$ drift in minutes and are refit on the target, with tilt, thickness, dose scale, background and descan. The thermal widths, a property of the material, come from a prior fit, tables or simulation. Repeating the calibration scan after the target bounds the drift.
  \item Image: scan the target at low dose and fit only the positions, a smooth scan model, and the drifting terms of step 1, steps (a)--(b), with the calibrated parameters held fixed except one Gaussian position jitter. A Gaussian spread of probe positions adds to $\sigma_{\mathrm{src}}$ in quadrature. With the instrument calibration transferred, this one per-scan term catches session blur (stage vibration, drift jitter) that would otherwise widen every atom. It acts on the probe, so unlike a blur of the potential it is not degenerate with the mean of the per-atom widths.
  \item Bound: compute the error bars of the positions, step (d), with the fixed parameters as known, and the bias a calibration error would cause.
\end{enumerate}
A pixel reconstruction estimates every pixel of the object and the probe from the low-dose counts; here they determine about three numbers per atom. The bounds assume the model is right; experiments at these doses are the test.

\paragraph{(g) Optimizer ablation.} The solver has three parts: Adam on mini-batches, a full-batch L-BFGS polish, and the scaling of parameters in that polish. On the simulated bilayer the ablation holds the model, data, start and held-out set fixed and changes only the optimizer: Adam alone for the same wall time; Adam then L-BFGS in raw units; and the production solver (Adam, then L-BFGS with per-group scales). Each arm is scored by training objective, held-out $D_H$ and error against truth, against wall time.
Two effects are already measured. On real data, Fisher scaling decides whether per-atom widths move at all: unscaled, L-BFGS leaves 98\% of the 1{,}702 widths within 1\% of their start after 400 iterations; scaled, 66\% of the interior widths move within 40 iterations and the held-out deviance falls (\S S3). On the simulated bilayer, 10 Adam epochs (1.96\,h) reach 3.3\,pm in-plane but leave depth near its start, and the full-batch polish moves each atom toward its likelihood minimum in depth (\S\S S6, S11).

\section*{S18. Exact Foldy--Lax reference}
\paragraph{Method.} The reference is Foldy--Lax multiple scattering with partial-wave T-matrices. Each atom carries the thermally smeared Peng potential of \cref{eq:atompot}, tapered to zero at 7\,\AA, in the relativistically corrected scalar wave equation. Its phase shifts $\delta_l$ come from the variable-phase equation (fourth-order Runge--Kutta), giving a T-matrix diagonal in $l$, $T_l=(e^{2i\delta_l}-1)/2$. The optical theorem holds to nine digits, and \code{ELSEPA}~\cite{salvat2005elsepa} agrees on the same potential to $3.6\times10^{-4}$ in the dark field. Foldy--Lax~\cite{foldy1945multiple,lax1951multiple} then makes the wave incident on each atom of a column the beam plus the outgoing waves of all others, coupled by outgoing-to-regular translations.
GMRES, preconditioned by the forward-only sweep, solves it to a residual of $10^{-12}$ in 0--36 iterations; dense LU agrees to $2\times10^{-15}$.

\paragraph{Back-scattering and multislice.} The backward-hemisphere fraction is below $10^{-14}$ over all columns for the tabulated potential (evaluated four times beyond its fit range). A Coulomb atom gives $1.6\times10^{-4}$\,\AA$^2$ for W at 80\,kV, and the conclusion stands. Dropping the backward coupling changes the dark-field amplitude by at most $3\times10^{-13}$. Forward-only, the system is lower triangular, solved in one sweep: multislice with one atom per slice.
\paragraph{Agreement.} Dark-field amplitude error of the multislice against the exact reference, relative $L_2$, 80\,kV, mean-field atoms unless marked, reference sampling 0.05\,\AA:
\begin{center}\small
\resizebox{\columnwidth}{!}{\begin{tabular}{@{}lcccc@{}}\toprule
column & 1\,\AA\ grid & 0.25\,\AA\ grid & slab, 0.0625\,\AA & eikonal-exact\\\midrule
W (2 atoms, 6.49\,\AA) & 3.3\% & 0.61\% & 0.93\% & 0.18\%\\
Se (4 atoms) & 1.0\% & 0.30\% & 0.39\% & 0.17\%\\
W, one static atom ($U=0$), one projected slice & \multicolumn{4}{c}{2.2\% (0.58\% with the thermal width on)}\\\bottomrule
\end{tabular}}
\end{center}
For the bilayer's columns the multislice is accurate to 1\% (0.3\% eikonal). With frozen phonons at the production sampling, against \code{abTEM}, one slice per atomic plane errs by 2.7\% (mean field) and 6.0\% (static configurations, dark-field power 4--7\% low).

\paragraph{Potential.} The Peng \etal~\cite{peng1996robust} factors are a fit. \code{ELSEPA}'s Dirac--Fock atom with a finite nucleus is the physical reference. In the dark field they differ for W by $+1.6\%$, $-3.2\%$ and $-4.8\%$ at 80, 200 and 300\,kV, and for Se by $-1.0\%$ to $-1.2\%$. At 80\,kV this error is comparable to slicing error.
\section*{Data and code}
The public bilayer and simulated datasets are on Zenodo~\cite{nguyen2024uncorrected,huang2026moire3d}, and the PrScO$_3$ data on PARADIM (CC BY-NC-ND)~\cite{paradim_ssmm2j11}. The 25\,mrad data are those of Zhang \etal~\cite{zhang2025phasons}; the 18.4\,mrad scan is unpublished data from the same microscope. Code will be released under a suitable research license on acceptance of the paper.

\section*{Limitations (full list)}

\begin{itemize}
  \item Bounds assume the model is right. Thermal diffuse scattering and partial coherence are the remaining model errors.
  \item Fits are slower than a learned network (minutes against seconds). The learned output is a good warm start.
  \item The detector blur $\sigma_d^2$ of \cref{tab:params} is an effective blur. On specimen data it can also absorb small-angle inelastic scattering, which the model omits.
  \item Se bounds assume full occupancy. About 5\% of Se sites are vacant in these samples, beam-induced at 80\,kV and confined to the outer Se planes~\cite{huang2026moire3d}. A vacancy trades against a wider Se site, so Se widths need an occupancy parameter per site (W unaffected).
  \item Ellipse (anisotropy) bounds assume a rigid scan and a flat layer. Drift is about 1\,\AA\ per minute, \ie\ 0.3\,\AA\ over a 16\,s scan (0.5\% of the field), and moir\'e corrugations bend each layer~\cite{huang2026moire3d}. Both can mimic anisotropy. Per-region tilt and drift parameters would absorb both and are not fitted here.
  \item Per-site widths on the public 10.3\,mrad data are effective column widths. The W and Se$_2$ sublattices are not told apart, and pairs closer than 0.4\,\AA\ merge into one column.
  \item The simulated benchmark is static and coherent. It tests positions and the instrument model, not thermal widths, which are checked against molecular dynamics and thermometry.
  \item Atoms are fitted only in a thin specimen (0.15 depths of field). The thick specimens (PdH, PrScO$_3$, PZT, 4.9--12.8 depths of field) are fitted with a pixel object. Thick specimens matter: specimens cut from bulk are tens of nanometers thick, and there multislice ptychography reached the lattice-vibration limit~\cite{chen2021latticevib}. Atoms are harder there: a column holds tens of atoms, the wave channels along it, and its atoms lie ten times closer in depth than the depth of field (3.95 against 41\,\AA\ for PZT), so the counts barely fix one atom's depth. A column must then be parameterized jointly, \eg\ as a lattice with displacements.
\end{itemize}

\end{document}